\documentclass[11pt]{article}
\usepackage{comment}
\usepackage{hyperref}
\usepackage{enumitem}
\usepackage{amsmath,amssymb,epsf,cite,graphicx,subfigure}
\numberwithin{equation}{section}
\newcommand{\beq}{\begin{equation}}
\newcommand{\eeq}{\end{equation}}
\newcommand{\bea}{\begin{eqnarray}}
\newcommand{\eea}{\end{eqnarray}}
\newcommand{\bdm}{\begin{displaymath}}
\newcommand{\edm}{\end{displaymath}}

\def\<{\langle}
\def\>{\rangle}

\newcommand{\hiddensubsubsection}[1]{\stepcounter{subsubsection}\subsubsection*{\arabic{section}.\arabic{subsection}.\arabic{subsubsection}\hspace{1em}{#1}}}

\def\a{\alpha}

\def\d{\delta}
  
\def\g{\gamma}

\def\s{\sigma}                                   

\def\p{\partial}

\newcommand{\dph}{{\delta \varphi}}

\newcommand{\tr}{{\rm{tr}}}

\def\N{{\mathcal{N}}}

\def\We{\wp}

\def\sren{S_{\textrm{ren}}}

\def\Dv{\vec{D}}
\def\Bv{\vec{B}}
\def\Ev{\vec{E}}
\def\Hv{\vec{H}}

\begin{document}
\baselineskip=15.5pt
\pagestyle{plain}
\setcounter{page}{1}

\begin{flushright}
{CCTP-2012-18, DAMTP-2012-63\\ Imperial/TP/2012/JE/01, SU/ITP-12/25}
\end{flushright}

\vskip 0.8cm

\begin{center}
{\LARGE \bf Holographic Wilson Loops, Dielectric Interfaces, and Topological Insulators}
\vskip 0.5cm

John Estes$^{1,2,a}$, Andy O'Bannon$^{3,b}$, Efstratios Tsatis$^{4,c}$ and Timm Wrase$^{5,d}$

\vspace{0.1cm}

{\it ${}^1$ Blackett Laboratory, Imperial College, London, SW7 2AZ, United Kingdom \\}

\vspace{0.1cm}

{\it ${}^2$ Institute of Theoretical Physics, KULeuven, Celestijnenlaan 200D B-3001 Leuven, Belgium  \\}

\vspace{0.1cm}

{\it ${}^3$ Department of Applied Mathematics and Theoretical Physics, University of Cambridge, \\
Cambridge CB3 0WA, United Kingdom \\}

\vspace{0.1cm}

{\it ${}^4$ Department of Engineering Sciences, University of Patras, 26110 Patras, Greece
 \\}

\vspace{0.1cm}

{\it ${}^5$ Stanford Institute for Theoretical Physics, Stanford University, \\
Stanford, CA 94305, United States \\}

\vspace{0.1cm}

{\tt  ${}^a$johnaldonestes@gmail.com, ${}^b$A.OBannon@damtp.cam.ac.uk,\\ ${}^c$etsatis@upatras.gr,
${}^d$timm.wrase@stanford.edu} \\

\medskip

\end{center}

\vskip0.3cm

\begin{center}
{\bf Abstract}
\end{center}
We use holography to study (3+1)-dimensional $\N=4$ supersymmetric $SU(N_c)$ Yang-Mills theory (SYM) in the large-$N_c$ and large coupling limits, with a (2+1)-dimensional interface where the Yang-Mills coupling or $\theta$-angle changes value, or ``jumps.'' We consider interfaces that either break all supersymmetry or that preserve half of the $\N=4$ supersymmetry thanks to certain operators localized to the interface. Specifically, we compute the expectation values of a straight timelike Wilson line and of a rectangular Wilson loop in the fundamental representation of $SU(N_c)$. The former gives us the self-energy of a heavy test charge while the latter gives us the potential between heavy test charges. A jumping coupling or $\theta$-angle acts much like a dielectric interface in electromagnetism: the self-energy or potential includes the effects of image charges. $\N=4$ SYM with a jumping $\theta$-angle may also be interpreted as the low-energy effective description of a fractional topological insulator, as we explain in detail. For non-supersymmetric interfaces, we find that the self-energy and potential are qualitatively similar to those in electromagnetism, despite the differences between $\N=4$ SYM and electromagnetism. For supersymmetric interfaces, we find dramatic differences from electromagnetism which depend sensitively on the coupling of the test charge to the adjoint scalars of $\N=4$ SYM. In particular, we find one special case where a test charge has vanishing image charge.

\newpage

\tableofcontents

\section{Introduction and Summary} \label{intro}

$\N=4$ supersymmetric $SU(N_c)$ Yang-Mills (SYM) theory in (3+1)-dimensions is a conformal field theory (CFT), meaning the theory is invariant under the (3+1)-dimensional conformal group, $SO(4,2)$. In the space of deformations of $\N=4$ SYM, a special subset preserve (2+1)-dimensional conformal symmetry, \textit{i.e.}\ an $SO(3,2)$ subgroup of $SO(4,2)$. For example, consider $\N=4$ SYM on a half-space, that is, half of $\mathbb{R}^{3,1}$ with a boundary that is $\mathbb{R}^{2,1}$. Suppose we take two copies of this theory, with different values of $g$, and glue them together along their boundaries. We thus obtain a (3+1)-dimensional theory with a coupling that jumps in one spatial direction, say the $x_3$ direction. We will call the location of the jump, which we may set to $x_3=0$, an interface. Clearly for such a system translational symmetry in $x_3$ is broken, nevertheless, the theory preserves translational symmetry along the interface and, with appropriate boundary conditions, an entire $SO(3,2)$ subgroup of $SO(4,2)$, namely the subgroup that leaves the (2+1)-dimensional interface invariant. We thus obtain a \textit{conformal} interface, or, following standard parlance, an ``Interface CFT''~\cite{Bak:2003jk,Clark:2004sb,D'Hoker:2006uv,Azeyanagi:2007qj,Gaiotto:2008sd}.\footnote{A CFT in which a defect preserves some conformal symmetry is called a ``defect CFT''~\cite{Azeyanagi:2007qj}. Generically some degrees of freedom may be localized to the defect. An example of such a defect CFT in string theory is the low-energy theory describing the (2+1)-dimensional intersection of D3-branes with D5- and/or NS5-branes, $\N=4$ SYM with constant (non-jumping) coupling coupled to (2+1)-dimensional hypermultiplets that preserve an $SO(3,2)$ subgroup of $SO(4,2)$~\cite{Karch:2000gx,DeWolfe:2001pq,Erdmenger:2002ex,D'Hoker:2007xz}. Interface CFTs are special cases of defect CFTs in which a coupling constant jumps at an interface, and no degrees of freedom are localized to that interface.}

Examples of interface CFTs include $\N=4$ SYM theory with a jumping coupling, as well as related theories obtained by S-duality, or more generally by $SL(2,\mathbb{R})$ transformations, including for example $\N=4$ SYM with a constant coupling but a jumping $\theta$-angle. A jumping coupling breaks all supersymmetry (SUSY)~\cite{Clark:2004sb}, and hence its $SL(2,\mathbb{R})$-duals also do. Various amounts of SUSY may be restored by adding operators of the $\N=4$ SYM theory localized entirely at the interface~\cite{Clark:2004sb,D'Hoker:2006uv,Kim:2008dj,Gaiotto:2008sd,Kim:2009wv}. Notice that adding interface-localized operators does not introduce any new propagating degrees of freedom at the interface.

What physical effects does a jumping coupling have? To gain some intuition, consider (3+1)-dimensional Maxwell theory. Let us consider the Maxwell action first in non-relativistic form, with electric and magnetic fields $E^i$ and $B^i$, with $i = 1,2,3$, and electric permittivity $\epsilon$ and magnetic permeability $\mu$. In general, $\epsilon$ and $\mu$ may each jump independently, but if we impose relativistic invariance, so that the speed of light $c = 1/\sqrt{\epsilon \mu} \equiv 1$ is constant, and convert to relativistic notation, introducing the field strength $F^{\mu\nu}$ with $F^{i0} \equiv E^i$ and $F^{ij} \equiv - \epsilon^{ijk} B^k$, then we find
\beq
\frac{\epsilon}{8\pi} E^i E^i - \frac{1}{8\pi \mu} B^iB^i = \frac{\epsilon}{8\pi} \left [ E^iE^i - B^iB^i \right] = -\frac{1}{4g^2} F^{\mu\nu} F_{\mu\nu},
\eeq
where $1/g^2 = \epsilon/4\pi$. A jumping coupling $g$ in the relativistic theory thus represents an $\epsilon$ and $\mu$ that jump simultaneously in such a way that $c$ remains constant. A jumping $\epsilon$ occurs at the interface between two materials with different dielectric constants. The main physical effect of a dielectric interface is to induce image charges: a test charge on one side of the interface will experience a potential equivalent to that produced by a fictitious image charge on the opposite side.

Remarkably, the holographic duals for many conformal interfaces in large-$N_c$, strongly-coupled $\N=4$ SYM have been found~\cite{Bak:2003jk,Clark:2005te,D'Hoker:2006uu,Gomis:2006cu,D'Hoker:2007xy,D'Hoker:2007xz,Aharony:2011yc,Assel:2011xz,Suh:2011xc}, and generically are deformations of the $AdS_5 \times S^5$ solution of type IIB supergravity \cite{Maldacena:1997re,Witten:1998qj,Gubser:1998bc} that have only an $SO(3,2)$ isometry in the non-compact directions, \textit{i.e.}\ an $AdS_4$ factor. For example, $\N=4$ SYM with a jumping coupling is dual to type IIB supergravity formulated on the so-called ``Janus'' spacetime~\cite{Bak:2003jk}.\footnote{To be more precise, $\N=4$ SYM with a jumping coupling and a particular interface-localized operator is dual to the Janus solution, as we discuss in detail in section~\ref{interfaces}.} The Janus solution breaks all SUSY but preserves an $AdS_4$ factor as well as the entire $S^5$, and includes a non-trivial dilaton whose value jumps at the AdS boundary, which is the holographic representation of the jumping coupling. $SL(2,\mathbb{R})$ transformations produce solutions with an axion that jumps at the boundary, dual to $\N=4$ SYM with a jumping $\theta$-angle. All half-BPS conformal defect solutions of type IIB supergravity were found in refs.~\cite{Gomis:2006cu,D'Hoker:2007xy,D'Hoker:2007xz,Aharony:2011yc,Assel:2011xz}, including a SUSY version of Janus.\footnote{Recall that here a half-BPS solution preserves eight Poincar\'e supercharges and eight superconformal generators. 1/8-BPS Janus solutions are also known~\cite{Clark:2005te,D'Hoker:2006uu,Suh:2011xc}. In what follows ``SUSY Janus'' always refers to the half-BPS Janus solution.}

Our goal in this paper is to compute the self-energy of a single heavy test charge, as well as the potential between heavy test charges, in four interface CFTs: $\N=4$ SYM with a jumping coupling or a jumping $\theta$-angle (but never both), with and without SUSY.

To be more precise, we will compute the expectation values of Wilson loops in the fundamental representation of $SU(N_c)$, which represent the phase acquired by an infinitely heavy test quark traversing the loop. We will consider two kinds of loops. The first is the straight time-like Wilson line (technically a loop only if we include the ``point at infinity''), which in the infinite-time limit gives the self-energy of a single heavy test quark. The second is the rectangular Wilson loop, representing a heavy quark and anti-quark, which in the infinite-time limit gives the potential between the quark and anti-quark. To be still more precise, we will actually compute Maldacena loops~\cite{Rey:1998ik,Maldacena:1998im}, which involve not only the $\N=4$ SYM gauge fields but also the adjoint scalars. We will refer to the Maldacena loop as a Wilson loop, unless stated otherwise. In all cases, we expect the interface to modify the result of the undeformed $\N=4$ SYM theory. For example, we expect a jump in the coupling to act much like a dielectric interface in ordinary classical electrodynamics, in the sense that a self-energy or potential will include effects from image charges.

In $\N=4$ SYM without an interface, the (renormalized) expectation value of the straight Wilson line is trivial, indicating that a single heavy test quark has vanishing self-energy. This very special property of $\N=4$ SYM follows from SUSY, specifically the fact that the straight Wilson line is half-BPS, which prevents radiative corrections to its expectation value~\cite{Drukker:1999zq,Erickson:2000af}.

As for the rectangular Wilson loop in $\N=4$ SYM without an interface, the distance $L$ between the quark and anti-quark is the only scale in the problem, hence dimensional analysis dictates that the potential be Coulombic, $V(\lambda,L) = f(\lambda)/L$, and the non-trivial information is the dependence on $\lambda$, \textit{i.e.}\ the function $f(\lambda)$.\footnote{At leading order in the large-$N_c$ expansion the potential will not depend on $N_c$.} At large $\lambda$, the Wilson loop is described holographically by a string hanging down into $AdS_5$ with both endpoints at the boundary. The on-shell action of that string gives the Wilson loop expectation value~\cite{Rey:1998ik,Maldacena:1998im}, and hence the potential:
\beq
\label{eq:maldacenapotential}
V(\lambda,L) = - \frac{4 \pi^2}{\Gamma\left(1/4\right)^4} \frac{\sqrt{2\lambda}}{L}, \qquad \textrm{$\lambda \gg 1$ holographic result.}
\eeq
At small $\lambda$, the Wilson loop expectation value may be computed straightforwardly in perturbation theory. In fact, the sum of all planar diagrams without internal vertices, the ladder diagrams, gives~\cite{Erickson:1999qv,Erickson:2000af} (in each case, only the leading term is shown)
\beq
\label{eq:ladderpotential}
V(\lambda,L) = \begin{cases} - \frac{1}{4\pi} \frac{2\lambda}{L}, & \lambda \ll 1 \\ & \\ - \frac{1}{\pi} \frac{\sqrt{2\lambda}}{L}, & \lambda \gg 1. \end{cases} \qquad \textrm{sum of ladder diagrams.}
\eeq
When $\lambda \gg 1$, the dependence on $\lambda$ is the same as the holographic result, although the numerical coefficient is different.\footnote{As emphasized in refs.~\cite{Erickson:1999qv,Erickson:2000af}, in the ladder summation at leading order in the $\lambda \gg 1$ limit, the existence of a $\sqrt{\lambda}$ factor is gauge-independent, however the coefficient of this factor is gauge-dependent. The result in eq.~\eqref{eq:ladderpotential} is quoted in Feynman gauge.} In other words, the sum of ladder diagrams contain some, but not all, contributions to the potential at strong coupling. The leading behavior in $\lambda$ changes from the weak-coupling factor of $\lambda$ to the strong-coupling factor of $\sqrt{\lambda}$ due to screening effects~\cite{Rey:1998ik,Maldacena:1998im}.

For $\N=4$ SYM with a conformal interface, we can use the $SO(3,2)$ symmetry to constrain the forms of the self-energy and potential. The $SO(3,2)$ is a \textit{subgroup} of $SO(4,2)$, so in particular the $SO(3,2)$ dilatation generator is that of the original $SO(4,2)$, and as such acts on all of $x_1$, $x_2$, and $x_3$. The interface spans the $x_1$ and $x_2$ directions, and sits at $x_3=0$.

In the presence of the interface, the expectation value of a single straight Wilson line will depend only on its position in the $x_3$ direction, $x_3=L_3$. We expect a single test charge to induce an image charge on the opposite side of the interface, a distance $|L_3|$ from the interface. We can then express the self-energy of the test charge as a potential $V$ between the test charge and its image. Due to scale invariance, $V \propto 1/(2|L_3|)$, with a coefficient that depends on the strength of the image charge, and that in our case must go to zero if the jump in the coupling or $\theta$-angle goes to zero.

\begin{figure}[ht!]
\begin{center}
\subfigure[]{\fbox{\includegraphics[height=2.2in]{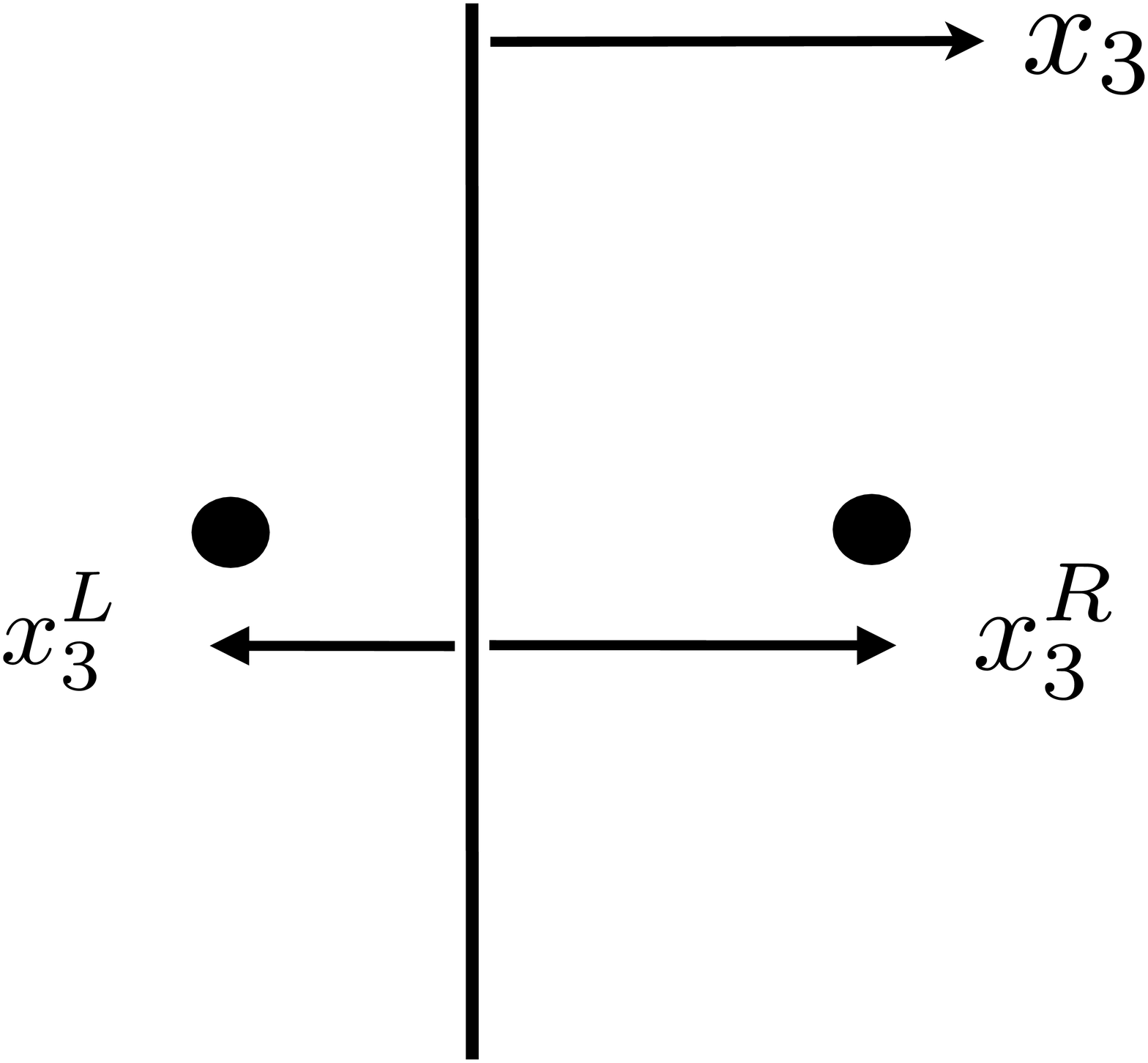}}}\hspace{5mm}
\subfigure[]{\fbox{\includegraphics[height=2.2in]{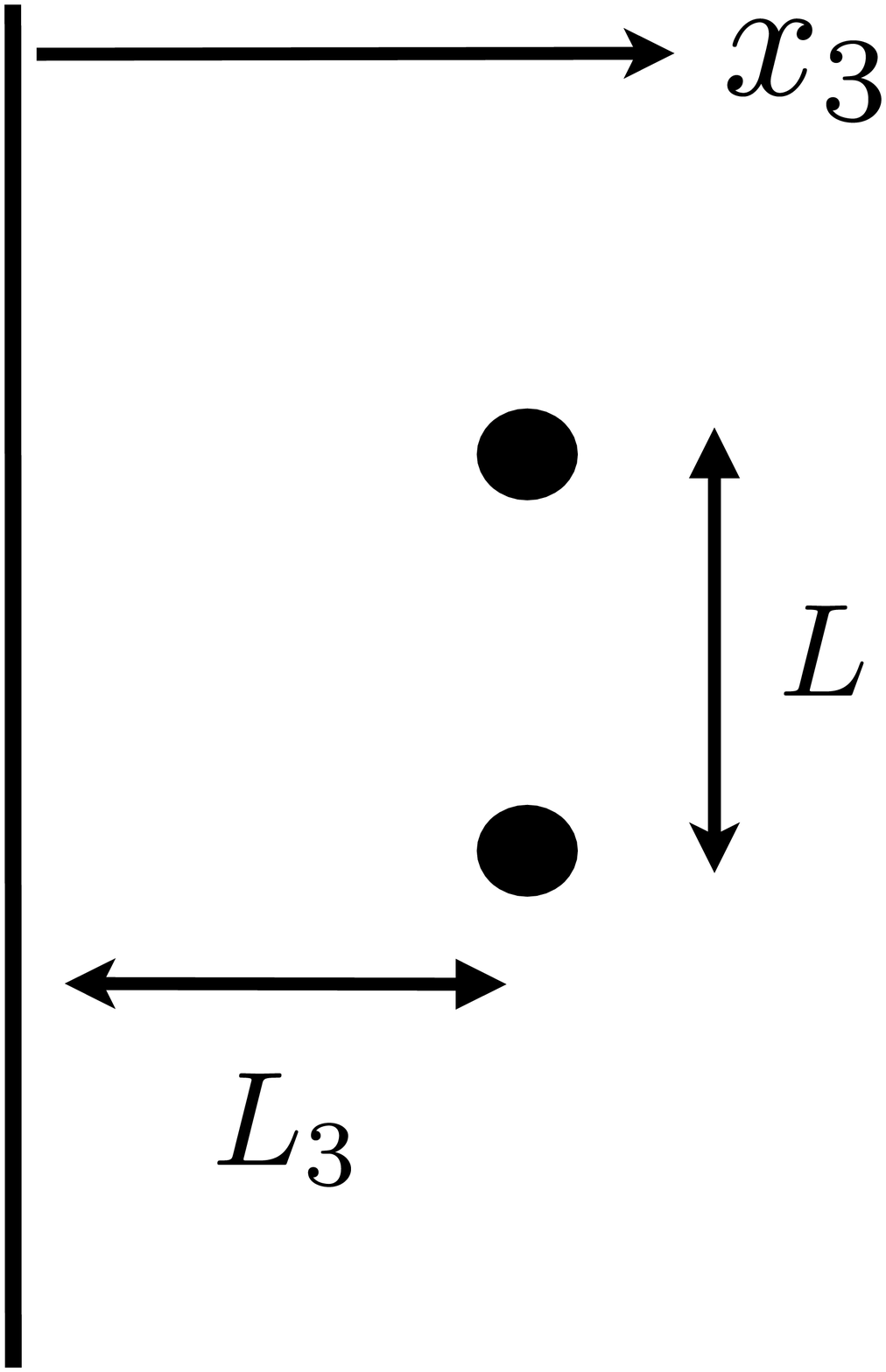}}}
\end{center}
\caption{Depictions of the orientations of the rectangular Wilson loops in the two cases we consider. The vertical axis is one of the directions $x_1$ or $x_2$, the horizontal axis is $x_3$. The solid black vertical line represents the interface where the coupling or $\theta$-angle jumps, which is at $x_3=0$. The quark and anti-quark, represented by the solid black dots, define a line that is either parallel or perpendicular to the defect. (a) The perpendicular case, where the quark and anti-quark sit at positions $x_3^{\textrm{L}}$ and $x_3^{\textrm{R}}$ from the interface. We consider only $x_3^L<0$ and $x_3^R>0$. (b) The parallel case, where $L$ is the distance between the quark and anti-quark, which are located at $x_3=L_3$.}
\label{fig:configs}
\end{figure}

For the rectangular Wilson loop in the presence of the interface, $L$ is no longer the only scale in the problem. The quark and anti-quark provide two points that define a line. That line may be parallel to the interface, perpendicular, or some linear combination of the two. For simplicity we will consider only the perpendicular and parallel cases, as depicted in figure~\ref{fig:configs}. Moreover, in the perpendicular case we will only consider test charges on opposite sides of the interface, rather than two test charges on the same side. In each case the potential $V$ can depend on two scales. For the perpendicular case, depicted in fig.~\ref{fig:configs} (a), we parameterize the two scales as the $x_3$ positions of the two test charges, denoted $x_3^{\textrm{L}}$ and $x_3^{\textrm{R}}$, where the superscripts L and R denote ``left'' ($x_3 <0$) and ``right'' ($x_3>0$). The distance between the test charges is $L = x_3^{\textrm{R}} - x_3^{\textrm{L}}$. If we define an ``average distance'' $L_{\textrm{av}} \equiv \frac{1}{2} \left(x_3^{\textrm{R}} + x_3^{\textrm{L}}\right)$, then dimensional analysis and scale invariance constrain the potential to take the form $V(\lambda,L,L_{\textrm{av}}) = \frac{1}{L}f(\lambda,L_{\textrm{av}}/L)$. For the parallel case, depicted in fig.~\ref{fig:configs} (b), we parameterize the two scales as the distance between the quark and anti-quark, $L$, and their position in the $x_3$ direction, $x_3=L_3$. Dimensional analysis and scale invariance constrain the potential to take the form $V(\lambda,L,L_3) = \frac{1}{L}f(\lambda,L_3/L)$. In each case, our objective is to extract the non-trivial information contained in the function $V L = f$.

Wilson lines or loops located precisely on the interface, at $x_3=0$, are special cases, and raise several questions. First is a question about interface CFTs in general (regardless of any test charges): what are the values of $g$ and the $\theta$ exactly at the interface? Do we use the values from $x_3<0$, from $x_3>0$, some kind of average values, or some more complicated functions? Second, for a single test charge at finite $x_3$, we have $V \propto 1/(2|L_3|)$, but if we send $L_3\to 0$ does $V$ diverge? In physical terms, at $x_3=0$, doesn't the test charge sit exactly on top of its own image charge? Can we sensibly define $V$ in such circumstances? To address questions such as these, we will include in our analysis Wilson lines and loops sitting precisely at $x_3=0$.

For all four interface CFTs that we consider we obtain results in the large-$N_c$, large coupling limits using holography, where the Wilson loop is described by a string in Janus spacetime~\cite{Rey:1998ik,Maldacena:1998im,Bak:2003jk}. Previous holographic calculations of Wilson loop expectation values in $\N=4$ SYM with conformal defects appear in refs.~\cite{Bak:2003jk,Drukker:2010jp,Nagasaki:2011ue}. For the non-SUSY interface with a jumping coupling, we also consider finite $N_c$ and small coupling, and obtain results using perturbation theory.

\subsection{Motivation}

Why bother computing Wilson loop expectation values in $\N=4$ SYM theory with conformal interfaces? Many reasons come to mind, but we will emphasize only two.

First, Wilson loops provide a valuable check of the AdS/CFT correspondence. In $\N=4$ SYM with constant $g$, the expectation value of a circular Wilson loop may be computed exactly, for all values of $N_c$ and $\lambda$, from a Gaussian matrix model~\cite{Erickson:2000af,Drukker:2000rr,Pestun:2007rz}. In the large-$N_c$ and large-$\lambda$ limits, the exact result agrees precisely with the holographic result, providing one of the many direct tests of the AdS/CFT correspondence.

The existence of a matrix model for the circular Wilson loop expectation value can be deduced in various ways. One way begins with the expectation value of the straight Wilson line, which is trivial, as mentioned above. A conformal transformation maps the straight line to a circle, upon including one additional point, the point at infinity~\cite{Berenstein:1998ij}. The circular Wilson loop has a non-trivial expectation value, suggesting that the only contribution to the expectation value must come from a single point, the point at infinity, which in turn suggests the existence of a matrix model. Notice that the main ingredients here were 1) the triviality of the straight Wilson line expectation value and 2) conformal symmetry.

Wilson loops may also provide direct tests for $\N=4$ SYM with various conformal interfaces and their holographic duals. Our work is one step in that direction.

As a second motivation, analysis of $\N=4$ SYM with a jumping $\theta$ angle may shed light on certain (3+1)-dimensional topological insulators (TIs). One definition of a TI is an insulator with topologically-protected gapless edge modes that give rise to quantized, dissipationless transport. Since 1980, the canonical example of a TI has been the integer quantum Hall state (QHS), a (2+1)-dimensional insulating state where the gapless modes are (1+1)-dimensional chiral fermions giving rise to an integer-quantized Hall conductivity. The integer QHS breaks time reversal symmetry, T. Since 2005 a number of (3+1)-dimensional TIs have been discovered that preserve T, including $\textrm{Bi}_2\textrm{Se}_3$, $\textrm{Bi}_2\textrm{Te}_3$, and $\textrm{Sb}_2\textrm{Te}_3$, where the edge modes are (2+1)-dimensional Dirac fermions, giving rise to $\mathbb{Z}_2$-quantized magneto-electric response. To date, the edge modes of some (3+1)-dimensional T-invariant TIs have been observed directly, however the $\mathbb{Z}_2$-quantized magneto-electric response has not yet been observed. For reviews of TIs, see for example refs.~\cite{Hasan:2010xy,Moore:2010rev,2011RvMP...83.1057Q,2011ARCMP...2...55H}.

All of the above TIs admit low-energy effective descriptions in terms of free fields: for the integer QHS the effective description is Chern-Simons theory~\cite{doi:10.1142/S0217979292000037}, while for the (3+1)-dimensional T-invariant TIs the effective description is Maxwell electrodynamics with a $\theta$-angle that takes one of the two values allowed by T-invariance, zero or $\pi$ (mod $2\pi$)~\cite{2008PhRvB..78s5424Q,2008arXiv0810.2998E}. These topological terms reproduce the dissipationless, quantized transport properties. The boundary between a (3+1)-dimensional T-invariant TI and the vacuum appears in the effective description as, for example, a $\theta$-angle that jumps from $\pi$ to zero.

In the presence of a jumping $\theta$-angle, a test electric charge will induce an image charge with both electric and magnetic charge, \textit{i.e.}\ an image dyon. Experiments to observe (the effects of) these image dyons were proposed in ref.~\cite{Qi27022009}. TIs may thus allow the first observation of objects with nonzero magnetic monopole charge, at least at the level of an effective field theory.

In contrast to the integer QHS, the fractional QHS cannot be described in terms of free fields. An open question is whether (3+1)-dimensional T-invariant \textit{fractional} TIs exist, whose signature would be fractionally-quantized magneto-electric response, and if so, what low-energy (interacting) theory describes their transport properties? One proposal for such an effective theory exploits non-Abelian gauge fields~\cite{Maciejko:2010tx,Swingle:2010rf,Maciejko:2011ed}. We emphasize, however that to date no T-invariant fractional TI has been observed experimentally. An obvious question is: could a fractional TI be detected experimentally via the effects of image charges, in analogy with non-fractional TIs~\cite{Qi27022009}? The nature of the image charges may depend sensitively on the interactions of the low-energy theory.

As proposed in ref.~\cite{Karch:2009sy}, and as we will explain in detail in section~\ref{topo}, we can think of $\N=4$ SYM with a $\theta$-angle that jumps from zero to $\pi$ (mod $2\pi$) as the low-energy effective description of a very special (3+1)-dimensional T-invariant fractional TI. $\N=4$ SYM is unlikely to be realized experimentally, but does have the advantage that strong-coupling calculations are tractable via AdS/CFT. We can therefore address the question: how does strong coupling affect image charges in a fractional TI? Do any unexpected or exotic effects occur at large coupling? Our work is a first step towards answering these and other related questions.

\subsection{Summary and Outlook}

In sections~\ref{nonsusyjanus} and~\ref{susyjanus} we review non-SUSY and SUSY Janus, respectively, and in section~\ref{topo} we explain how $\N=4$ SYM with a jumping $\theta$-angle describes a fractional TI (clarifying the original proposal of ref.~\cite{Karch:2009sy}). In section~\ref{wilsonfieldtheory} we turn to the calculation of Wilson loop expectation values, and the self-energy or potential $V$. In section~\ref{em} we compute $V$ in ordinary electromagnetism with a jumping coupling and $\theta$-angle. In section~\ref{sunc} we compute the expectation value for a rectangular Wilson loop parallel to a non-SUSY interface where the coupling jumps in $\N=4$ SYM, using perturbation theory. With one special choice of boundary conditions on the scalars of $\N=4$ SYM, for which the self-energy of a single test charge vanishes, and in the large-$N_c$ and large coupling limits, we also calculate the contribution to this Wilson loop expectation value from the sum of ladder diagrams. In section~\ref{wilsonholo} we turn to the holographic calculation of $V$. In sections~\ref{sec:straight} through~\ref{intloops} we present our results for $V$, which are mostly numerical, but include some exact results for a single test charge in $\N=4$ with a SUSY interface with a small jump in the coupling or $\theta$-angle. We collect some useful technical results in two appendices.

For all non-SUSY interfaces that we consider we find that our holographic results for $V$ are qualitatively similar to the analogous results in electromagnetism. For example, consider a single test charge. In electromagnetism with a jumping coupling, a test charge is always attracted to the side of the interface with smaller coupling, as we review in section~\ref{em}. With a jumping $\theta$-angle, the test charge is always attracted towards the interface. Electromagnetism is a linear theory, so the interaction potential $V$ between two test charges follows simply by linear superposition. Our holographic results for $\N=4$ SYM with a non-SUSY interface indicate that with a jumping coupling or $\theta$-angle the test charge is attracted to the side with smaller coupling or to the interface, respectively. In other words, the induced image charge has the same sign as in electromagnetism. This similarity is surprising, given that electromagnetism is a free theory of gauge fields alone, while $\N=4$ SYM is an interacting theory of gauge fields, fermions, and scalars, and moreover we study the large-coupling limit and we study test charges that couple to both the gauge fields and the scalars. Additionally, we find that our results for the interaction potential between two test charges are qualitatively similar to those of electromagnetism: compare for example our holographic result for the potential between two test charges along a line perpendicular to a non-SUSY interface, shown in fig.~\ref{fig:perpendicularns}, with the analogous results in electromagnetism, shown in fig.~\ref{fig:empotperpfigs}. This similarity is striking because $\N=4$ SYM is not a linear theory.

For large-$N_c$, strongly-coupled $\N=4$ SYM with a SUSY interface, the results for the self-energy of a test charge depend sensitively on the coupling of the charge to the adjoint scalars. For example, for a SUSY interface where the coupling jumps we find one case where the induced image charge vanishes! For a SUSY interface where the $\theta$-angle jumps, we find cases where the image charge has the ``wrong'' sign, as compared to our intuition from electromagnetism, and cases where the image charge either diverges or goes to zero as the jump in the $\theta$-angle goes to infinity, in dramatic contrast to electromagnetism. The interface-localized terms that are the main difference between the non-SUSY and SUSY interfaces play a decisive role in determining image charges.

For two test charges located precisely on an interface in large-$N_c$, strongly-coupled SYM, we find a (holographic) renormalization scheme that subtracts the infinite self-energy of the test charges, allowing us to define a finite interaction potential between them. Within that renormalization scheme, we can thus define an effective 't Hooft coupling precisely on the interface. For an interface where the coupling jumps, either non-SUSY or SUSY, we find surprising similarity between our numerical results for the effective 't Hooft coupling and an effective coupling for the analogous system in electromagnetism, defined by simply deleting any self-interaction terms from the potential.

Our results raise a number of questions for future research. For example, as mentioned above, with a SUSY interface where the coupling jumps, we find one kind of test charge with vanishing image charge \textit{i.e.} we find a trivial timelike Wilson line. Our system has conformal symmetry, so we may map that line to a circle. Does that circular Wilson loop have non-zero expectation value? If so, can we reproduce that expectation value from a matrix model? Indeed, in ref.~\cite{Drukker:2010jp} a matrix model was derived for a circular Wilson loop sitting exactly on the interface. Our results suggest that something similar may be possible for a circular Wilson loop away from the interface.

Our results for $\N=4$ SYM with a non-SUSY jumping $\theta$-angle demonstrate that in fractional TIs strong coupling does not necessarily produce dramatic or exotic effects: qualitatively, we observe physics very similar to electromagnetism. We have only performed the simplest ``experiments,'' however, probing the system with one or two static test charges. $\N=4$ SYM with a jumping $\theta$-angle may have much more to teach us about fractional TIs.

What about dyonic test charges in $\N=4$ SYM, holographically dual to $(p,q)$-strings? What kinds of image charges do they have in the presence of non-SUSY and SUSY interfaces? What about defects in $\N=4$ SYM that support localized degrees of freedom, as arise for example in the (2+1)-dimensional intersection of D3-branes with D5- and/or NS5-branes~\cite{Karch:2000gx,DeWolfe:2001pq,Erdmenger:2002ex,D'Hoker:2007xz}? How do those affect the self-energy of a test charge?

As a caution to the reader: we use a Lorentzian-signature metric in sections~\ref{interfaces} and~\ref{em} and a Euclidean-signature metric in sections~\ref{sunc} and~\ref{wilsonholo}, and in the appendices. Throughout the paper we work exclusively in the Einstein frame of type IIB supergravity.

\section{Holographic Conformal Interfaces}
\label{interfaces}

We will study $\N=4$ SYM with conformal interfaces, which preserve an $SO(3,2)$ subgroup of the $SO(4,2)$ conformal group. The holographic duals will therefore include an $AdS_4$ subspace, so let us recall the $AdS_4$ foliation, or $AdS_4$ ``slicing,'' of $AdS_5$. We begin with the metric of Lorentzian-signature $AdS_5$, in Poincar\'e slicing,
\beq
\label{poincareslicing}
ds^2_{AdS_5} = R^2 \frac{dr^2}{r^2} + \frac{r^2}{R^2} \left( -dt^2 + dx_1^2 + dx_2^2 + dx_3^2 \right),
\eeq
where $R$ is the $AdS_5$ radius of curvature, which is related to the string length squared $\alpha'$ as (in Einstein frame) $R^4 = 4 \pi N_c \, \alpha'^2$, and $r$ is the $AdS_5$ radial coordinate. The Poincar\'e horizon is at $r=0$, while the boundary is at $r \to \infty$. Introducing new coordinates $u$ and $x$ via
\beq
\label{uxcoords}
r = R^2 u \cosh x, \qquad x_3 = \frac{\tanh x}{u},
\eeq
puts the metric into $AdS_4$ slicing,
\beq
\label{ads4slicing}
ds^2_{AdS_5} = R^2 \left( dx^2 + \cosh^2 x \, ds^2_{AdS_4}\right),
\eeq
where the unit-radius $AdS_4$ metric is
\beq
\label{ads4metric}
ds^2_{AdS_4} = \frac{du^2}{u^2} + u^2 \left( -dt^2 + dx_1^2 + dx_2^2 \right).
\eeq
In $AdS_4$ slicing, we can approach the Poincar\'e horizon by fixing $x$ and taking $u\rightarrow 0$, which in eq.~\eqref{uxcoords} sends $r \rightarrow 0$ and $x_3 \rightarrow \infty$. We can approach the boundary in three ways. If we take $x \to \pm\infty$ with $u$ fixed, then from eq.~\eqref{uxcoords} we see that $r \to \infty$ with $x_3 > 0$ or $x_3<0$, respectively. If we fix $x$ and take $u \to \infty$, (the boundary of the $AdS_4$ subspace), then $r \to \infty$ and $x_3=0$.

\subsection{Non-Supersymmetric Janus}
\label{nonsusyjanus}

The non-SUSY Janus solution of type IIB supergravity is a one-parameter dilatonic deformation of the $AdS_5 \times S^5$ solution in which only the metric, dilaton, and Ramond-Ramond (RR) five-form are non-trivial~\cite{Bak:2003jk,D'Hoker:2006uu}. We begin by writing the $AdS_5 \times S^5$ solution in $AdS_4$ slicing, as in eq.~\eqref{ads4slicing}, but changing to a new $x$ coordinate,
\beq
x \to - \tanh^{-1} x,
\eeq
so that eq.~\eqref{uxcoords} becomes
\beq
r = \frac{R^2 u}{\sqrt{1-x^2}}, \qquad x_3 = -\frac{x}{u},
\eeq
and the boundary regions formerly at $x \to \pm \infty$ now correspond to $x \to \mp 1$. The $AdS_5 \times S^5$ metric then takes the form
\beq
\label{ads5inx}
ds^2 = R^2 \left(\frac{1}{(1-x^2)^2} \, dx^2 + \frac{1}{1-x^2} \, ds^2_{AdS_4}\right) + R^2 ds^2_{S^5},
\eeq
with $ds^2_{S^5}$ the metric of a unit-radius $S^5$. The dilaton, $\phi$, is constant: $\phi = \phi_0$.

To write the metric and dilaton of the non-SUSY Janus solution, we need some special functions. First, we need the Weierstrass elliptic function $\We(x)$, defined by the equation
\beq
\left( \partial_x \We \right)^2 = 4 \We^3 - g_2 \We - g_3,
\eeq
with periods $g_2$ and $g_3$. Next, we need the Weierstrass sigma- and zeta-functions, $\sigma(x)$ and $\zeta(x)$ respectively, which are related to $\We(x)$ via
\beq
\We(x) = - \zeta'(x), \qquad \zeta(x) = \frac{\sigma'(x)}{\sigma(x)}.
\eeq

The metric of the non-SUSY Janus solution is~\cite{Bak:2003jk,D'Hoker:2006uu}
\beq
\label{eq:nonsusyjanusmetric}
ds^2 = R^2 \left( \gamma^{-1} h(x)^2 dx^2 + h(x) ds^2_{AdS_4} \right)+ R^2 ds^2_{S^5},
\eeq
with warp factor
\beq
h(x) = \g \left( 1 + \frac{4 \g - 3}{\We(x) + 1 - 2\g} \right),
\eeq
where $\We(x)$ has periods
\beq
g_2 = 16 \g (1-\g), \qquad g_3 = 4(\g-1).
\eeq
The dilaton of the non-SUSY Janus solution, $\phi(x)$, is
\beq
\label{eq:nonsusyjanusdilaton}
\phi(x) =  \phi_0 + \sqrt{6(1-\gamma)} \left( x + \frac{4 \gamma - 3}{\We'(\chi)} \left( \ln \frac{\sigma(x+\chi)}{\sigma(x-\chi)} - 2 \zeta(\chi) x\right)\right),
\eeq
where the constant $\chi$ is defined by $\We(\chi) = 2(1-\g)$.

Non-SUSY Janus also has a non-trivial RR five-form, with $N_c$ units of flux on the $S^5$~\cite{Bak:2003jk,D'Hoker:2006uu}. To compute Wilson loops we will introduce strings into non-SUSY Janus. In Einstein frame the string action involves the dilaton and the pull-backs of the metric and Neveu-Schwarz (NS) two-form, therefore we will not need the RR five-form, so we will omit it from our review.

The non-SUSY Janus solution is completely specified by the four real constants $R$, $\phi_0$, $N_c$, and $\g$. We can constrain the value of $\g$ as follows. When $\gamma = 3/4$ the warp factor $h(x)=1$, and the solution is a product geometry $AdS_4 \times \mathbb{R} \times S^5$ with a linear dilaton. Solutions with $\g < 3/4$ are generally singular, so we will impose $\g > 3/4$. On the other hand, to maintain reality of the dilaton we must impose $\g \leq 1$. When $\g=1$, the warp factor becomes $h(x) = \frac{1}{1-x^2}$, so the metric becomes that of eq.~\eqref{ads5inx}, for $AdS_5 \times S^5$ with radius of curvature $R$. Additionally, when $\gamma=1$ the dilaton becomes constant, $\phi(x) = \phi_0$. In summary, we take $\g \in (3/4,1]$.

The boundary is defined as the location in $x$ where the metric diverges. Clearly that will occur where $h(x)$ has a pole, which occurs at points with $x=x_0$ obeying $\We(x_0) = 2 \g - 1$. When $\gamma=1$, we find $x_0 = \pm 1$, as expected for $AdS_5 \times S^5$. More generally, since $h(x)$ is an even function the geometry will have two asymptotic boundaries at $x = \pm x_0$. To make the asymptotic boundaries explicit, let us change variables. In the $x \to \pm x_0$ limits, we take
\beq
x \equiv \pm x_0 \mp 2 \sqrt{\g} \, e^{\mp 2 \hat{x}},
\eeq
so that for large $\hat{x}$ the Janus metric approaches
\beq
ds^2 = R^2 \left( d\hat{x}^2 + \frac{e^{2\hat{x}}}{4} ds^2_{AdS_4} + ds^2_{S^5} \right),
\eeq
which we recognize as the asymptotic form of the $AdS_4$-sliced $AdS_5 \times S^5$ metric in eq.~\eqref{ads4slicing}. Taking $x \to \pm x_0$ brings us to points on the boundary with $x_3<0$ or $x_3>0$, respectively. The geometry thus has two asymptotic $AdS_5 \times S^5$ regions at $x = \pm x_0$, corresponding to the two halves of space in the dual CFT. Crucially, however, when $\gamma \neq 1$, the value of the dilaton is not the same in the two regions. Let
\beq
\phi_{\pm} \equiv \lim_{x \to \mp x_0} \phi(x), \qquad \delta \phi \equiv \phi_+ - \phi_-,
\eeq
so that the dilaton approaches $\phi_{\pm}$ at boundary points with $x_3>0$ or $x_3<0$, respectively. A formula for the jump in the dilaton $\delta \phi$ as a function of $\g$ is straightforward to obtain but unilluminating, so we will omit it. We will note however that $\delta \phi = 0$ when $\g=1$ (the $AdS_5 \times S^5$ solution) and $\delta \phi$ increases as $\g$ decreases, with $\delta \phi \to \infty$ as $\g \to 3/4$ (where the dilaton is linear in $x$). We can thus obtain a $\delta \phi$ of any size within the region $\g \in (3/4,1]$. In short, non-SUSY Janus has one more free parameter than $AdS_5 \times S^5$, $\g$, which controls the size of the jump in $\phi$ at the boundary.

Clearly the field theory dual to the Janus solution is a deformation of $\N=4$ SYM with a $g$, and hence a $\lambda = g^2 N_c$, whose value jumps at $x_3=0$. To make a precise statement, we need the dictionary between $\phi$ and $g$. We can fix the dictionary using the $SL(2,\mathbb{R})$ transformation properties of type IIB supergravity and large-$N_c$ $\N=4$ SYM. On the supergravity side, the axio-dilaton,
\beq
\tau \equiv C_{(0)} + i e^{-2\phi},
\eeq
where $C_{(0)}$ is the axion, transforms covariantly under $SL(2,\mathbb{R})$,
\beq
\tau \to \frac{a \tau + b}{c \tau + d},
\eeq
where $a,b,c,d \in \mathbb{R}$ and $ad-bc=1$. Notice that our $\phi$ gives the open string coupling, and hence is related to the closed string coupling, $g_s$, via $e^{2\phi} = g_s$.

To find the corresponding $SL(2,\mathbb{R})$-covariant object in $\N=4$ SYM, we need to fix the normalization of the fields in the classical Lagrangian. The field content of $\N = 4$ $SU(N_c)$ SYM consists of a gauge field, $\hat{A}_\mu$, six real scalars $\hat{\Phi}^I$ with $I=1,2,...,6$, and four Weyl fermions, all in the adjoint representation of $SU(N_c)$. The hats on $\hat{A}_{\mu}$ and $\hat{\Phi}^I$ are meant to indicate that these objects are matrices valued in the Lie algebra of $SU(N_c)$. The scalars are in the \textbf{6} and the fermions are in the \textbf{4} of the $SO(6)$ R-symmetry. In what follows we will ignore the fermions. In the $\N=4$ SYM Lagrangian, ${\cal L}_{\N=4}$, the terms involving only the gauge fields and scalars are,\footnote{Our choice of orientation is $\epsilon^{0123} = +1$.}
\beq
\label{N=4action}
{\cal L}_{\N=4} \supset - \frac{1}{4g^2} \tr \hat{F}_{\mu \nu} \hat{F}^{\mu \nu} + \frac{\theta}{32 \pi^2} \epsilon^{\mu \nu \rho \sigma} \tr \hat{F}_{\mu \nu} \hat{F}_{\rho \sigma} - \frac{1}{2g^2} \tr (D^\mu \hat{\Phi}^I D_\mu \hat{\Phi}^I) + \frac{1}{4g^2} \tr( [\hat{\Phi}^I, \hat{\Phi}^J][\hat{\Phi}^I, \hat{\Phi}^J]),
\eeq
where $D_{\mu}$ is the gauge-covariant derivative and the traces are over color indices. We have normalized the fields so that all terms (not involving $\theta$) have an overall $1/g^2$. With that choice of normalization, the $SL(2,\mathbb{R})$-covariant coupling of $\N=4$ SYM is
\beq
\tau \equiv \frac{\theta}{2\pi} + i \frac{2 \pi}{g^2},
\eeq
hence we obtain the dictionary: in each asymptotic $AdS_5 \times S^5$ region, we have
\beq
C_{(0)} = \frac{\theta}{2\pi}, \qquad e^{2\phi} = \frac{g^2}{2\pi},
\eeq
and by extension $\lambda = g^2 N_c = 2 \pi N_c e^{2\phi}$. Given the asymptotic values $\phi_{\pm}$ we can thus determine the values of the coupling on the two sides of the interface, $g_{\pm}$ or $\lambda_{\pm}$. Recalling that in the 't Hooft limit $g^2$ is $\mathcal{O}\left(1/N_c\right)$, the natural scaling for $\theta$ in the 't Hooft limit is $\mathcal{O}\left(N_c\right)$, so that both $\textrm{Re} \, \tau$ and $\textrm{Im}\,\tau$ have the same $\mathcal{O}\left(N_c\right)$ scaling. Notice that with those scalings, $g^2 \theta$ is $\mathcal{O}\left(N_c^0\right)$.

For the moment let us set $\theta=0$, and ask the following question: what is the classical Lagrangian of the field theory dual to the non-SUSY Janus solution~\cite{Clark:2004sb,D'Hoker:2006uv}? To construct the most general classical Lagrangian that includes a jumping coupling and that is consistent with the $SO(3,2) \times SO(6)$ symmetry of the non-SUSY Janus solution, the first step is to promote $g$ to a smooth function of $x_3$ in eq.~\eqref{N=4action}. Generically, such a function will introduce at least one dimensionful scale, so to recover scale invariance we will ultimately take a limit in which $g$ becomes a step function. With $g$ a smooth function of $x_3$, we next write the classical action as a sum of all possible operators consistent with the $SO(3,2) \times SO(6)$ symmetry, including operators involving powers of $\partial_3 g$. Via field redefinitions and integration by parts, the action can be reduced to the sum of two terms, namely the $\N=4$ SYM Lagrangian plus an additional non-trivial term involving the scalars and fermions, which we call ${\cal L}_{\rm int}$, to which the scalars' contribution is
\beq
\label{intextra}
{\cal L}_{\rm int} \supset \kappa \frac{\partial_3 g}{g^3} \hat{\Phi}^I D_3 \hat{\Phi}^I,
\eeq
with some real constant $\kappa$~\cite{Clark:2004sb,D'Hoker:2006uv}. The fermions' contribution to ${\cal L}_{\rm int}$ introduces no other constants beyond $\kappa$. ${\cal L}_{\rm int}$ becomes localized to the interface when $g$ becomes a step function. One ambiguity remains: the symmetries do not fix the value of $\kappa$. No corresponding free parameter appears in the Janus solution, so presumably the field theory dual to Janus has a specific value of $\kappa$. What that value is remains an open question. The value $\kappa=1$ is special, however. When $\kappa=1$, integrating the scalars' bulk kinetic terms by parts produces an interface term that exactly cancels the term in eq.~\eqref{intextra}. Moreover, as argued in ref.~\cite{Clark:2004sb}, $\kappa=1$ might be the only value of $\kappa$ for which $SO(3,2)$ conformal symmetry will be preserved at the quantum level.

What is the physical meaning of $\kappa$? For test charges that couple to the scalars, $\kappa$ controls the size of image charges. In the presence of a jumping $g$, a test charge that couples to the scalars will interact with an image charge via scalar exchange. ${\cal L}_{\rm int}$ is quadratic in the scalars and so modifies the scalar propagator, hence $\kappa$ will influence the size of the image charge, as we will see in detail in section~\ref{sunc}. In particular, we will see that when $\kappa=0$, the gauge and scalar propagators are of the same form, and as a result their contributions to the image charges cancel, so that effectively a test charge has no image charge. Our holographic calculation of the interaction potential will show that test charges have nonzero image charges, however. That in turn suggests that the field theory dual to the non-SUSY Janus solution must have non-zero $\kappa$. In other words, our holographic calculation will provide suggestive evidence \textit{excluding} a single value, $\kappa=0$.

If we begin with a jumping $g$, then $SL(2,\mathbb{R})$ transformations can generate a jumping $\theta$. For the theory with a jumping $\theta$, all possible interface-localized operators consistent with $SO(3,2)$ and an $SO(3) \times SO(3)$ subgroup of the $SO(6)$ symmetry were written in ref.~\cite{Gaiotto:2008sd}. These operators will then appear in the action with coefficients that, in the absence of SUSY, are undetermined \textit{a priori}.

\subsection{Supersymmetric Janus}
\label{susyjanus}

The SUSY Janus solution is a deformation of the $AdS_5 \times S^5$ solution in which the metric, dilaton, RR five-form, and RR and NS three-forms are non-trivial~\cite{D'Hoker:2007xy}. We begin with the $AdS_4$ foliation of $AdS_5$ with radial coordinate $x$ in eq.~\eqref{ads4slicing} (not the $x$ coordinate in eq.~\eqref{ads5inx}). Next we rewrite the $S^5$ as a pair of $S^2$'s fibered over a line segment with coordinate $y$,
\beq
\label{s5metric}
ds^2_{S^5} = dy^2 + \cos^2y \, ds^2_{S^2} + \sin^2 y \, ds^2_{S^2},
\eeq
with $ds^2_{S^2}$ the metric of a unit-radius $S^2$ and $y \in [0,\pi/2]$. The coordinates $x$ and $y$ together describe a Riemann surface, which may be parameterized by a complex coordinate $v \equiv x + i y$. The SUSY Janus geometry is a fibration of the $AdS_4$ and the two $S^2$'s over the Riemann surface. Notice that the Riemann surface has the topology of a strip:
\beq
\textrm{Re}(v)=x \in (-\infty,\infty), \qquad \textrm{Im}(v) = y \in \left [0,\pi/2 \right].
\eeq

The metric and dilaton of the SUSY Janus solution are~\cite{D'Hoker:2007xy}
\begin{subequations}
\beq
\label{eq:susyjanusmetric}
ds^2 = f_4^2 \, ds^2_{AdS_4} + f_1^2 \, ds^2_{S^2} + f_2^2 \, ds^2_{S^2} + \rho^2 \, dv d\bar{v},
\eeq
\beq
\label{eq:susyjanusdilaton}
e^{4 \phi} = N_2/N_1,
\eeq
\end{subequations}
where $ds^2_{AdS_4}$ is the $AdS_4$ metric in eq.~\eqref{ads4metric}. The functions $f_4^2$, $f_1^2$, $f_2^2$, $\rho^2$, $N_1$, and $N_2$ all depend on $v$ and $\bar{v}$. To write these functions, let us introduce $h_1(v,\bar{v})$ and $h_2(v,\bar{v})$, two real, harmonic functions on the Riemann surface, which obey the boundary conditions
\beq
\left . h_1 \right |_{y=0} = \left . \partial_y h_2\right |_{y=0} = 0, \qquad \left . h_2 \right |_{y=\pi/2} = \left . \partial_y h_1 \right |_{y=\pi/2} = 0.
\eeq
For SUSY Janus these harmonic functions are~\cite{D'Hoker:2007xy}
\beq
h_1(v,\bar{v}) = - i \alpha_1 \sinh \left(v - \frac{\d \phi}{2}\right) + c.c., \qquad h_2(v,\bar{v}) = \a_2 \cosh\left(v+\frac{\d \phi}{2}\right) + c.c.,
\eeq
which are completely specified by the three constants $\a_1$, $\a_2$ and $\d \phi$. The functions appearing in the SUSY Janus solution are completely determined by $h_1(v,\bar{v})$ and $h_2(v,\bar{v})$ as follows~\cite{D'Hoker:2007xy}:
\begin{subequations}
\beq
w \equiv \partial_v h_1 \, \partial_{\bar{v}} h_2 + \partial_{\bar{v}} h_1 \, \partial_v h_2,
\eeq
\beq
N_1 = 2 h_1 h_2 \, |\partial_v h_1 |^2 - h_1^2 \, w, \qquad N_2 = 2 h_1 h_2 \, |\partial_v h_2 |^2 - h_2^2 \, w,
\eeq
\beq
f_4^8 = 16 \, \frac{N_1 N_2}{w^2}, \qquad \rho^8 = \frac{2^8 \, N_1 N_2 \, w^2}{h_1^4 \, h_2^4},
\eeq
\beq
f_1^8 = 16 \, h_1^8 \, \frac{N_2 \, w^2}{N_1^3}, \qquad f_2^8 = 16 \, h_2^8 \, \frac{N_1 \, w^2}{N_2^3}.
\eeq
\end{subequations}
The geometry has two asymptotically $AdS_5 \times S^5$ regions, where $\textrm{Re}(v) = x \to \pm \infty$. To be explicit, a change of coordinates as $x \to \pm \infty$,
\beq
x = \hat{x} \pm \frac{1}{2} \ln \cosh (\d \phi),
\eeq
puts the asymptotic SUSY Janus metric in the same form as the $AdS_4$-sliced $AdS_5 \times S^5$ metric in eq.~\eqref{ads4slicing}. The values of the $AdS_5$ radius and the dilaton in the two asymptotic regions are
\beq
R^4 = 16 \, |\a_1 \a_2| \cosh (\d \phi), \qquad e^{2 \phi_{\pm}} = \left| \frac{\a_2}{\a_1}\right| e^{\pm \d \phi}.
\eeq
We thus see that the three parameters $\a_1$, $\a_2$, and $\d \phi$, map to the asymptotic $AdS_5$ radius, the overall background value of the dilaton, and the size of the jump in the dilaton at the boundary, $\delta \phi = \phi_+ - \phi_-$. To recover $AdS_5 \times S^5$, we simply take $\d \phi = 0$ in all of the above.

The SUSY Janus solution also includes a non-trivial RR five-form with $N_c$ units of flux on the internal space as well as a non-trivial RR two-form with flux on one $S^2$ and a non-trivial NS two-form with flux on the other $S^2$~\cite{D'Hoker:2007xy}. To compute Wilson loops we will introduce strings into the SUSY Janus spacetime. In Einstein frame the string action involves the dilaton and the pull-backs of the metric and NS two-form. We will thus not need the RR five-form or two-form, so we will omit them from our review. Furthermore, for the strings we will study in section~\ref{wilsonholo} the pull-back of the NS two-form will vanish, hence we will omit the NS two-form also.

What is the classical Lagrangian of the field theory dual to SUSY, jumping-coupling Janus? As shown in ref.~\cite{D'Hoker:2006uv} the classical Lagrangian is ${\cal L}_{\N=4}$, with $\theta=0$ and with $g$ promoted to a function of $x_3$, plus a term ${\cal L}_{\rm int}$, to which the scalars' contribution is
\beq
\label{susyint}
{\cal L}_{\rm int} \supset \frac{\p_3 g}{g^3} \tr \bigg( -\frac{2}{3} i \epsilon^{IJK} \hat{\Phi}^I [\hat{\Phi}^J, \hat{\Phi}^K] + D_3 (\hat{\Phi}^I \hat{\Phi}^I)\bigg) - \frac{(\p_3 g)^2}{g^4} 2 \tr \hat{\Phi}^I \hat{\Phi}^I, \qquad I,J,K = 1,2,3,
\eeq
in the limit where $g$ approaches a step function. Here $I,J,K$ run over $1,2,3$, so that ${\cal L}_{\rm int}$ involves only three of the six scalars, and thus breaks the $SO(6)$ R-symmetry down to $SO(3) \times SO(3)$, which matches the isometry of the two $S^2$'s of the SUSY Janus solution. Notice that SUSY fixes the coefficients of the terms in ${\cal L}_{\rm int}$. Performing an $SL(2,\mathbb{R})$ transformation will generate a jumping $\theta$-angle, as well as new interface-localized terms whose coefficients are again fixed by SUSY~\cite{Gaiotto:2008sd}.

\subsection{Janus and Topological Insulators}
\label{topo}

For both non-SUSY and SUSY Janus, we can generate new solutions with non-vanishing axion using $SL(2,\mathbb{R})$ transformations. An $SL(2,\mathbb{R})$ transformation of the Janus solution leaves the Einstein metric unchanged, but produces a new dilaton, $\phi'$, and a non-trivial axion, $C_{(0)}$:
\beq
\label{eq:postsl2r}
C_{(0)} = \frac{b \, d + a \, c \, e^{-4 \phi}}{d^2 + c^2 \, e^{-4\phi}}, \qquad e^{2 \phi'} = d^2 e^{2\phi} + c^2 e^{-2\phi},
\eeq
where $a,b,c,d \in \mathbb{R}$ and $ad-bc=1$. The transformation appears to introduce three new parameters (four real numbers with a constraint), but by an appropriate constant shift of $\phi$, which was a symmetry of the $C_{(0)}=0$ solution, we can always set either $c$ or $d$ to one. The $SL(2,\mathbb{R})$ transformation thus introduces only two new parameters, the two asymptotic values of $C_{(0)}$. We can easily construct solutions with the dilaton constant at the boundary but with a jumping axion by demanding $e^{2\phi'_+} = e^{2 \phi'_-}$, or equivalently
\beq
\label{eq:postsl2rconstantdilaton}
d^2 e^{2\phi_+} + c^2 e^{-2\phi_+} = d^2 e^{2\phi_-} + c^2 e^{-2\phi_-} \quad \Rightarrow \quad \frac{c^2}{d^2} = \frac{e^{2 \phi_+} - e^{2\phi_-}}{e^{-2\phi_-}-e^{-2\phi_+}}.
\eeq
For the $C_{(0)}$ in eq.~\eqref{eq:postsl2r} to be finite and for a solution to eq.~\eqref{eq:postsl2rconstantdilaton} to exist, $c$ and $d$ must both be non-zero, which implies that $e^{2 \phi'}$ satisfies a bound, $e^{2\phi'} \geq 2 |cd|$. Recall that the string coupling is $g_s = e^{2\phi'}$. If we work with $SL(2,\mathbb{Z})$, so that $c$ and $d$ are integers, then because of this bound the string coupling will be order one or larger. To remain in the weakly-coupled regime, we thus move outside of $SL(2,\mathbb{Z})$, to $SL(2,\mathbb{R})$, so that we can adjust $c$ and $d$ to keep $g_s$ sufficiently small. Although we are then not guaranteed quantized charges in the full quantum theory, type IIB string theory, at the classical level $SL(2,\mathbb{R})$ transformations give us perfectly valid solutions of supergravity. We can achieve a $\d C_{(0)}$ of any size by suitable adjustments of $a$, $b$, $c$, and $d$.

As mentioned in the introduction, we can think of $\N=4$ SYM with a constant coupling and a $\theta$-angle jumping from zero to $\pi$ (mod $2\pi$) as the low-energy effective description of a (3+1)-dimensional T-invariant fractional TI, as we will now explain in detail.

A TI is a physical system with
\begin{enumerate}
\item a $U(1)$ symmetry for which we will study transport, for example the $U(1)$ gauge invariance of electromagnetism,
\item a mass gap in the sector charged under the $U(1)$ (hence an insulator), and
\item a topological quantum number distinct from the vacuum, which appears at low energies as a quantized, dissipationless transport coefficient. Here ``topological'' means invariant under any continuous deformation that preserves all symmetries and does not close the mass gap.
\end{enumerate}
Since 1980, the canonical example of a TI has been the integer QHS, which breaks time reversal symmetry, T, and where the topological quantum number is the Hall conductivity. Since 2005 a number of TIs have been discovered that preserve T. One example is $\textrm{HgTe}$, for which electronic transport is effectively (2+1)-dimensional. In (3+1) dimensions, examples include $\textrm{Bi}_2\textrm{Se}_3$, $\textrm{Bi}_2\textrm{Te}_3$, and $\textrm{Sb}_2\textrm{Te}_3$. For reviews of TIs, see for example refs.~\cite{Hasan:2010xy,Moore:2010rev,2011RvMP...83.1057Q,2011ARCMP...2...55H}.

The physics of all the TIs mentioned above is simplest to understand via two levels of effective field theory. We start at the shortest length scales, or highest energies, with a lattice Hamiltonian and associated band structure, which by assumption is that of an insulator, with a valence band separated from a conduction band by a band gap. Here the topological quantum number appears as a topological invariant associated with a Berry's connection defined over momentum space (the Brillouin zone) using the electronic Bloch wave-functions. In the integer QHS the topological quantum number is the first Chern number of this (Abelian) Berry's connection, summed over all occupied bands~\cite{Thouless:1982zz} (the TKNN invariant), which is $\mathbb{Z}$-valued. For the (3+1)-dimensional T-invariant TIs, the topological quantum number is the $\mathbb{Z}_2$-valued integral of the (non-Abelian) Chern-Simons form built from the Berry's connection~\cite{2007PhRvB..75l1306M,2009PhRvB..79s5322R,2007PhRvL..98j6803F,2008PhRvB..78s5424Q,2008arXiv0810.2998E}.

Integrating out all bands except the top-most band, we reach the first level of effective field theory: a free Dirac Hamiltonian whose mass matrix, which may involve non-trivial Dirac matrices, encodes the information about the symmetries of the state. For the integer QHS, the effective theory is some number of Dirac fermions with real masses, which break T. Here the $\mathbb{Z}$-valued topological invariant appears simply as the number of Dirac fermions. For the (3+1)-dimensional T-invariant TIs, the effective theory is a single Dirac fermion with a complex mass. Here the $\mathbb{Z}_2$-valued topological invariant appears as the phase of that mass, which can be either of two values allowed by T-invariance, zero or $\pi$ (mod $2\pi$). In other words, the mass is real but can be positive or negative.

Integrating out the massive Dirac fermion(s), we obtain the second, and lowest, level of effective field theory, a topological term for external $U(1)$ electric and magnetic fields.\footnote{If the $U(1)$ is gauged, and the associated dynamical gauge field is gapless, then the effective description at any scale must include gauge field kinetic terms, \textit{i.e.}\ a Maxwell action. For simplicity, we will neglect the Maxwell contribution to the low-energy effective action, so that all electric and magnetic fields are external/non-dynamical.} Here the topological invariant appears as the coefficient of this term, and thus determines a dissipationless, quantized transport coefficient associated with the $U(1)$ charge. In the integer QHS, the ultimate low-energy physics is described by Chern-Simons theory~\cite{PhysRevLett.51.2077,PhysRevLett.52.18,doi:10.1142/S0217979292000037}, whose coefficient determines the Hall conductivity, which is thus $\mathbb{Z}$-quantized. For the (3+1)-dimensional T-invariant TIs, the low-energy physics is a $\theta$-angle term, where T-invariance demands that the value of $\theta$ be either zero or $\pi$ (mod $2\pi$)~\cite{2008PhRvB..78s5424Q,2008arXiv0810.2998E} (the phase of the mass in the previous level of effective description). The $\theta$-angle term determines the magneto-electric response~\cite{2008PhRvB..78s5424Q,2008arXiv0810.2998E}: an applied magnetic field generates an electric polarization and an applied electric field generates a magnetization. In a (3+1)-dimensional T-invariant TI, the magneto-electric response is thus $\mathbb{Z}_2$-quantized.

An analysis of quantum anomalies in TI states reveals that gapless fermionic modes must necessarily exist at the boundary between a TI and the vacuum, which we will call ``edge modes''~\cite{Ryu:2010ah}.\footnote{More generally, such edge modes exist at the boundary between two TIs with different topological quantum numbers. Notice that the anomaly analysis proves that the two definitions of TIs, the one we gave in the introduction, in terms of gapless edge modes, and the one in this section, are equivalent.} Given that the bulk of the material is insulating, these edge modes are what give rise to quantized $U(1)$ response. Indeed, topology and symmetry forbid the edge modes from becoming localized, hence they produce dissipationless $U(1)$ response even in the presence of arbitrarily strong disorder, so long as that disorder does not alter the symmetries~\cite{PhysRevLett.99.146806}. In the integer QHS the edge modes are (1+1)-dimensional chiral fermions~\cite{PhysRevB.23.5632,PhysRevB.25.2185}. In (3+1)-dimensional T-invariant TI states, the edge modes are (2+1)-dimensional Dirac fermions~\cite{2007PhRvL..98j6803F,2007PhRvB..76d5302F}.

The edge modes appear in the first level of effective description as a Dirac mass that varies in space with a zero at the location of the boundary, indicating the presence of boundary-localized zero modes~\cite{2007PhRvB..76d5302F,2008PhRvB..78s5424Q}. Reaching the lowest level of effective description requires giving the edge modes a mass, for example by breaking some or all of the symmetry ``locally'', at the boundary only, \textit{i.e.}\ somehow suppressing the transmission of symmetry-breaking effects to the bulk~\cite{2007PhRvB..76d5302F,2008PhRvB..78s5424Q}. Integrating out the now-massive edge modes, we reach the lowest level of effective description, a topological term whose coefficient jumps at the boundary. For example, for (3+1)-dimensional T-invariant TIs, the ultimate low-energy description of the boundary is a $\theta$-angle term where $\theta$ jumps from zero to $\pi$ (mod $2\pi$) at the boundary~\cite{2008PhRvB..78s5424Q}.

Remarkably, a complete classification exists for all TI states that can be described by Hamiltonians in the same universailty classes as the free Dirac Hamiltonians (\textit{i.e.}\ connected by adiabatic continuity)~\cite{PhysRevB.78.195125,Ryuetalrev,Kitaev:2009mg,Ryu:2010zza,Ryu:2010ah,LeClair:2012qc}. What TI states are possible for Hamiltonians in other universality classes remains an open question. To date, the only examples realized experimentally are fractional QHSs, in which the Hall conductivity is quantized to be a rational number times the electron's charge (in units of Planck's constant). A natural question is whether fractional T-invariant TIs exist.

To our knowledge, no precise definition of fractionalization exists.\footnote{For a recent attempt to define an order parameter for fractionalization, see ref.~\cite{Hartnoll:2012ux}.} The concept of fractionalization is simple, however: in an interacting many-electron system, the low-energy emergent degrees of freedom can carry a fraction of the electron's charge. Indeed, that is what happens in (1+1)-dimensional interacting many-electron systems, Luttinger liquids~\cite{1995RPPh...58..977V}, where charge and spin propagate independently, and is also what happens in fractional QHSs~\cite{1983PhRvL..50.1395L}.

In one approach to fractionalization, which goes by various names, the ``partonic,'' ``projective,'' or ``slave particle'' approach~\cite{PhysRevB.40.8079,doi:10.1142/S0217984991000058,PhysRevLett.66.802,Blok1992615,doi:10.1142/S0217979292000840,PhysRevB.60.8827,doi:10.1142/S0217979292000037,2010PhRvB..81o5302B}, we begin by writing the electron creation operator as a product of some number of other operators, \textit{i.e.}\ some constituent ``partons.'' The statistics and $U(1)$ charges of these partons must be arranged such that their product is fermionic and has a net $U(1)$ charge equal to the electron's. We must also enlarge the Hilbert space to account for these new degrees of freedom, with a constraint that only electrons appear in physical states. Of course, such a description, which introduces additional, redundant degrees of freedom, is always possible, but usually is an unnecessary complication. A fractionalized phase, however, corresponds to a ``deconfined'' phase for the partons, \textit{i.e.}\ a phase in which the partons are eigenstates of the Hamiltonian, rather than the electrons.

Suppose for example we want the electron to fractionalize into $n$ constituent partons. We can achieve that by extending the symmetry group from $U(1)$ to  $U(1) \times SU(n)$, with the partons in the fundamental representation of $SU(n)$. An electron is then a singlet of $SU(n)$ but charged under the $U(1)$. At the first level of effective field theory we obtain not a free Dirac Hamiltonian for the electron but a theory of partons interacting via emergent $SU(n)$ gauge fields, which in this context are called ``statistical gauge fields''~\cite{doi:10.1142/S0217979292000037,Maciejko:2011ed}. A fractionalized state then is a deconfined, or at least non-confined, state of this non-Abelian gauge theory. To describe a partonically fractionalized T-invariant TI~\cite{Maciejko:2010tx,Swingle:2010rf,Maciejko:2011ed}, the partons must have complex masses whose phases are either zero or $\pi$ (mod $2\pi$). Integrating out the partons, at the lowest level of effective theory we obtain $SU(n)$ gauge fields with a $\theta$-angle equal to zero or $\pi$ with periodicity $2\pi$, and a $U(1)$ $\theta$-angle equal to zero or $\pi/n$ with a periodicity $2\pi/n$, which guarantees T-invariance~\cite{Maciejko:2010tx}. The fractional $U(1)$ $\theta$-angle will clearly produce fractional magneto-electric response. Besides deconfined non-Abelian statistical gauge fields (with gauge group $SU(n)$ or otherwise), (3+1)-dimensional partonically fractionalized T-invariant TIs can be described by non-Abelian statistical gauge fields with the gauge group Higgsed to a discrete subgroup or by Abelian statistical gauge fields~\cite{Maciejko:2011ed}.

What are the edge modes in a (3+1)-dimensional partonically fractionalized T-invariant TI? They are (2+1)-dimensional fermions in ``half'' of a fractional QHS, \textit{i.e.}\ the Hall conductivity is quantized to be $1/2$ times a rational number times the electron's charge~\cite{Maciejko:2010tx}. Gapping these edge modes and integrating out both the partons and the edge modes, we reach a description of the edge at the lowest level of effective field theory, with an $SU(n)$ $\theta$-angle that jumps from zero to $\pi$ (mod $2\pi$) and a $U(1)$ $\theta$-angle that jumps from zero to $\pi/n$ (mod $2\pi/n$). In other words, on one side is the vacuum, a state that is non-fractionalized and topologically trivial, while on the other side is the TI, which is fractionalized and topologically non-trivial.

We can now state our main proposal: we can think of $\N=4$ SYM with a $\theta$-angle that jumps from zero to $\pi$ as the low-energy effective description of a (3+1)-dimensional fractional T-invariant TI, where the statistical gauge fields are the $SU(N_c)$ gauge fields. Notice that $\N=4$ SYM with a jumping $\theta$-angle actually describes an interface between a state that is fractionalized and topologically non-trivial and a state that is fractionalized and topologically trivial, \textit{i.e.}\ \textit{both} states are fractionalized. Said another way, our topologically-trivial vacuum state is the conformally-invariant vacuum of $\N=4$ SYM, which, being non-confining, we interpret as fractionalized.

Our proposal raises several important questions. Where is the $U(1)$ symmetry that will have fractional $\theta$-angle, \textit{i.e.}\ the $U(1)$ mentioned in part 1 of our definition of TIs above?\footnote{We are not the first to propose that $\N=4$ SYM with a jumping $\theta$-angle can be interpreted as the low-energy effective description of a fractional TI. A similar proposal was made in ref.~\cite{Karch:2009sy}, however there the $SU(N_c)$ was identified as the generalization of the $U(1)$ of electromagnetism, rather than as the gauge group of the statistical gauge fields. Clearly that cannot be the case. In a TI the $U(1)$ of electromagnetism has gauge-invariant currents from which we can measure transport coefficients. The $SU(N_c)$ gauge currents are not gauge-invariant, hence we have no obvious way to define gauge-invariant transport coefficients from them.} What kind of edge modes did the system have, before we gapped them and integrated them out to obtain $\N=4$ SYM with a jumping $\theta$-angle? We will not be able to answer these questions completely, but we will take some first steps.

In previous constructions of holographic duals to (3+1)-dimensional fractional TIs~\cite{HoyosBadajoz:2010ac,Ammon:2012dd} the $U(1)$ was a global flavor symmetry. In particular, consider $\N=4$ SYM coupled to $N_f$ $\N=2$ SUSY hypermultiplets in the fundamental representation of $SU(N_c)$, \textit{i.e.}\ flavor fields. The $\N=2$ hypermultiplet includes a Dirac fermion, a ``quark'', and two complex scalars, the ``squarks''.

$\N=2$ SUSY admits a constant complex hypermultiplet mass. We will assume that the mass matrix in flavor space is proportional to the identity, so that the theory has a global $U(N_f)$ symmetry which we may decompose as $U(1) \times SU(N_f)$. We will call the $U(1)$ factor baryon number, and henceforth ignore the $SU(N_f)$ factor.

If $N_f \ll N_c$ then we can take the probe limit, where we expand all observables in $N_f/N_c \ll 1$ and retain terms only up to order $N_f N_c$. Physically, the probe limit corresponds to discarding quantum effects due to the flavor fields. In diagrammatic perturbation theory in $\lambda$, the probe limit consists of discarding diagrams with quark or squark loops.

The holographic dual of large-$N_c$, strongly-coupled $\N=4$ SYM coupled to $N_f$ hypermultiplets, in the probe limit, is type IIB supergravity in $AdS_5 \times S^5$ with $N_f$ probe D7-branes extended along $AdS_5 \times S^3$~\cite{Karch:2002sh}. Roughly speaking, the magnitude and phase of the complex hypermultiplet mass are dual to the two wordlvolume scalars on the D7-branes that describe the position of the D7-branes in the two transverse directions on the $S^5$. The conserved baryon number current is dual to the D7-brane $U(1)$ worldvolume gauge field~\cite{Nakamura:2006xk,Kobayashi:2006sb}.

A nonzero, real $\N=2$ SUSY-preserving mass $M$ appears in the bulk, in the probe limit, as a D7-brane that ``ends'' at some radial position~\cite{Karch:2002sh}. More precisely, at the $AdS_5$ boundary the D7-branes wrap an equatorial $S^3 \subset S^5$, but as they extend into the bulk the $S^3$ shrinks and eventually collapses to a point at some radial position, where the D7-brane ``ends''. The position where the D7-brane ends corresponds to the mass scale $M/\sqrt{\lambda}$~\cite{Kruczenski:2003be}.

In the language of partonic fractionalization, the hypermultiplet fields are the partons and the $SU(N_c)$ gauge fields are the statistical gauge fields. Giving the hypermultiplets a mass whose phase varies in one spatial direction as a step function, jumping from zero to $\pi$, will produce an edge between fractionalized T-invariant TI states, as described above. Integrating out the partons, and also gapping and integrating out the edge modes, we reach the lowest level of effective theory, in which the baryon number $U(1)$ acquires a $\theta$-angle that jumps from zero to $\pi/N_c$ while the $SU(N_c)$ gauge fields will acquire a $\theta$-angle that jumps from zero to $\pi$. In the probe limit, however, the jump in the $SU(N_c)$ $\theta$-angle is a sub-leading effect, and will not be apparent.

Remarkably, holographic duals of such an edge between fractional TI states have been found, using probe D7-branes~\cite{HoyosBadajoz:2010ac,Ammon:2012dd}. These D7-branes have a more complicated embedding than the simple ``D7-brane that ends,'' so we will not attempt to describe the embeddings in detail, rather we will just highlight a few salient points. Two solutions have been found so far, one that breaks all SUSY, in which only the worldvolume scalars are active~\cite{HoyosBadajoz:2010ac}, and another solution that preserves four real supercharges of the $AdS_5 \times S^5$ background, and in which both the worldvolume scalars and the gauge field are active~\cite{Ammon:2012dd}. The baryon number $U(1)$'s fractional $\theta$-angle can be extracted from the probe D7-brane Wess-Zumino term~\cite{HoyosBadajoz:2010ac}. The probe D7-brane does not affect any of the background supergravity fields, including the axion, which indicates that in the field theory the $SU(N_c)$ $\theta$-angle does not jump, as expected in the probe limit.

Most importantly, the edge modes are apparent in the bulk: near the interface, the D7-branes bend and become extended along $AdS_4 \times S^4$~\cite{HoyosBadajoz:2010ac}. Such D7-branes along $AdS_4 \times S^4$ (in the absence of worldvolume gauge field flux) break all SUSY, and are dual to (2+1)-dimensional fermions alone, with no scalar superpartners, and with couplings only to the $\N=4$ SYM gauge field and one of the six real $\N=4$ SYM scalars, both restricted to the (2+1)-dimensional interface~\cite{Davis:2008nv,Myers:2008me,PTPS.177.128,Wapler:2009tr,Wapler:2009rf,Kutasov:2011fr}. These (2+1)-dimensional fermions are the edge modes of the holographic fractional TI states described above. A combination of perturbation theory in $\lambda$ and holographic calculations reveals that these fermions exhibit a phase transition as a function of coupling~\cite{Kutasov:2011fr}: at small $\lambda$, the vacuum preserves parity and $T$, but above some critical $\lambda$ the fermions pair, such that the Dirac mass operator acquires a nonzero expectation value, spontaneously breaking parity and $T$.

The crucial observation for us is that the non-SUSY jumping-axion Janus solution carries D7-brane charge, localized at the boundary at the point where the axion jumps, as can be detected by the monodromy of the axion about that point. The Janus geometry actually has a conical singularity there, associated with an angular excess~\cite{Bak:2003jk}, as expected for D7-branes. The fact that non-SUSY jumping-axion Janus carries both D3- and D7-brane charge suggests that such a solution could be constructed from an intersection of D3- and D7-branes, an obvious candidate being the non-SUSY intersection of ref.~\cite{HoyosBadajoz:2010ac}. To go from the probe D7-brane solutions of ref.~\cite{HoyosBadajoz:2010ac} to non-SUSY jumping-axion Janus, we would need to leave the probe limit, allowing the D7-branes to back-react on the fields of supergravity, and require the D7-branes to end at (or near) the boundary, so that in the field theory both the partons and the edge modes are, in effect, infinitely heavy, and can be integrated out. If we performed those steps then we would expect the resulting solution to look like non-SUSY jumping-axion Janus deep in the bulk (at very low energies).

Many things about such a construction are unclear and confusing to us, on both sides of the correspondence, however. For example, on the supergravity side, the back-reaction of the D7-branes will break the $SO(6)$ isometry of the $S^5$ to a subgroup, since as probes the D7-branes wrap only an $S^3 \subset S^5$. How can the full $SO(6)$ isometry be restored deep in the bulk? On the field theory side, all SUSY and part of the R-symmetry are broken, so if we integrate out the hypermultiplets on both sides of the interface, and gap the edge modes and integrate them out, we have little reason to expect that the resulting effective Lagrangian would be precisely that of $\N=4$ SYM with a jumping $\theta$-angle. Such issues may be clearer if SUSY is restored: a D-brane construction of SUSY jumping-axion Janus may be more straightforward, especially on the field theory side, where the SUSY should provide greater control over allowed terms in a low-energy effective action.\footnote{The SUSY probe D7-brane solution of ref.~\cite{Ammon:2012dd} cannot give rise to the SUSY jumping-axion Janus solution reviewed in section \ref{susyjanus}, since the former preserves only four real supercharges while the latter preserves sixteen real supercharges of the $AdS_5 \times S^5$ solution.}

If we could determine an intersection of D3-branes and D7-branes that gives rise to non-SUSY jumping-axion Janus, then we could determine the edge modes in the dual fractional TI. If we assume that such a D-brane construction exists, then we can identify the ``missing'' $U(1)$: the $U(1)$ with fractional $\theta$-angle is the baryon number $U(1)$. When the D7-branes are in a probe limit, that $U(1)$ appears in the open string sector in the bulk, as the $U(1)$ worldvolume D7-brane gauge field. Including the back-reaction of the D7-branes, and requiring that the D7-branes end at the boundary, we expect to find a description in terms of closed strings alone, with the bulk degrees of freedom that are dual to the $U(1)$ localized at the boundary, at the point where the axion jumps. Such modes would be very much like singletons, in that they would be non-dynamical, boundary-localized modes. Indeed, such modes would be similar to a familiar singleton: recall that the worldvolume theory on a stack of $N_c$ D3-branes is actually $U(N_c)$, not just $SU(N_c)$, where the ``extra'' $\N=4$ $U(1)$ vector multiplet appears in holography as a non-dynamical, boundary-localized singleton mode (for a review, see section 3.1 of ref.~\cite{Aharony:1999ti}). In our case, since the baryon number $U(1)$ is global (unlike the $U(1) \in U(N_c)$), any effective low-energy description will include only the jumping fractional $\theta$-angle, and not a Maxwell term.

To summarize, our proposal is the following. Non-SUSY Janus with an axion that jumps from zero to $1/2$ (mod one) is the holographic dual of an interface CFT that we can interpret as the low-energy effective description of an interface between two fractionalized states, a topologically trivial vacuum and a T-invariant TI. More precisely, the interface CFT is $\N=4$ SYM with a $\theta$-angle that jumps from zero to $\pi$ (mod $2\pi$). We interpret the $SU(N_c)$ gauge fields as statistical gauge fields. The $U(1)$ with fractional $\theta$-angle is a global symmetry, namely a baryon number $U(1)$ associated with the partons and edge modes that we integrated out to obtain the interface CFT. We leave several open questions for future research, chief among them being: what were the edge modes that were integrated out?

\section{Wilson Loops: Field Theory Calculation}
\label{wilsonfieldtheory}

Our goal is to calculate the potential $V$ representing the self-energy of a single test charge or the potential between heavy test charges in $\N=4$ SYM with a conformal interface, with gauge group $SU(N_c)$ and in the large-$N_c$ limit. In subsection \ref{em} we review the analogous calculation in ordinary electromagnetism. In section~\ref{sunc} we turn to the calculation of $V$ in $\N=4$ SYM using perturbation theory in the 't Hooft coupling. In section \ref{wilsonholo} we will compare our results for $V$ in electromagnetism and in perturbative $\N=4$ SYM to our holographic results.

\subsection{Electromagnetism}
\label{em}

The Lagrangian of (3+1)-dimensional electromagnetism, in the absence of sources, is
\beq
{\cal L}_{EM} = - \frac{1}{4 g^2} F_{\mu \nu} F^{\mu \nu} - \frac{\theta}{32 \pi^2} \epsilon^{\mu \nu \rho \sigma} F_{\mu \nu} F_{\rho \sigma},
\eeq
where $F_{\mu\nu}$ is the $U(1)$ field strength. In order to study an interface, we will allow $g$ and $\theta$ to be functions of position. Let us define the canonical momentum $G^{\mu \nu}$ as
\beq
\label{eq:defG}
G_{\mu \nu} \equiv -2 \frac{\delta {\cal L}_{EM}}{\delta F^{\mu \nu}} = \frac{1}{g^2} F_{\mu \nu} + \frac{\theta}{8 \pi^2} \epsilon_{\mu \nu \rho \sigma} F^{\rho \sigma},
\eeq
which is also known as a ``constitutive relation.'' Next we introduce electric and magnetic source currents,
\beq
J^{\mu}_e = 2\pi( \rho_e,j_e^i), \qquad J_m^{\mu} = (\rho_m,j_m^i),
\eeq
respectively, where the time components of the currents, $\rho_e$ and $\rho_m$, are the electric and magnetic charge densities, while $j_e^i$ and $j^i_m$ are the spatial electric and magnetic currents, with $i=1,2,3$. The factor of $2\pi$ in our definition of $J^{\mu}_e$ is for later convenience. Including the sources, the equation of motion and Bianchi identity may be written as
\beq
\partial_{\mu} G^{\mu \nu} = J_e^{\nu}, \qquad \frac{1}{2} \epsilon^{\mu \nu \rho \sigma} \partial_{\mu} F_{\rho \sigma} = J_m^\nu.
\eeq
Next, let us introduce the electric field $E^i$, magnetic field $B^i$, electric displacement $D^i$, and magnetic displacement $H^i$:
\beq
\label{ebdhdef}
F^{i0} \equiv E^i, \quad F^{ij} \equiv - \epsilon^{ijk} B^k, \quad G^{i0} \equiv D^i, \quad G^{ij} \equiv - \epsilon^{ijk} H^k.
\eeq
The explicit expressions for the fields $\Dv$ and $\Hv$ in terms of $\Ev$ and $\Bv$ are
\beq
\label{DBexp}
\Dv = \frac{1}{g^2} \Ev + \frac{\theta}{4 \pi^2} \Bv, \qquad \Hv = \frac{1}{g^2} \Bv - \frac{\theta}{4 \pi^2} \Ev \ .
\eeq
For later use, let us also define a matrix ${\cal M}$ by rewriting the constitutive relation:
\beq
\label{mdef}
\left(\begin{array}{c} 2 \pi \Dv \\ \Bv \end{array} \right) = {\cal M} \left(\begin{array}{c} \Ev \\ 2 \pi \Hv \end{array} \right), \qquad {\cal M} \equiv \left( \begin{array}{cc} \frac{2\pi}{g^2} + \frac{g^2 \theta^2}{8 \pi^3} & \frac{g^2 \theta}{4\pi^2} \\ \frac{g^2 \theta}{4 \pi^2} & \frac{g^2}{2\pi} \\ \end{array}\right).
\eeq
Using $\Ev$, $\Bv$, $\Dv$, and $\Hv$, the equation of motion and Bianchi identity take familiar forms,
\begin{align}
\label{eq:meom}
{\vec \nabla} \cdot \Dv &=  \rho_e, \qquad  \qquad \qquad {\vec \nabla} \cdot \Bv = \rho_m, \nonumber\\
{\vec \nabla} \times \Hv &= \frac{\p \Dv}{\p t}  + \vec j_e, \qquad - {\vec \nabla} \times \Ev = \frac{\p \Bv}{\p t} + \vec j_m.
\end{align}
The net electric and magnetic charges are defined from the above equations as
\beq
Q^e = \int d^3x \, \vec{\nabla} \cdot \Dv, \qquad Q^m = \int d^3x \, \vec{\nabla} \cdot \Bv.
\eeq

We will consider an interface at $x_3=0$, allowing both $g$ and $\theta$ to jump there, with values $g_-$ and $\theta_-$ in the $x_3<0$ region and values $g_+$ and $\theta_+$ in the $x_3>0$ region. We will calculate the potential $V$ in the two regions $x_3>0$ and $x_3<0$ using the method of \textit{image charges}.

\begin{figure}[ht!]
  \begin{center}
   \subfigure[]{\includegraphics[width=0.45\textwidth]{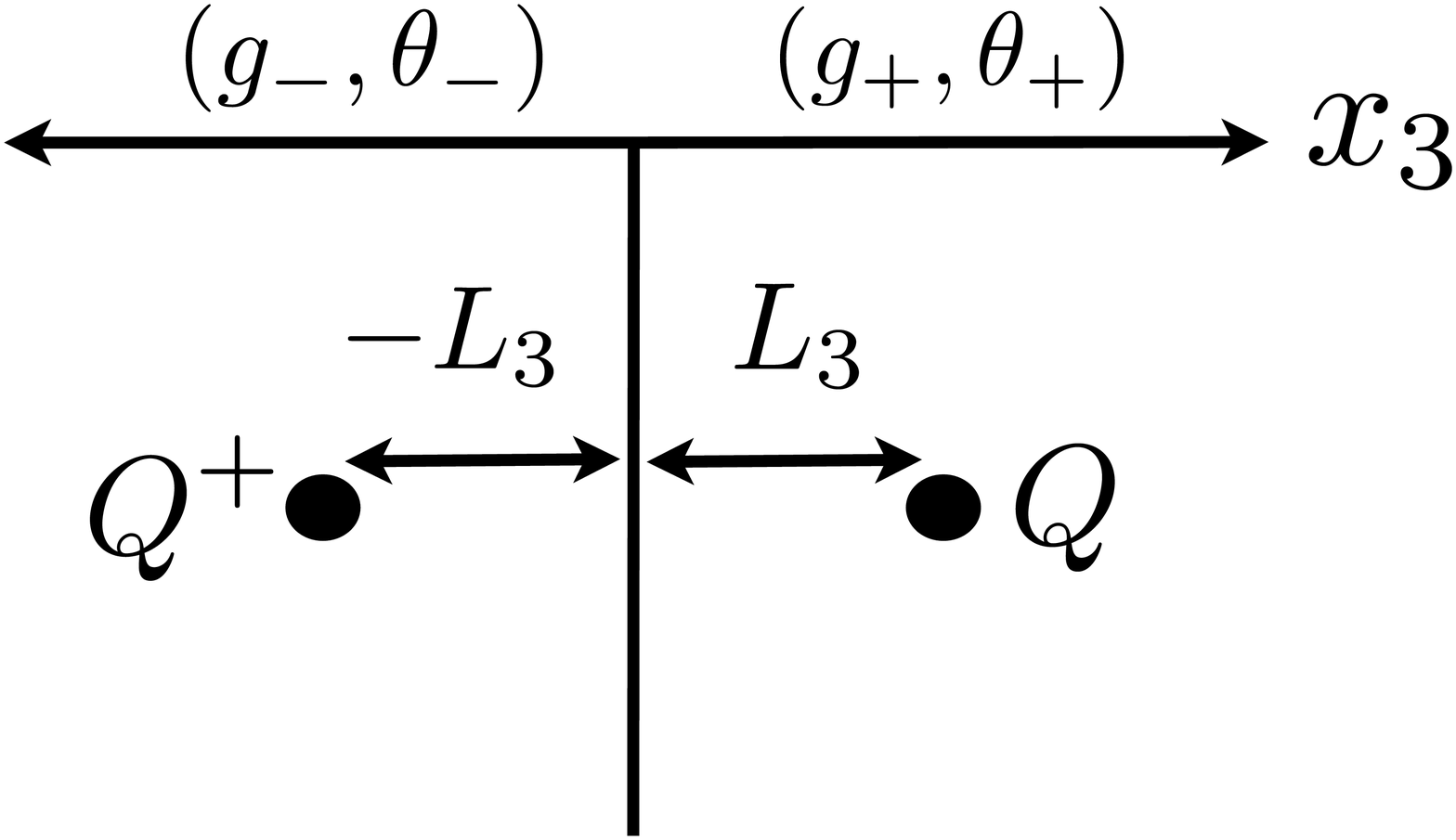}}\hspace{5mm}
     \subfigure[]{\includegraphics[width=0.45\textwidth]{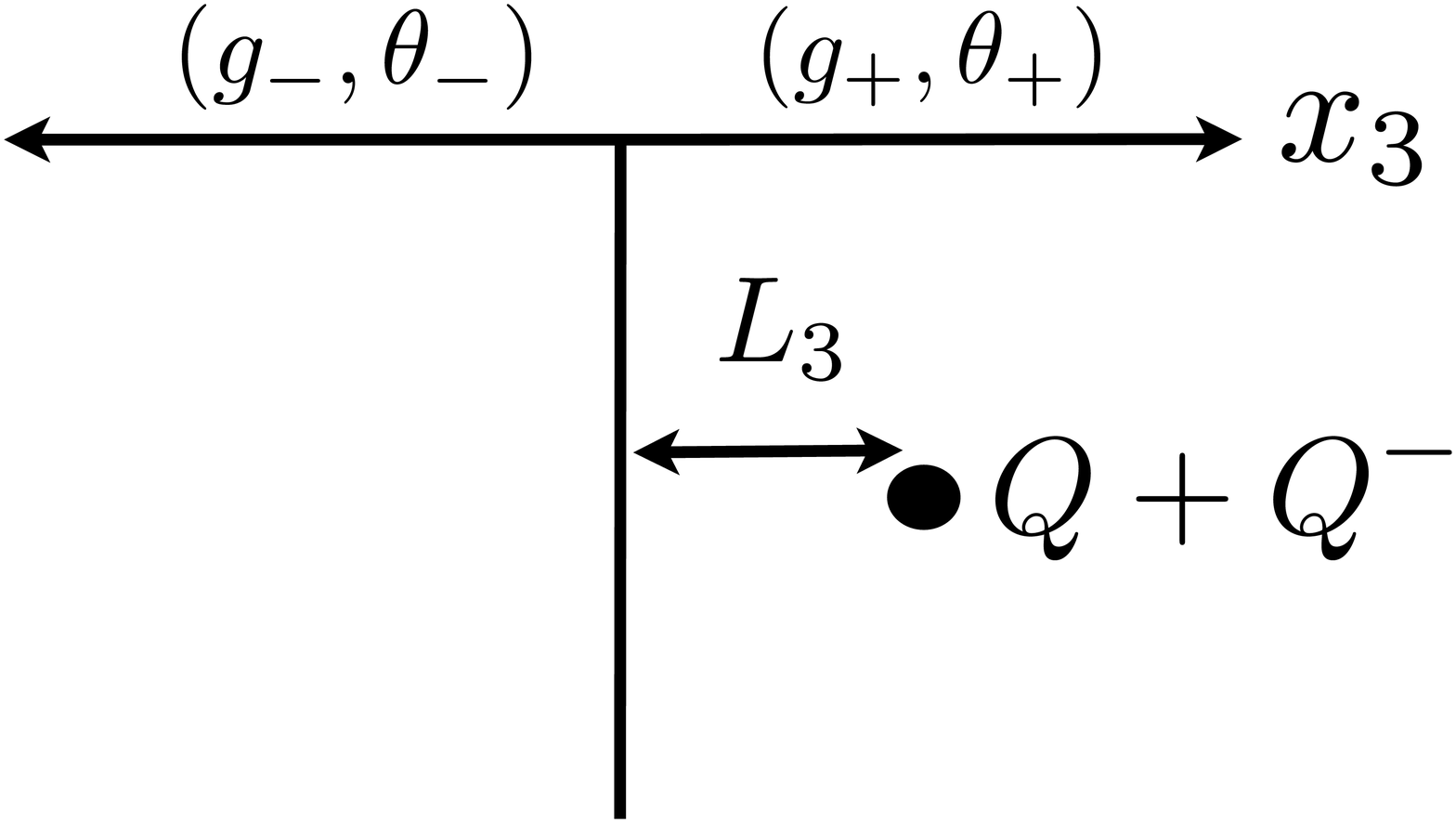}}
  \end{center}
\caption{Depiction of the image charges associated with a test charge $Q$ in electromagnetism with a jumping coupling and/or $\theta$-angle. The horizontal axis is the $x_3$ direction, with the interface at $x_3=0$: the coupling and $\theta$-angle have values $(g_-,\theta_-)$ for $x_3<0$ and $(g_+,\theta_+)$ for $x_3>0$. We place the test charge $Q$ (the black dot) in the $x_3>0$ region, at a position $x_3=L_3$. (a) Observers in the $x_3>0$ region will detect an image charge $Q^+$ in the $x_3<0$ region, a distance $L_3$ from the interface. (b) Observers in the $x_3<0$ region will detect an image charge $Q^-$ coincident with $Q$. Equivalently, in the $x_3<0$ region the interface effectively shifts the charge $Q$ as $Q \to Q+Q^-$.}
\label{fig:imagecharges1}
\end{figure}

We begin by introducing a single test charge and determining the fields $\Ev$, $\Bv$, $\Dv$, and $\Hv$ it produces in the presence of the interface. From the equations of motion in eq.~\eqref{eq:meom}, we can show that at the interface, the perpendicular components of $(2\pi \Dv,\Bv)$ and the parallel components of $(\Ev, 2 \pi \Hv)$ must be continuous. Explicitly, these matching conditions are
\begin{align}
\label{matchingcond}
D_3 |_{x_3 \to 0^-} = D_3 |_{x_3 \to 0^+},
\quad & \quad
B_3 |_{x_3 \to 0^-} = B_3 |_{x_3 \to 0^+}, & \nonumber \\
E_i|_{x_3 \to 0^-} = E_i|_{x_3 \to 0^+},
\quad & \quad
H_i|_{x_3 \to 0^-} = H_i|_{x_3 \to 0^+}, & i = 1,2.
\end{align}
We can reproduce these conditions by introducing image charges. To be concrete, consider a dyonic test charge with electric charge $Q^e$ and magnetic charge $Q^m$. We will arrange these into a column vector $Q = (2 \pi Q^e,Q^m)^T$. We will place our test dyon at position $\vec{x}' = (x_1',x_2',x_3')$, with $x_3' = L_3 >0$, which is the side with coupling and $\theta$-angle $g_+$ and $\theta_+$, as depicted in fig.~\ref{fig:imagecharges1}. Next we introduce electric and magnetic potentials $\Phi_e$ and $\Phi_m$, defined by
\beq
\label{DBpots}
\Dv = - \vec{\nabla} \Phi_e, \qquad \Bv = - \vec{\nabla} \Phi_m,
\eeq
and arrange these into a column vector also, $\Phi = (2 \pi \Phi_e, \Phi_m)^T$. We can then introduce the image charges $Q^+$ and $Q^-$ by writing
\beq
\label{imagedefs}
\Phi(\vec{x}) = \begin{cases} \frac{Q}{4 \pi}\frac{1}{r} + \frac{Q^+}{4 \pi} \frac{1}{\tilde{r}} &  x_3 > 0, \qquad \textrm{(same side as $Q$)} \\ & \\ \frac{Q}{4 \pi} \frac{1}{r} + \frac{Q^-}{4 \pi} \frac{1}{r} & x_3 < 0, \qquad \textrm{(side opposite to $Q$)} \end{cases}
\eeq
where
\beq
r = \left | \vec{x} - \vec{x}' \right |, \quad \tilde{r} = \left | \vec{x} - \vec{x}' + 2 L_3 \hat{x}_3\right|,
\eeq
where $\hat{x}_3$ is the unit vector in the $x_3$ direction. In the $x_3>0$ region, the image charge appears to be in the $x_3<0$ region, as depicted in fig.~\ref{fig:imagecharges1} (a), hence the distance to $Q$ is $r$ while the distance to $Q^+$ is $\tilde{r}$, which is larger than $r$. In the $x_3<0$ region the test charge $Q$ and its image $Q^-$ are coincident, hence only $r$ appears in $\Phi(\vec{x})$ in that region. In other words, in the $x_3<0$ region, the effect of the interface is to shift the charge $Q \to Q + Q^-$, as depicted in fig.~\ref{fig:imagecharges1} (b). A straightforward exercise shows that the continuity/matching conditions on $\Ev$, $\Bv$, $\Dv$, and $\Hv$ are satisfied if we choose~\cite{Karch:2009sy}
\beq
\label{imageresult}
Q^+ = \left({\cal M}_+{\cal M}_-^{-1}+1\right)^{-1}\left({\cal M}_+{\cal M}_-^{-1}-1\right)Q, \qquad Q^- = - Q^+,
\eeq
where ${\cal M}_{\pm}$ are the values of ${\cal M}$ on the two sides of the interface. The image charges are completely determined by $Q$ and by the values of $g$ and $\theta$ on each side of the interface. Notice that if $g$ and $\theta$ are the same on both sides of the interface, then ${\cal M}_+{\cal M}_-^{-1} = 1$ and hence $Q^+ = 0$ and $Q^-=0$.

Using the image charges $Q^+$ and $Q^-$ in eq.~\eqref{imageresult}, we can compute the potential energy between test charges in the presence of an interface. The magnetic contribution to the potential energy comes from $\Phi_m$, while the electric contribution comes not from $\Phi_e$ but from the potential $V$ defined in terms of the electrostatic force, $Q^e \Ev \equiv - \vec{\nabla} V$. From eqs.~\eqref{DBexp} and \eqref{DBpots} we have
\beq
\label{Epots}
\Ev = g^2 \Dv - \frac{g^2 \theta}{4 \pi^2} \Bv = - g^2 \vec{\nabla} \Phi_e + \frac{g^2 \theta}{4 \pi^2} \vec{\nabla} \Phi_m.
\eeq
Away from the interface, where $g$ and $\theta$ are constant and hence commute with the gradient $\vec{\nabla}$, we thus find
\beq
V = Q^e \left( g^2 \Phi_e - \frac{g^2 \theta}{4 \pi^2} \Phi_m \right).
\eeq
$V$ is the appropriate potential to compare to our Wilson loop results.

Let us compute $V$ first for the simplest case, with a single test charge $Q$ that is purely electric, $Q = (2\pi Q^e,0)^T$. We want to compute $V$ in the $x_3>0$ region, that is, we want to know the interaction energy of $Q^e$ with its own image charge. Using eq.~\eqref{imageresult}, we can easily compute the values of the image charges,
\begin{subequations}
\bea
\label{singletestimages}
Q^{e+} & = & Q^e \frac{16 \pi^4 (g_-^4 - g_+^4) + g_+^4 g_-^4 (\theta_+^2 - \theta_-^2)}{16\pi^4 (g_+^2 + g_-^2)^2 + g_+^4 g_-^4 (\theta_+ - \theta_-)^2},\\ Q^{m+} & = & Q^e \frac{8\pi^2 g_+^4 g_-^4 (\theta_+ - \theta_-)}{16\pi^4 (g_+^2 + g_-^2)^2 + g_+^4 g_-^4 (\theta_+ - \theta_-)^2}.
\eea
\end{subequations}
Notice that when $\theta_+ \neq \theta_-$ the image charge is dyonic (both $Q^{e+}$ and $Q^{m+}$ are nonzero). The electric and magnetic fields in the $x_3>0$ region are then
\begin{subequations}
\beq
\label{eq:imageE}
\Ev(\vec{x}) = \frac{g_+^2 Q^e}{4 \pi} \frac{\vec{x}-\vec{x}'}{\left| \vec{x}-\vec{x}'\right |^3} + \frac{g_+^2 \tilde{Q}^e}{4 \pi} \frac{\vec{x}-\vec{x}'+ 2 L_3 \, \hat{x}_3}{\left| \vec{x}-\vec{x}' + 2 L_3 \, \hat{x}_3\right |^3},
\eeq
\beq
\label{eq:imageB}
\Bv(\vec{x}) = \frac{Q^{m+}}{4 \pi} \frac{\vec{x}-\vec{x}'+ 2 L_3 \, \hat{x}_3}{\left| \vec{x}-\vec{x}' + 2 L_3 \, \hat{x}_3\right |^3}.
\eeq
\end{subequations}
On the right-hand side in eq.~\eqref{eq:imageE}, the first term is the field produced by $Q^e$ itself while the second term is the field produced by the image charges, which includes contributions from both the electric and magnetic image charges $Q^{e+}$ and $Q^{m+}$, via eq.~\eqref{Epots}. We have labeled the net contribution $\tilde{Q}^e$. Explicitly, $\tilde{Q}^e$ is given by
\beq
\label{tildeqdef}
\tilde{Q}^e =  Q^{e+} - \frac{\theta_+}{4 \pi^2} Q^{m+} = Q^e \left [ \frac{g_-^2 - g_+^2}{g_-^2 + g_+^2} - \frac{2 g_+^4 g_-^6 (\theta_+ - \theta_-)^2}{16 \pi^4 (g_-^2 + g_+^2)^3 + g_+^4 g_-^4 (g_-^2 + g_+^2) (\theta_+ - \theta_-)^2} \right].
\eeq
The potential energy $V$ between $Q^e$ and its image is thus simply
\beq
\label{eq:emsinglev}
V(L_3) = \frac{g_+^2 }{4 \pi} \frac{Q^e \tilde{Q}^e}{2L_3}.
\eeq
Essentially all the physical effects of the interface can be deduced from eq.~\eqref{eq:emsinglev}: since Maxwell electrodynamics is a linear theory, the potential produced by a distribution of charges will ultimately be some linear superposition of the potential for a single charge.

Consider two special cases. First consider a constant $\theta$-angle, $\theta_+ = \theta_-$, with a jumping coupling. If $g_+ > g_-$, that is, if we place $Q^e$ on the side with larger coupling, then from eq.~\eqref{tildeqdef} we see that $\tilde{Q}^e$ will have the \textit{opposite} sign to $Q^e$, hence $Q^e$ will be attracted to the interface. Alternatively, if $g_+ < g_-$, then $\tilde{Q}^e$ will have the \textit{same} sign as $Q^e$, hence $Q^e$ will be repelled from the interface. The general lesson is that $Q^e$ is always attracted to the side with smaller coupling. The second special case is a constant coupling, $g_+ = g_-$, with a jumping $\theta$-angle. In this case, the $\tilde{Q}^e$ in eq.~\eqref{tildeqdef} always has opposite sign to $Q^e$: the last term in eq.~\eqref{tildeqdef} is $Q^e$ times a strictly negative number. Whether $\theta_+ > \theta_-$ or vice-versa doesn't matter. We thus conclude that with constant coupling and jumping $\theta$-angle, $Q^e$ is always attracted towards the interface.

\begin{figure}[ht!]
\begin{center}
\subfigure[]{\includegraphics[width=.45\textwidth]{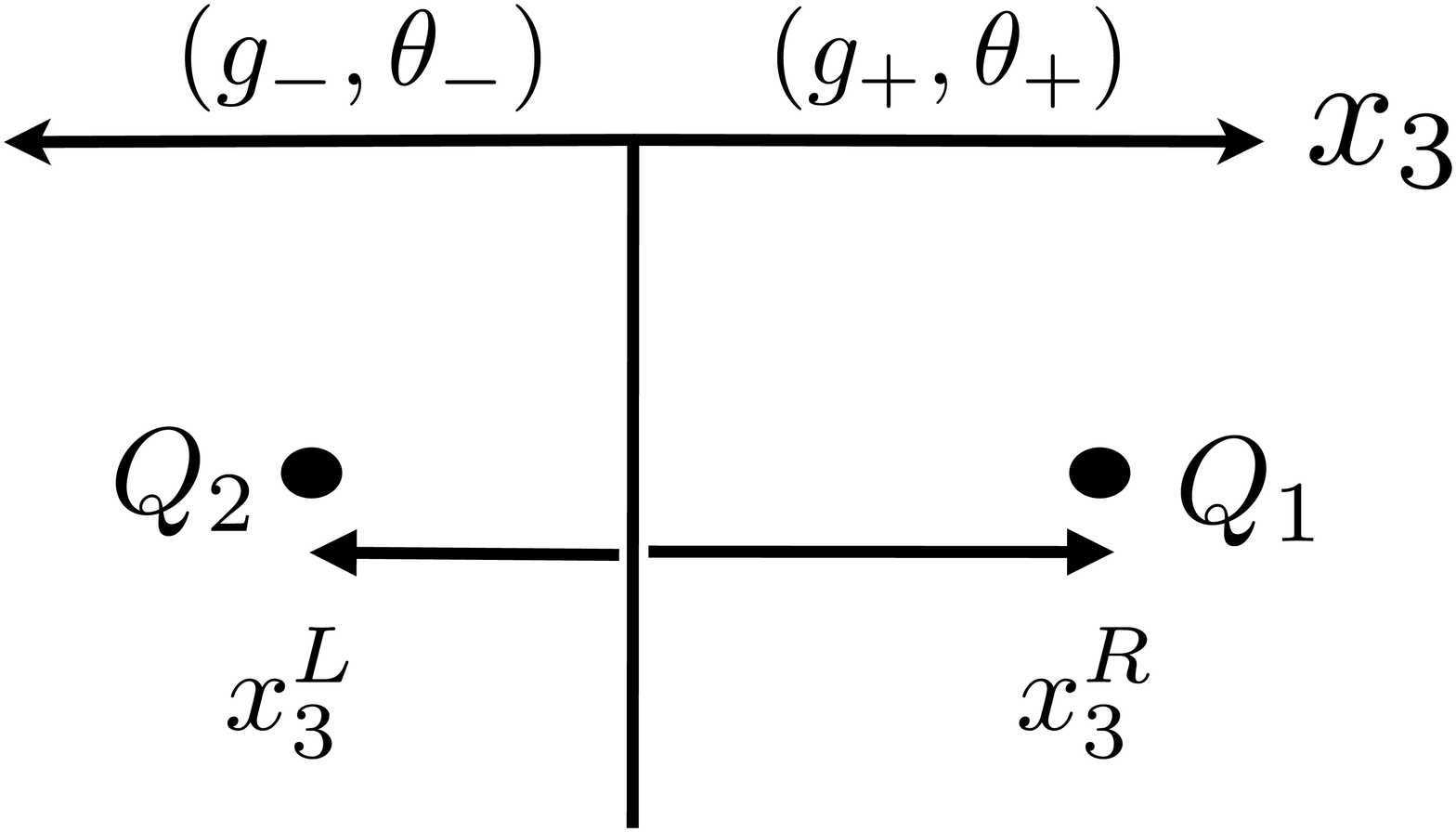}}\hspace{5mm}
\subfigure[]{\includegraphics[width=.45\textwidth]{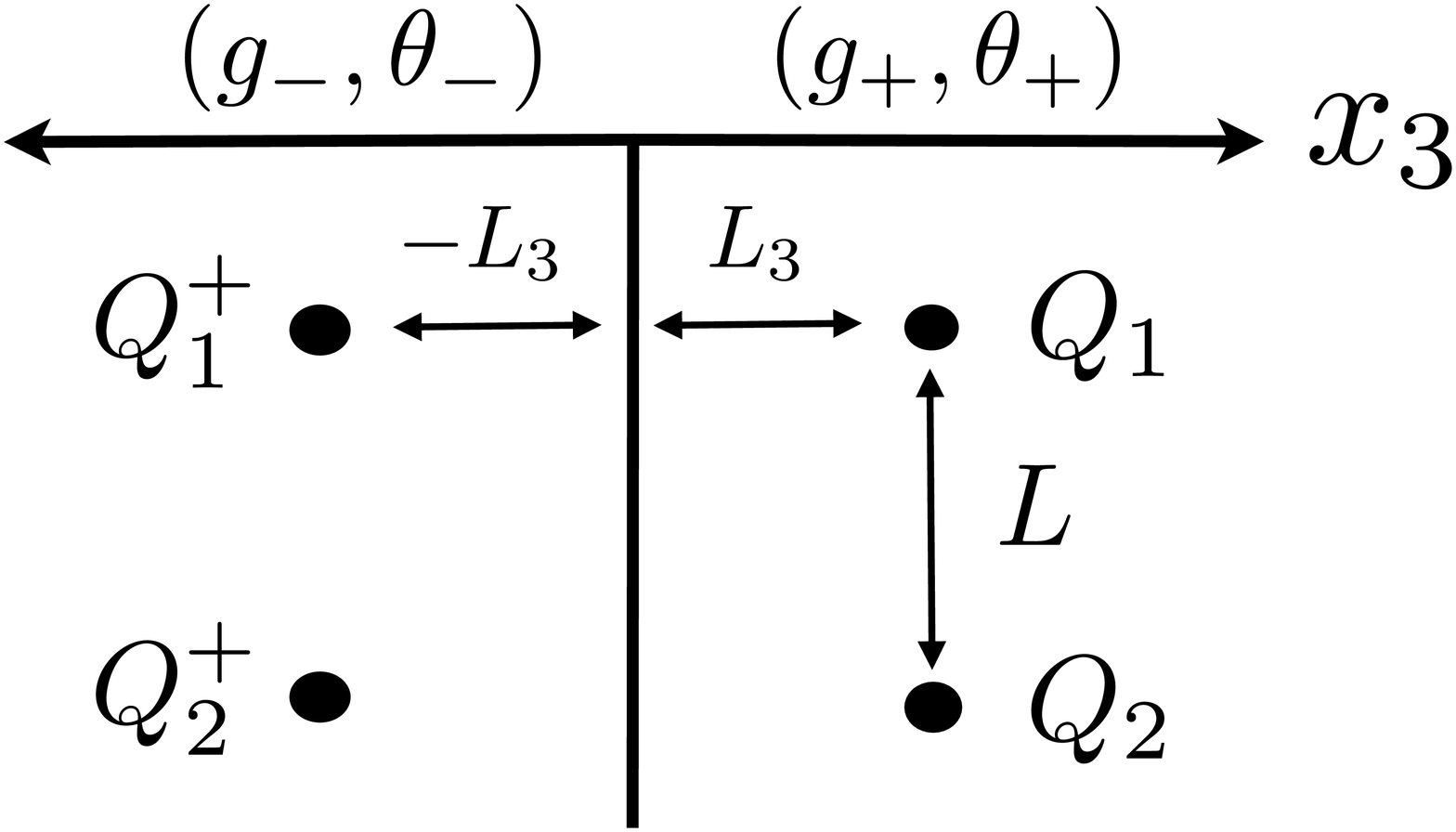}}
\end{center}
\caption{Depictions of the analogues in electromagnetism of the rectangular Wilson loops of $\N=4$ SYM shown in fig.~\ref{fig:configs}. Our conventions are the same as in fig.~\ref{fig:imagecharges1}. (a) The analogue of the perpendicular rectangular Wilson loop of fig.~\ref{fig:configs} (a): two test charges $Q_1$ and $Q_2$ along a line perpendicular to the interface, on opposite sides of the interface. For clarity, in this case we have not depicted the image charges. (b) The analogue of the parallel rectangular Wilson loop of fig.~\ref{fig:configs} (b): two test charges $Q_1$ and $Q_2$ along a line parallel to the interface, in the $x_3>0$ region. Also depicted are the corresponding image charges $Q^+_1$ and $Q^+_2$.}
\label{fig:imagecharges2}
\end{figure}

\begin{figure}[ht]
  \begin{center}
   \subfigure[Jumping coupling]{\includegraphics[width=0.45\textwidth]{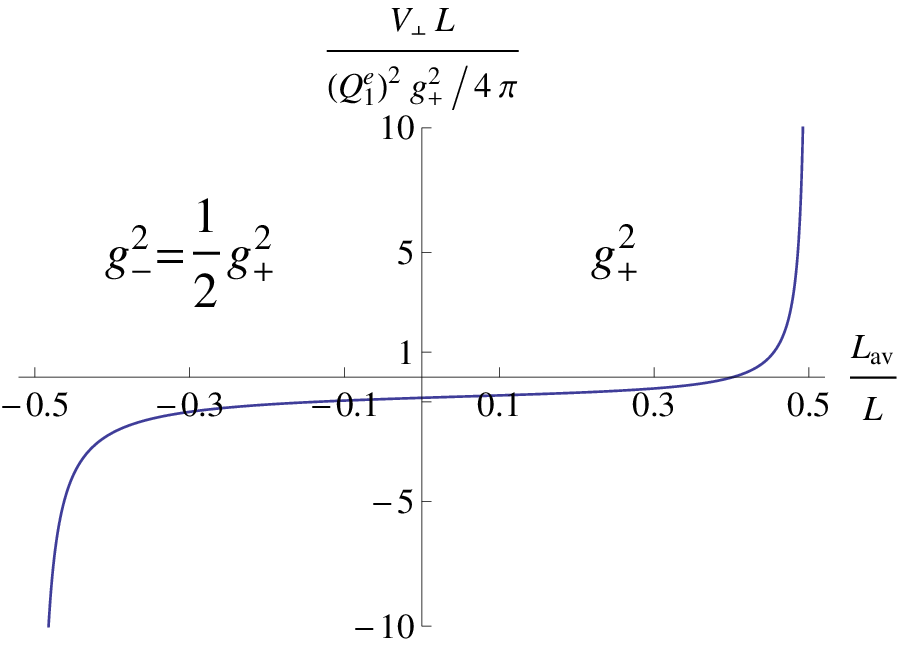}}\hspace{5mm}
     \subfigure[Jumping $\theta$-angle]{\includegraphics[width=0.45\textwidth]{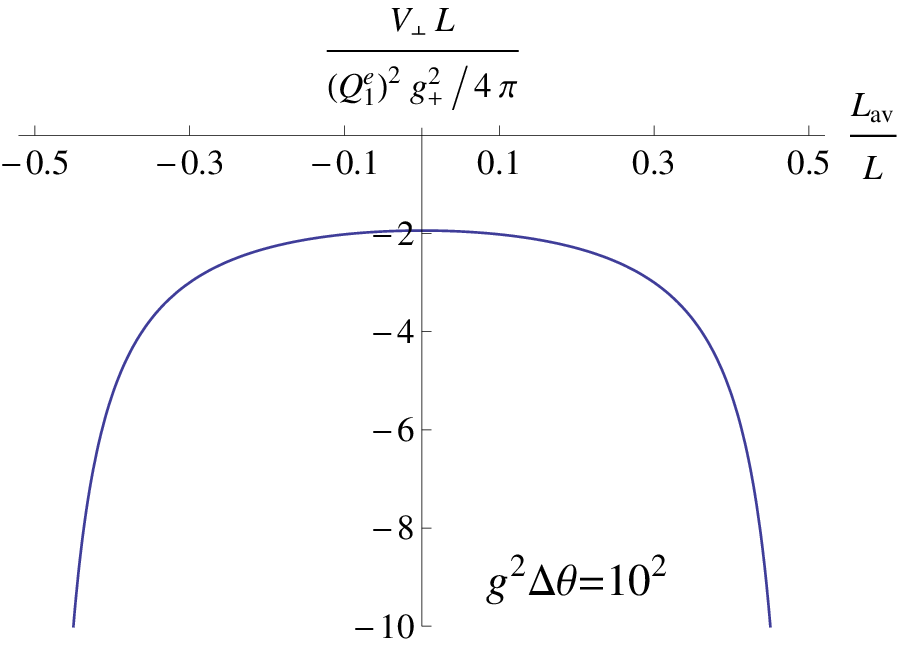}}
  \end{center}
  \caption{The potential $V_{\perp}(L,L_{\textrm{av}})$ of eq.~\eqref{emperpdipolepot} between two test electric charges along a line perpendicular to the interface, as depicted in fig.\ref{fig:imagecharges2} (a), multiplied by $L/(Q^e_1)^2 g_+^2 /4\pi$, as a function of $L_{\textrm{av}}/L$, for two cases: (a) constant $\theta$ and jumping $g$, with $g_-^2 = \frac{1}{2} g_+^2$, (b) constant $g$ and jumping $\theta$, with $g^2 |\theta_- -\theta_+| \equiv g^2 \Delta \theta = 10^2$. In each case we see the potential diverging as a test charge approaches its own image, as explained in detail in the accompanying text.}
\label{fig:empotperpfigs}
\end{figure}

We can also consider the analogues in electromagnetism of the rectangular Wilson loops depicted in fig.~\ref{fig:configs}. The analogue of fig.~\ref{fig:configs} (a) is two purely electric test charges $Q^e_1$ and $Q^e_2$ along a line perpendicular to the interface, on opposite sides of the interface, with $Q^e_2 = - Q^e_1$, as depicted in fig.~\ref{fig:imagecharges2} (a). We take $Q^e_1$ to be at position $x_3^R > 0$ and $Q^e_2$ to be at $x_3^L <0$, so that $L = x_3^R - x_3^L > 0$ is the distance between the charges. We will also define $L_{\textrm{av}} = \frac{1}{2} (x_3^R + x_3^L)$, such that
\beq
\frac{L_{\textrm{av}}}{L} = \frac{1}{2} \frac{x_3^R + x_3^L}{x_3^R - x_3^L} \in \left[-\frac{1}{2},+\frac{1}{2}\right],
\eeq
where $L_{\textrm{av}}/L \in [-1/2,0)$ means $x_3^R<-x_3^L$, so that the charge on the right is closer to the interface than the charge on the left, and $L_{\textrm{av}}/L \in (0,+1/2]$ means $x^3_R>-x_3^L$, so that the charge on the left is closer. $L_{\textrm{av}}/L = 0$ means $x_3^R = - x_3^L$, so that the two test charges sit perfectly symmetrically about the interface. The potential in this case is
\beq
\label{emperpdipolepot}
V_{\perp}(L,L_{\textrm{av}}) = \frac{g_+^2}{4\pi} \frac{Q^e_1 Q^e_2}{L} + \frac{g_+^2}{4\pi} \frac{Q^e_1 \tilde{Q}^e_2}{L} + \frac{g_+^2}{4 \pi} \,\frac{Q^e_1 \tilde{Q}^e_1}{L+2L_{\textrm{av}}} + \frac{g_-^2}{4 \pi} \, \frac{Q^e_2 \hat{Q}^e_2}{L-2L_{\textrm{av}}},
\eeq
where $\hat{Q}^e$ is simply $\tilde{Q}^e$ after swapping $g_+$ and $g_-$. On the right-hand-side of eq.~\eqref{emperpdipolepot}, the first term comes from the interaction of $Q^e_1$ with $Q^e_2$, which has nothing to do with the interface, the second term is the interaction of $Q^e_1$ with $Q^e_2$'s image charge, which is in fact coincident with $Q^e_2$, the third term comes from the interaction of $Q^e_1$ with its own image, and the fourth term comes from the interaction of $Q^e_2$ with its own image. If $L_{\textrm{av}}/L \to -1/2$, then $x_3^R/x_3^L \to 0$, so that $Q^e_1$ approaches the interface. From eq.~\eqref{emperpdipolepot} we can see that the dominant interaction (the smallest denominator) will be the third term. Conversely, if $L_{\textrm{av}}/L \to +1/2$ then $Q^e_2$ approaches the interface, and the dominant interaction is the fourth term. Physically, in each limit the potential diverges because a test charge sits exactly on top of its image. In each case we can invoke the lessons we learned for the single test charge to determine whether the potential diverges to $+\infty$ or $-\infty$ (repulsion or attraction, respectively). For example, if $\theta$ is constant and $g_+ > g_-$ then we expect $Q^e_1$ to be attracted to its image and $Q^e_2$ to be repelled, hence we expect $V_{\perp} \to \pm \infty$ as $L_{\textrm{av}}/L \to \pm 1/2$. With a constant $g$ and jumping $\theta$, we expect $V_{\perp} \to -\infty$ in both limits. In fig.~\ref{fig:empotperpfigs} we plot $V_{\perp}(L,L_{\textrm{av}})$ and observe the expected behavior in each case.

\begin{figure}[ht!]
  \begin{center}
   \subfigure[Jumping coupling]{\includegraphics[width=0.45\textwidth]{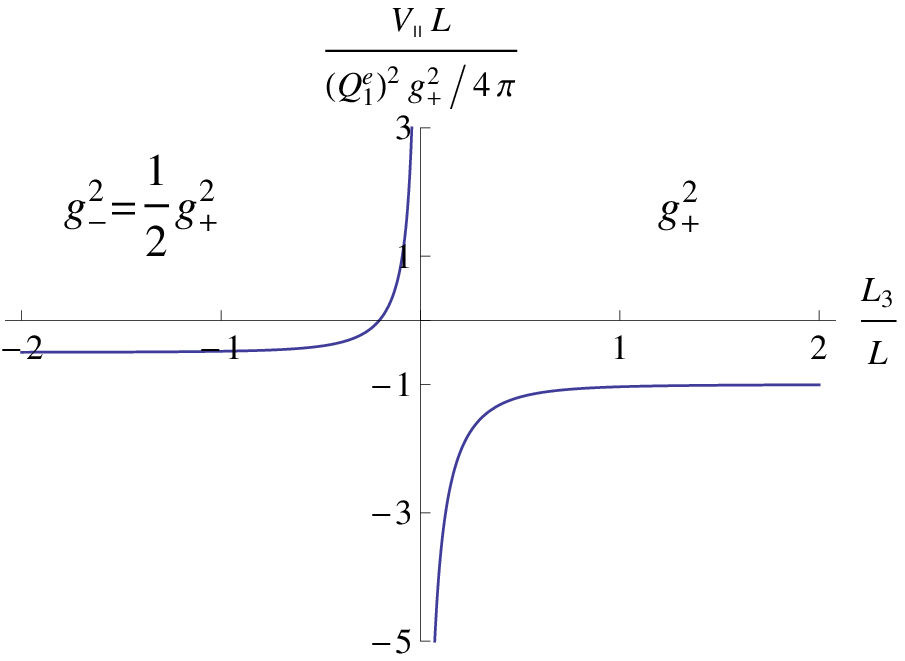}}\hspace{5mm}
     \subfigure[Jumping $\theta$-angle]{\includegraphics[width=0.45\textwidth]{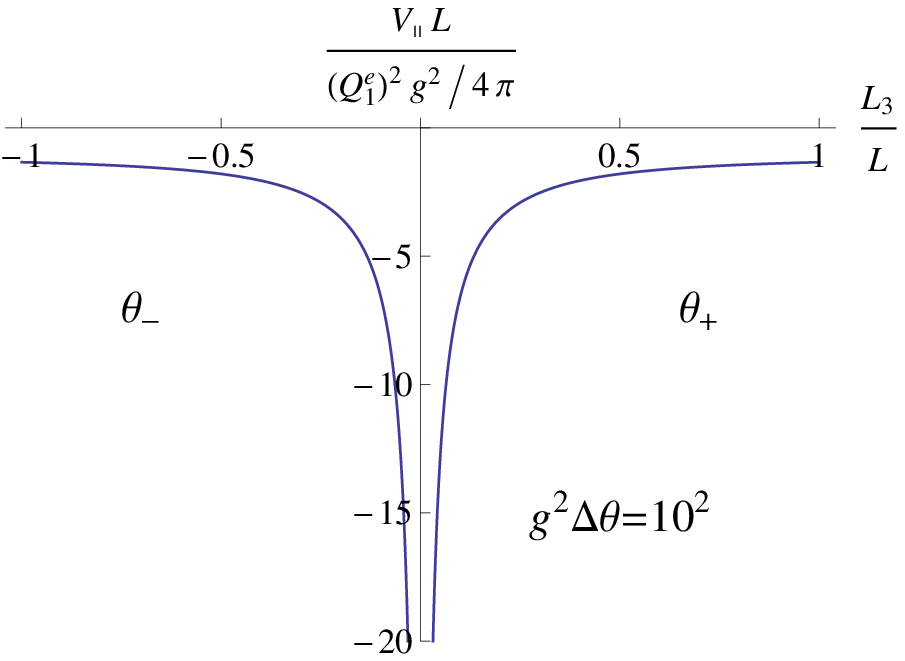}}
  \end{center}
  \caption{The potential $V_{\parallel}(L,L_3)$ of eq.~\eqref{emparadipolepot} of an electric dipole parallel to the interface at $x_3=L_3$, as depicted in fig.\ref{fig:imagecharges2} (b), multiplied by $L/(Q^e_1)^2 g_+^2 /4\pi$, as a function of $L_3/L$, for two cases. (a) Constant $\theta$ and jumping $g$, with $g_-^2 = \frac{1}{2} g_+^2$, where we see that the potential produces a force towards the left, \textit{i.e.}\ the dipole is attracted to the side with smaller coupling. When $L_3/L \to 0$, the potential diverges because the test charges sit exactly on top of their images. (b.) Constant $g$ and jumping $\theta$, with $g^2 \Delta \theta = 10^2$, where we see that the potential produces a force towards the interface, \textit{i.e.}\ the dipole is attracted to the interface.}
\label{fig:empotparafigs}
\end{figure}

The analogue of fig.~\ref{fig:configs} (b) is two purely electric test charges, $Q^e_1$ and $Q^e_2$, separated by a distance $L$ from each other and defining a line parallel to the interface at some position $x_3' = L_3 > 0$, as depicted in fig.~\ref{fig:imagecharges2} (b). Additionally, we take these charges to be equal and opposite, $Q^e_2 = - Q^e_1$, and will thus treat them together as a dipole of length $L$. Each of $Q^e_1$ and $Q^e_2$ will induce an image charge, $\tilde{Q}^e_1$ and $\tilde{Q}^e_2$ respectively, with values given by eq.~\eqref{tildeqdef}. The potential $V$ is then
\beq
\label{emparadipolepot}
V_{\parallel}(L,L_3) = \frac{g_+^2}{4 \pi} \left [ \frac{Q^e_1 Q^e_2}{L} + \frac{Q^e_1 \tilde{Q}^e_1}{2L_3} + \frac{Q^e_2 \tilde{Q}^e_2}{2L_3} + \frac{Q^e_1 \tilde{Q}^e_2}{\sqrt{L^2 + 4L_3^2}} \right].
\eeq
On the right-hand side in eq.~\eqref{emparadipolepot}, the first term comes from the interaction of $Q^e_1$ with $Q^e_2$, the second and third terms come from the interactions of $Q^e_1$ and $Q^e_2$ with their own image charges, respectively, and the fourth term comes from the interaction of $Q^e_1$ and $Q^e_2$ with each other's image charges. The fourth term on the right-hand-side is actually invariant under $Q^e_1 \leftrightarrow Q^e_2$ because $\tilde{Q}^e_2 \propto Q^e_2$ (recall eq.~\eqref{tildeqdef}). For a dipole parallel to the interface, each charge is closer to its own image than to the other charge's image: $\sqrt{L^2 + 4L_3^4}\geq2L_3$ (the hypotenuse of a right triangle is longer than either side). We thus expect the interaction of each charge with its own image to be larger than the interaction with the other charge's image. The net result is that the general rules we learned for the single test charge apply also for the dipole in this case, for example if the $\theta$-angle is constant but the coupling jumps, then the dipole will be attracted to the side with smaller coupling. In fig.~\ref{fig:empotparafigs} we plot $V_{\parallel}(L,L_3)$ in two cases, constant $\theta$ with jumping $g$, and constant $g$ with jumping $\theta$, and observe the expected behavior in each case.

\subsection{$SU(N_c)$ $\N=4$ SYM at large $N_c$}
\label{sunc}

In this subsection we compute the potential $V$ between two heavy test charges in $\N=4$ SYM with a conformal interface using perturbation theory in the 't Hooft coupling. We present explicit results for $V$ in perturbation theory only in the case of a non-SUSY interface where the coupling jumps, and only for test charges along a line parallel to the interface. In that case we show that for a special value of the interface parameter $\kappa$ defined in eq.~\eqref{intextra}, $\kappa=0$, the self-energy of a single test charge vanishes. Our results for $V$ in perturbation theory are valid for any $N_c$. In the large-$N_c$ limit and large-coupling limits, and for $\kappa=0$ only, we also compute the contribution to $V$ from the sum of ladder diagrams.

Following refs.~\cite{Maldacena:1998im,Drukker:1999zq}, we define the Wilson loop operator for $\N=4$ SYM with a conformal interface, in Euclidean signature,\footnote{We use Euclidean signature in the remainder of the paper, including the appendices.} as
\beq
\label{defwilson}
W_{\cal R}[C] \equiv \frac{1}{N_c} {\rm tr}_{{\cal R}} \, {\cal P} \exp \oint_{C} ds \left(i \hat{A}_\mu(x(s)) \dot x^\mu(s) + \hat{\Phi}_I(x(s)) \sqrt{\dot x^2(s)} \, \theta^I(s) \right),
\eeq
where on the right-hand side the factor of $1/N_c$ is our choice of normalization, the trace is taken in representation ${\cal R}$ of $SU(N_c)$, the ${\cal P}$ denotes path-ordering, $C$ denotes a closed path parameterized by $s$ (where dots denote $\frac{\partial}{\partial s}$), and the $\theta^I$ are the components of a unit-norm vector in $\mathbb{R}^6$. Physically, if we introduce an infinitely heavy test charge in representation ${\cal R}$ moving along $C$, the interaction of the test charge with the gauge fields and scalars will produce an insertion of $W_{\cal R}[C]$ into the path integral of $\N=4$ SYM. The expectation value $\langle W_{\cal R}[C]\rangle$ is completely determined by the choices of ${\cal R}$, $C$, and the $\theta^I$.

We will choose the $\theta^I$ to be $s$-independent constants. We will choose ${\cal R}$ to be the fundamental representation of $SU(N_c)$, so that the test charge is a quark. We will choose $C$ to be a rectangle, representing a source quark and anti-quark separated by a distance $L$ propagating for some (Euclidean) time $T$. Henceforth we will suppress the ${\cal R}$ and $C$ notation, so $W_{\cal R}[C] \to W$. The expectation value of the rectangular Wilson loop is related to the potential between the heavy test charges, $V$, as
\beq
\label{eq:vdic}
V = - \lim_{T \rightarrow \infty} \frac{1}{T} \log \langle W \rangle.
\eeq

When the 't Hooft coupling is small, $\lambda \ll 1$, we can compute $\langle W \rangle$ in perturbation theory. Expanding the Wilson loop expectation value, we find
\bea
\label{wilsonexpand}
\langle W \rangle = 1 & -  & \frac{N_c}{2} \oint_C ds \oint_C d\tilde{s} \, \dot{x}^\mu(s) \dot{x}^\nu(\tilde{s}) \langle A_\mu(x(s)) A_\nu(x(\tilde s)) \rangle \nonumber \\ & + & \frac{N_c}{2} \oint_C ds \oint_C d\tilde{s} \, |\dot{x}(s)| |\dot{x}(\tilde{s})| \theta^I \theta^J \langle \Phi^I(x(s)) \Phi^J(x(\tilde s)) \rangle + \ldots,
\eea
where we have stripped away the color factors by first writing $\hat{A}_\mu(x) = A_\mu^a(x) T^a$ with $SU(N_c)$ generators $T^a$, and similarly for $\hat{\Phi}^I$, then performing the trace using ${\rm tr} T^a T^b = \delta^{ab} N_c$, and then defining $A_{\mu}$ via
\beq
\langle A_{\mu}^a(x) A_{\nu}^b(\tilde x) \rangle \equiv \delta^{ab} \langle A_{\mu}(x) A_{\nu}(\tilde x) \rangle,
\eeq
and similarly for the scalars. The $\ldots$ in eq.~\eqref{wilsonexpand} represents all terms of higher order in perturbation theory. To compute $\langle W \rangle$ at leading order, we need the gauge and scalar propagators.

We will compute the gauge and scalar propagators of $\N=4$ SYM only for an interface with a jumping coupling. We take the $\theta$-angle to vanish on both sides of the interface. Our calculation will follow that of refs.~\cite{Erickson:1999qv,Erickson:2000af,Clark:2004sb} very closely.

First let us recall the form of the gauge and scalar propagators for constant $g$. The standard canonically-normalized scalar propagator, $\Delta(x,\tilde{x})$, which satisfies $- \p_\mu \p^\mu \Delta(x,0) = \delta^{(4)}(x)$, is
\beq
\Delta(x, \tilde x) =  \frac{1}{4 \pi^2} \frac{1}{(x - \tilde x)^2}.
\eeq
Following refs.~\cite{Erickson:1999qv,Erickson:2000af}, we choose the covariant gauge condition $\partial^{\mu} A_{\mu} = 0$ and Feynman gauge, in which case the scalar and gauge propagators are almost identical:
\beq
\label{constantgprops}
\langle \Phi^I(x) \Phi^J(\tilde{x}) \rangle = \delta^{IJ} g^2 \Delta(x,\tilde{x}), \qquad \langle A_\mu(x) A_\nu(\tilde x) \rangle = g^2 \delta_{\mu \nu} \Delta(x, \tilde x).
\eeq

Now let us introduce a jumping coupling,
\beq
g(x_3) = g_- \Theta(-x_3) + g_+ \Theta(x_3),
\eeq
where $\Theta(x_3)$ is the Heaviside step function,
\beq
\Theta(x_3) = \begin{cases} 0 &x_3<0, \\ +1 & x_3 \geq 0.\end{cases}
\eeq
Anticipating the presence of image charges, let us introduce the matrix $R_{\mu \nu}$,
\beq
R_{\mu \nu} = \rm{diag}(1,1,1,-1)_{\mu \nu},
\eeq
which implements a reflection in the $x_3$ direction.

To obtain the gauge propagator we follow the calculation of ref.~\cite{Clark:2004sb}. We know that $\langle A_\mu(x) A_\nu(\tilde{x}) \rangle$ must be a linear combination of the following terms:
\beq
\delta_{\mu\nu}\Delta(x, \tilde x) \Theta(\pm x_3)\Theta(\pm\tilde{x}_3), \quad R_{\mu\nu} \Delta(x, R \tilde x) \Theta(\pm x_3)\Theta(\pm \tilde{x}_3), \quad  \delta_{\mu\nu}\Delta(x, \tilde x) \Theta(\pm x_3)\Theta(\mp \tilde{x}_3).
\eeq
We can fix the coefficients of that linear combination as follows. In the presence of the jumping coupling, we can again define fields $\Ev$, $\Bv$, $\Dv$, and $\Hv$ as in eq.~\eqref{ebdhdef}. Since we work to leading order in the coupling, where we ignore the non-Abelian interactions, $\Ev$, $\Bv$, $\Dv$, and $\Hv$ will satisfy the equations of motion and Bianchi identity written in eq.~\eqref{eq:meom}, without sources. We can thus again derive the matching conditions in eq.~\eqref{matchingcond}, which we can easily translate into conditions on $A_{\mu}$ and hence on the $A_{\mu}$ propagator. Employing also the gauge constraint, we can write the final result for the gauge propagator as
\bea
\label{AApropagator}
\langle A_\mu(x) A_\nu(\tilde{x}) \rangle &=& g_+^2 \left( \delta_{\mu \nu} \Delta(x, \tilde x) + Q^+ R_{\mu \nu} \Delta(x, R \tilde x) \right) \Theta(x_3)\Theta(\tilde{x}_3) \nonumber \\
& + & \frac{2 g_+^2 g_-^2}{g_+^2 + g_-^2} \delta_{\mu \nu} \Delta(x, \tilde x) \left(\Theta(x_3)\Theta(-\tilde{x}_3) + \Theta(-x_3)\Theta(\tilde{x}_3) \right) \nonumber \\
& + & g_-^2 \left( \delta_{\mu \nu} \Delta(x, \tilde x) +Q^- R_{\mu \nu} \Delta(x, R \tilde x) \right) \Theta(-x_3)\Theta(-\tilde{x}_3),
\eea
where the image charges are given by
\beq
Q^+ = \frac{g_-^2 - g_+^2}{g_+^2 + g_-^2}, \qquad Q^- = \frac{g_+^2 - g_-^2}{g_+^2 + g_-^2}.
\eeq
The formula for $Q^+$ is simply eq.~\eqref{tildeqdef} with $\theta_+ = \theta_-$ and $Q^e = 1$. Our result for the gauge propagator agrees with that of ref.~\cite{Clark:2004sb}, and reduces to the result in eq.~\eqref{constantgprops} when $g_+ = g_-$.

For the scalar propagator, we start by writing the most general form consistent with the symmetries, using arguments similar to those given above for the gauge field propagator,
\bea
\label{scalprop}
\langle \Phi^I(x) \Phi^J(\tilde{x}) \rangle & = & \delta^{IJ} \bigg( g_+^2 \Delta(x, \tilde{x}) \Theta(x_3) + g_-^2 \Delta(x, \tilde{x}) \Theta(-x_3) \nonumber \\
& & + g_+^2 Q^{++} \Delta(x, R \tilde{x}) \Theta(x_3)\Theta(\tilde{x}_3) + g_-^2 Q^{--} \Delta(x, R \tilde{x}) \Theta(-x_3)\Theta(-\tilde{x}_3) \nonumber \\
& & + Q^{-+} \Delta(x, \tilde{x}) \Theta(-x_3) \Theta(\tilde{x}_3) + Q^{+-} \Delta(x, \tilde{x}) \Theta(x_3) \Theta(-\tilde{x}_3) \bigg),
\eea
where $Q^{\pm\pm}$ and $Q^{\pm\mp}$ are image charges whose values are determined by the interface-localized terms quadratic in the scalars, eqs.~\eqref{intextra} and~\eqref{susyint}.

For the non-SUSY interface, $Q^{\pm\pm}$ and $Q^{\pm\mp}$ are determined by the value of $\kappa$, the coefficient of the interface-localized term in eq.~\eqref{intextra}. For example, consider $\kappa=0$. In that case, the gauge field and the scalars obey the same equation of motion, and thus must obey the same matching conditions. With our choice of gauge, they must thus have the same propagator, up to factors of $\delta_{\mu\nu}$, $R_{\mu\nu}$, and $\delta^{IJ}$, so when $\kappa=0$ we have $Q^{++} = Q^{+-} = Q^+$ and $Q^{--} = Q^{-+} = Q^-$. More generally, we can think of the interface term as a boundary term that imposes a boundary condition on the propagator. The generic form of the propagator is that of eq.~\eqref{scalprop}, but in order to satisfy the boundary condition $Q^{\pm\pm}$ and $Q^{\pm\mp}$ must take certain values, and so must obviously depend on $\kappa$. The general lesson is that $\kappa$ controls the size of the image charges, along with $g_+$ and $g_-$.

For the SUSY interface, the interface-localized term in eq.~\eqref{susyint} involves only three of the six real scalars. The three scalars that do not appear in eq.~\eqref{susyint} will have propagators of the same form as the non-SUSY case, with $\kappa=0$. For the three scalars that do appear in eq.~\eqref{susyint}, the interface-localized terms will modify the values of $Q^{\pm\pm}$ and $Q^{\pm\mp}$ relative to the non-SUSY, $\kappa=0$ case. As a result, the Wilson loop expectation value will depend on the choice of the $\theta^I$ in eq.~\eqref{wilsonexpand}, since these determine the scalars to which the test charges couple. For a SUSY interface, we will only compute Wilson loop expectation values holographically, and we will indeed find that the results depend sensitively on the choice of $\theta^I$.

For the non-SUSY interface where the coupling jumps, we will now show that when $\kappa=0$ a test charge has no image charge. To do so, we will focus on a rectangular Wilson loop parallel to the interface, $W_{\parallel}$, depicted in fig.~\ref{fig:configs} (b). For such test charges, we will compute the expectation value $\langle W_{\parallel}\rangle$, and from that extract $V_{\parallel}(L,L_3)$ via eq.~\eqref{eq:vdic}, in perturbation theory for arbitrary $\kappa$ and then show that when $\kappa=0$ the image charge vanishes thanks to a cancellation between the gauge fields and scalars. Moreover, for the special case $\kappa=0$, and in the large-$N_c$ limit, we will extend the result for $V_{\parallel}(L,L_3)$ beyond perturbation theory by performing the sum of ladder diagrams, following refs.~\cite{Erickson:1999qv,Erickson:2000af}.

For a rectangular Wilson loop parallel to the interface, we parameterize the path $C$ as
\begin{align}
&x(s) = \{s,0,0,L_3\},& -&T/2 \leq s \leq T/2,& \nonumber \\
&x(s) = \{T-s,L,0,L_3\},& &T/2 \leq s \leq 3T/2,&
\end{align}
with $L>0$, and we will neglect the segments $\{\pm T/2,s,0,L_3\}$ with $0 \leq s \leq L$ (the short ends of the rectangle) since to obtain $V_{\parallel}(L,L_3)$ we will ultimately take the limit $T \gg L,L_3$.  The contribution to the expectation value $\langle W_{\parallel} \rangle$ from the gauge fields is
\begin{align}
\label{eq:gaugecontribution}
-\frac{N_c}{2}\oint_C ds \oint_C d\tilde{s} & \left[ \dot{x}^\mu(s) \dot{x}^\nu(\tilde{s}) \langle {\cal P} A_\mu(x(s)) A_\nu(x(\tilde s)) \rangle \right] \nonumber \\ &= -\frac{N_c}{2} \, \frac{T}{2 \pi}\left[ \lim_{\epsilon \rightarrow 0}\frac{g_{\pm}^2}{\epsilon} - \frac{g_{\pm}^2}{L} + \frac{g_{\pm}^2 \, Q^{\pm}}{2L_3} - \frac{g_{\pm}^2 \, Q^{\pm}}{\sqrt{L^2 + 4 L_3^2}} \right] + {\cal O}\left(\frac{L}{T},\frac{L_3}{T}\right) \ ,
\end{align}
where the $+$ sign is for $L_3>0$ and the $-$ sign is for $L_3<0$. We have introduced a cutoff, $\epsilon$, to regulate the divergent contribution to the self-energy of the test charges from the gauge fields, \textit{i.e.}\ the first term on the right-hand-side of eq.~\eqref{eq:gaugecontribution}. Ignoring that divergent self-energy term and applying eq.~\eqref{eq:vdic}, we find that $V_{\parallel}(L,L_3)$ is precisely that of electromagnetism, eq.~\eqref{emparadipolepot}, times a factor of $N_c$, as expected. The contribution to $\langle W_{\parallel} \rangle$ from the scalar fields is, for arbitrary $\kappa$,
\begin{align}
+\frac{N_c}{2}\oint_C ds \oint_C d\tilde{s} & \left[ |\dot{x}(s)| |\dot{x}(\tilde{s})| \theta^I \theta^J \langle {\cal P} \Phi^I(x(s)) \Phi^J(x(\tilde s)) \rangle \right] \nonumber \\ &= +\frac{N_c}{2} \, \frac{T}{2 \pi}\left[ \lim_{\epsilon \rightarrow 0}\frac{g_{\pm}^2}{\epsilon} + \frac{g_{\pm}^2}{L} + \frac{g_{\pm}^2 \, Q^{\pm\pm}}{2L_3} + \frac{g_{\pm}^2 \, Q^{\pm\pm}}{\sqrt{L^2 + 4 L_3^2}} \right] + {\cal O}\left(\frac{L}{T},\frac{L_3}{T}\right).
\end{align}
The scalars also produce a divergent contribution to the self-energy, but with precisely the right coefficient to cancel the divergent contribution from the gauge fields, as we expect for $\N=4$ SYM without an interface, where a single test charge has vanishing self-energy. The final result for $V_{\parallel}(L,L_3)$ is finite,
\beq
\label{eq:pertparapot}
V_{\parallel}(L,L_3) = - \frac{N_c}{4\pi} \left [ \frac{2g_{\pm}^2}{L} + \frac{g_{\pm}^2 \left(-Q^{\pm} + Q^{\pm\pm}\right)}{2L_3} + \frac{g_{\pm}^2\left(Q^{\pm}+Q^{\pm\pm}\right)}{\sqrt{L^2+4L_3^2}} \right].
\eeq
The first term on the right-hand-side of eq.~\eqref{eq:pertparapot} is simply the $\lambda \ll 1$ result for $\N=4$ SYM in eq.~\eqref{eq:ladderpotential}, representing the interaction of the two test charges with one another. The other two terms on the right-hand-side of eq.~\eqref{eq:pertparapot} represent interactions with image charges. If we send $L \to \infty$, effectively sending one test charge to infinity, then we isolate the self-energy of a single test charge in the presence of the interface, the second term on the right-hand-side of eq.~\eqref{eq:pertparapot}. That term will be non-zero if $\left(-Q^{\pm} + Q^{\pm\pm}\right) \neq0$, in which case the potential diverges when $L_3 \to 0$ because the test charge sits on top of its own image charge. As explained below eq.~\eqref{scalprop}, if $\kappa=0$ then $Q^{\pm\pm} = Q^{\pm}$, in which case $\left(-Q^{\pm} + Q^{\pm\pm}\right) =0$: the interaction of each test charge with its own image vanishes. We have thus proven that the cancellation of gauge field and scalar contributions to the self-energy that occured in $\N=4$ SYM without an interface, due to SUSY, persists to $\N=4$ SYM with a jumping coupling that \textit{breaks} SUSY, in the special case that $\kappa=0$.

\begin{figure}[ht!]
  \begin{center}
  \includegraphics[width=0.5\textwidth]{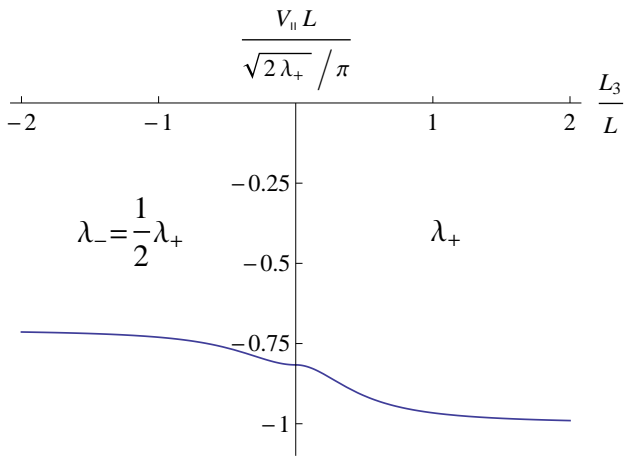}
  \end{center}
  \caption{The contribution from the sum of ladder diagrams to the potential $V_{\parallel}(L,L_3)$, eq.~\eqref{eq:ladder}, between two heavy test charges along a line parallel to a non-SUSY interface where the coupling jumps in large-$N_c$, strongly-coupled $\N=4$ SYM, for the special value $\kappa=0$ of the constant defined in eq.~\eqref{intextra}. More precisely, we chose $\lambda_-=\lambda_+/2$ and plotted $V_{\parallel}(L,L_3) L / (\sqrt{2\lambda_+}/\pi)$ as a function of $L_3/L$. In stark contrast to the analogous result in electromagnetism, shown in fig.~\ref{fig:empotparafigs} (a), here $V_{\parallel}(L,L_3)$ remains finite at $L_3/L=0$, and the dipole is attracted to the side with $\textit{larger}$ coupling (to the right), because the self-energy of each test charge vanishes.}
\label{fig:ladder}
\end{figure}

We can demonstrate that this effect persists beyond leading order in perturbation theory by computing the contribution from the sum of ladder diagrams to the Wilson loop expectation value, $\langle W_{\textrm{ladder}} \rangle \equiv\Gamma$, in the large-$N_c$ limit. For large-$N_c$ $\N=4$ SYM without an interface, $\Gamma$ obeys a recursion relation, as shown in ref.~\cite{Erickson:1999qv,Erickson:2000af}. The proof of that fact relies crucially on the vanishing self-energy of a single test charge in $\N=4$ SYM without an interface. For $\N=4$ SYM with a non-SUSY interface where the coupling jumps, in the special case $\kappa=0$ the self-energy again vanishes. In that case, a straightforward exercise shows that $\Gamma$ again obeys a recursion relation,
\beq
\label{eq:gammarecursion}
\Gamma(T,T') = 1 +\int_0^T ds \int_0^{T'} d\tilde{s} \, \frac{g_+^2 N}{2 \pi^2} \left[ \frac{1}{(s-\tilde{s})^2 +L^2} + \frac{Q^+}{(s-\tilde{s})^2 +L^2+4L_3^2}\right] \Gamma(s,\tilde{s}),
\eeq
which is of precisely the same form as in $\N=4$ SYM without an interface. In particular, the term in brackets on the right-hand-side of eq.~\eqref{eq:gammarecursion} is the sum of gauge and scalar propagators. The ladder sum $\Gamma(T,T')$ obeys the boundary conditions
\beq
\Gamma(T,0)=\Gamma(0,T')=1.
\eeq
Taking two derivatives of eq.~\eqref{eq:gammarecursion}, we obtain
\beq
\frac{\partial^2 \Gamma(T,T')}{\partial T \partial {T'}} = \frac{g_+^2 N}{2 \pi^2} \left[ \frac{1}{(T-T')^2 +L^2} + \frac{Q^+}{(T-T')^2 +L^2+4L_3^2}\right] \Gamma(T,T') \ .
\eeq
As in refs.~\cite{Erickson:1999qv,Erickson:2000af}, at large $T=T'$ and with large $g_+^2 N_c$, the solution is of the form
\beq
\Gamma\left(T,T'\right) \propto e^{\frac{T}{L}\sqrt{\frac{2 g_+^2 N_c}{\pi^2} \left( 1 + \frac{Q^+}{1+4(L_3/L)^2}\right)}}.
\eeq
We thus find
\beq\label{eq:ladder}
V_{\parallel}(L,L_3) = -\frac{1}{\pi}\frac{\sqrt{2 g_+^2 N_c}}{L} \sqrt{1 + \frac{Q^+}{1+4(L_3/L)^2}}, \qquad \textrm{sum of ladder diagrams.}
\eeq
When $g_+ = g_-$, so that $Q^+=0$, we recover the result in eq.~\eqref{eq:ladderpotential} for $\N=4$ SYM without an interface. In fig.~\ref{fig:ladder} we show the $V_{\parallel}(L,L_3)$ in eq.~\eqref{eq:ladder} for the case $\lambda_-=\lambda_+/2$. Notice the striking difference from the analogous result in electromagnetism, shown in fig.~\ref{fig:empotparafigs} (a). In particular, in electromagnetism, $V_{\parallel}(L,L_3)$ diverges as $L_3/L \to 0$ because the test charges approach their image charges, whereas the result in eq.~\eqref{eq:ladder} remains finite as $L_3/L\to0$ because the self-energy of each test charge vanishes. Moreover, fig.~\ref{fig:ladder} clearly shows that the dipole of test charges is attracted to the side with \textit{larger} coupling, again because the self-energy of each test charge vanishes, in stark contrast to electromagnetism.

\section{Wilson Loops: Holographic Calculation}
\label{wilsonholo}

To compute the expectation value of a Wilson loop in the fundamental representation holographically, we must introduce a fundamental string in the bulk with its endpoints on the boundary, tracing out the loop~\cite{Rey:1998ik,Maldacena:1998im}. Our Wilson loops, and hence our strings, will be static, with time extent $T$. We solve the string equations of motion with the given boundary conditions and evaluate the string action on the resulting solution. Generically, the on-shell action of a string with endpoints on the boundary is divergent due to the infinite distance to the boundary. The procedure to render the on-shell action finite is called holographic renormalization~\cite{Henningson:1998gx,Balasubramanian:1999re,deHaro:2000xn,Skenderis:2002wp}. We will denote the renormalized on-shell string action as $\sren$. According to the AdS/CFT dictionary, the expectation value of the Wilson loop, $\langle W \rangle$, is then given by
\beq
\label{eq:wilsondic}
\langle W \rangle \propto e^{-S_{\textrm{ren}}}.
\eeq
We can extract a potential $V$ from either a straight timelike Wilson line or a rectangular Wilson loop via eq.~\eqref{eq:vdic}
\beq
\label{eq:vdicholo}
V = - \lim_{T \rightarrow \infty} \frac{1}{T} \log \langle W \rangle = \lim_{T \to \infty} \frac{S_{\textrm{ren}}}{T}.
\eeq

In appendix~\ref{stringeoms} we write the action for strings in Janus spacetimes and describe our Ans\"atze and numerical methods for solving the equations of motion. In appendix~\ref{holorg} we perform the holographic renormalization of the on-shell action of these strings. In sections \ref{sec:straight}, \ref{sec:perpendicular}, and \ref{sec:parallel} we present our results for $V$ for straight, perpendicular and parallel Wilson lines/loops located away from the interface (at finite $x_3$), respectively. In each case we discuss both a jumping coupling and a jumping $\theta$-angle, with and without SUSY. In section \ref{intloops} we consider the special case of Wilson loops located exactly on the interface, where we can obtain analytic (rather than numerical) results for $V$.

\subsection{Straight Wilson Lines}
\label{sec:straight}

In this subsection we present our results for the expectation values of straight timelike Wilson lines in $\N=4$ SYM with a jumping coupling or $\theta$-angle, with and without SUSY. More precisely, we will present the self-energy $V$ of a single heavy test charge, obtained from the Wilson line expectation value via eqs.~\eqref{area1} and~\eqref{eq:vdicholo}.
In subsection 4.1.1 we present our numerical results, and in subsection 4.1.2 we present some exact results for the SUSY case at leading order in a small jump in the coupling or $\theta$-angle.

\hiddensubsubsection{Numerical Results}

In fig.~\ref{fig:straight} we present our numerical results for $\N=4$ SYM with a jumping coupling:

\begin{figure}[ht!]
  \begin{center}
   \subfigure[Non-SUSY interface, jumping coupling]{
        \includegraphics[width=0.45\textwidth]{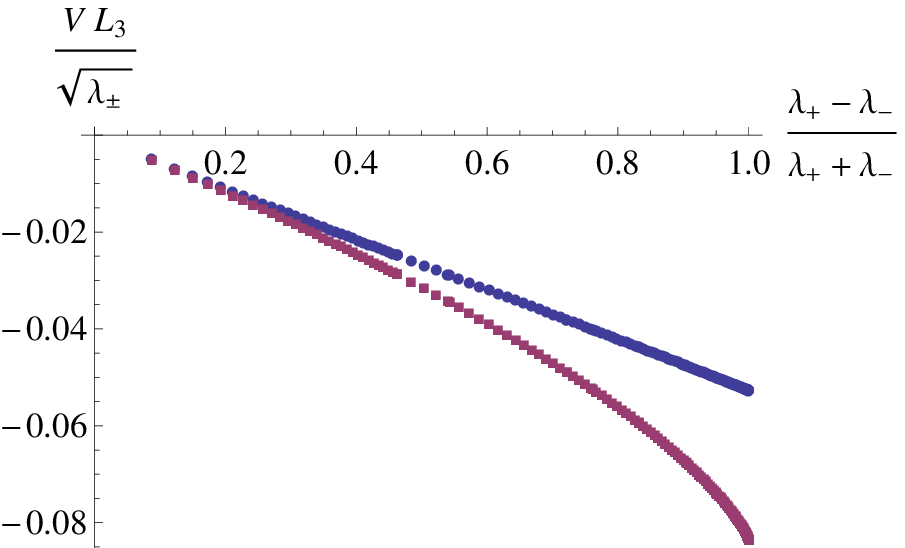}}\hspace{5mm}
     \subfigure[SUSY interface, jumping coupling]{
        \includegraphics[width=0.45\textwidth]{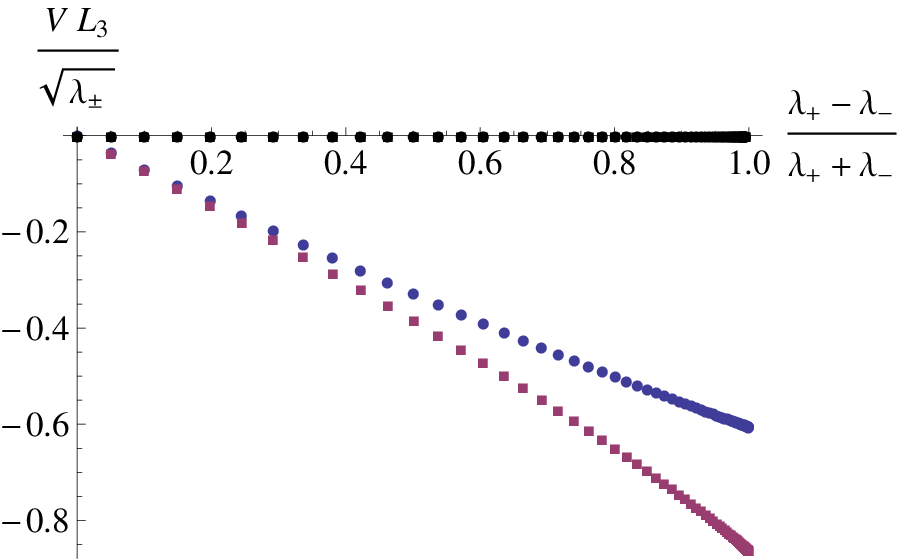}}
  \end{center}
  \caption{The self-energy $V$, expressed as $V L_3/\sqrt{\lambda_{\pm}}$, as a function of $\frac{\lambda_+ - \lambda_-}{\lambda_+ + \lambda_-}=\frac{g^2_+ - g^2_-}{g^2_+ + g^2_-}$ for a single heavy test charge in large-$N_c$, strongly-coupled $\N=4$ SYM with a jumping coupling.  (a) Our results for the non-SUSY interface, dual to non-SUSY Janus.  The blue dots correspond to a test charge to the left of the interface, $L_3<0$, which has smaller coupling, $\lambda_-<\lambda_+$, and for which we plot $V L_3/\sqrt{\lambda_-}$.  The purple squares are our numerical results for a test charge with $L_3>0$, where we plot $V L_3/\sqrt{\lambda_+}$.
  (b) Our results for the SUSY interface, dual to SUSY Janus. The blue dots and purple squares have the same meaning as in (a), and come from a straight string at $y=\pi/2$. The black dots are for a straight string at $y=0$ on either side of the interface, for which $V=0$.} \label{fig:straight}
\end{figure}

\begin{itemize}[leftmargin=0.2in]
\item \textbf{Non-SUSY, jumping coupling (fig.~\ref{fig:straight} (a)):} For any jump in the coupling we find a non-zero result for $V$, indicating that a non-zero image charge is present.  Recalling our discussions from sections~\ref{nonsusyjanus} and~\ref{sunc}, this indicates that the dual field theory has $\kappa \neq 0$, in agreement with the claim of ref.~\cite{Clark:2004sb}.  Moreover, we find the same qualitative behavior as in electromagnetism: the test charge is always attracted to the side with smaller coupling.  In fact, for a test charge on the side of the interface with smaller coupling (the blue dots in fig.~\ref{fig:straight} (a)) we find a linear dependence on the jump in the coupling-squared, which is the same behavior as in electromagnetism (see eqs.~\eqref{tildeqdef} and~\eqref{eq:emsinglev} with $\theta_+ = \theta_-$). Such linear dependence is non-trivial and surprising, given that $\N=4$ SYM is a non-linear (\textit{i.e.}\ interacting) theory.

\item \textbf{SUSY, jumping coupling (fig.~\ref{fig:straight} (b)):}  Here we find the result for $V$ depends sensitively on $y$, the location of the straight string in the internal space, dual to the couplings $\theta^I$ in the definition of the Wilson loop in eq.~\eqref{defwilson}. For $y=\pi/2$, the results are qualitatively similar to those in the non-SUSY case, although for a test charge on the side with smaller coupling, $V$'s dependence on the jump in the coupling-squared is no longer linear.

Our most surprising result comes from the string at $y=0$, represented by the black dots in fig.~\ref{fig:straight} (b), where $V=0$ exactly. To dispel any fear that this result may be a numerical artifact, in the next subsection we consider a small jump in the coupling and, working to leading order in that small jump, we show that $V=0$ analytically.  Additionally, the numerical results for a Wilson loop perpendicular to the interface presented in section~\ref{sec:perpendicular} will provide another, independent, consistency check.  Apparently for $y=0$, the test charge has no image charge!

As discussed in the introduction, the vanishing of $V$ for $y=0$ has a number of potential implications.  First, for the rectangular Wilson loop, the sum of ladder diagrams may be relatively easy to compute due to the vanishing self-energy, as discussed in section~\ref{sunc}. Secondly, for a theory with conformal symmetry we may map, by a conformal transformation, a straight Wilson line with a trivial expectation value to a circular Wilson loop. A non-zero expectation value for the latter may imply the existence of a matrix model, as in $\N=4$ SYM with constant coupling~\cite{Erickson:2000af,Drukker:2000rr,Pestun:2007rz}. Indeed, in ref.~\cite{Drukker:2010jp} a matrix model was derived for a circular Wilson loop sitting exactly on the interface in Euclidean $\N=4$ SYM on $S^4$. Our results suggests that something similar may be possible for a circular Wilson loop away from the interface.
\end{itemize}

\begin{figure}[ht!]
  \begin{center}
   \subfigure[Non-SUSY interface, jumping $\theta$-angle]{
        \includegraphics[width=0.45\textwidth]{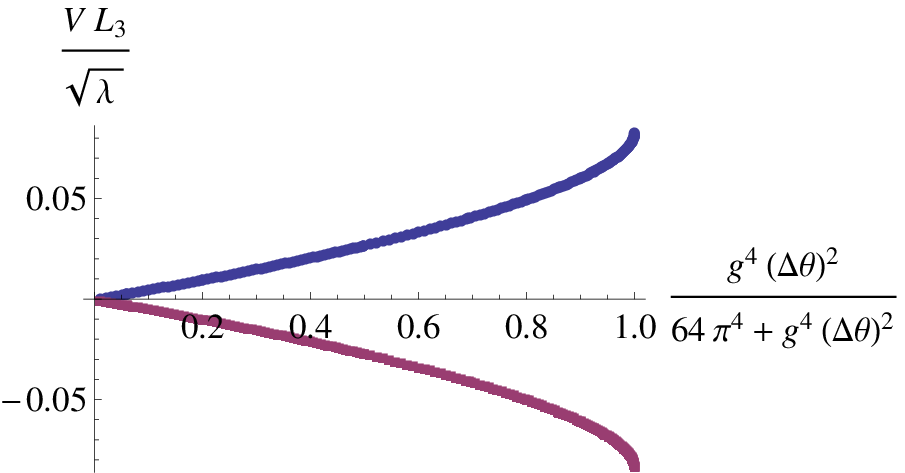}}\hspace{5mm}
   \subfigure[SUSY interface, jumping $\theta$-angle, $y=0$]{
        \includegraphics[width=0.45\textwidth]{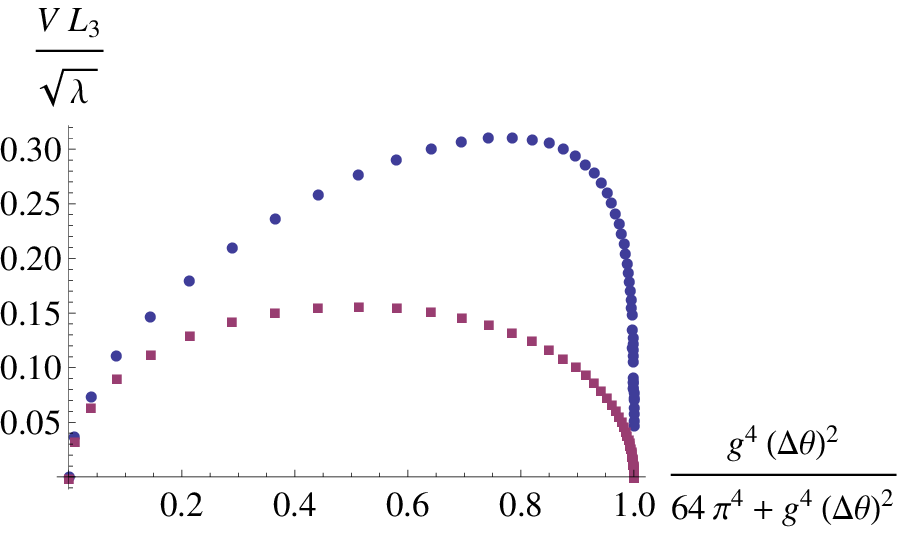}} \\
   \subfigure[SUSY interface, jumping $\theta$-angle, $y=\pi/2$]{
        \includegraphics[width=0.45\textwidth]{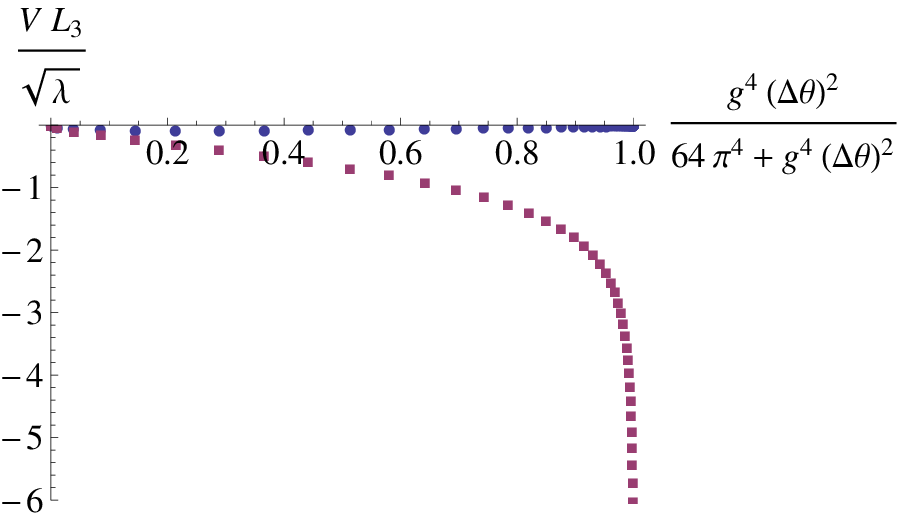}} \hspace{5mm}
   \subfigure[SUSY interface, jumping $\theta$-angle, $y=\pi/2$ (close-up)]
        {\includegraphics[width=0.45\textwidth]{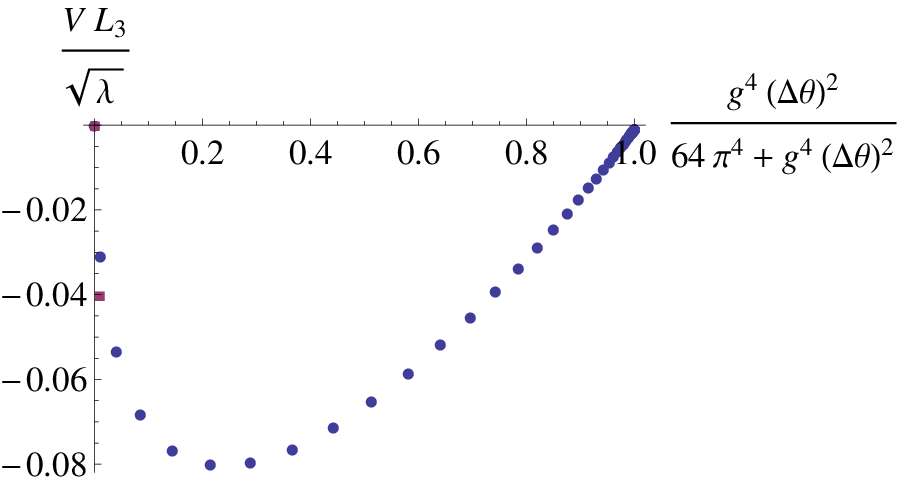}}
  \end{center}
  \caption{The self-energy $V$, expressed as $V L_3/\sqrt{\lambda}$, for a single heavy test charge in large-$N_c$, strongly-coupled $\N=4$ SYM with a jumping $\theta$-angle, computed from a straight string in Janus spacetimes.  The blue dots are for a test charge to the left of the interface, $L_3<0$, while the purple squares are for a test charge with $L_3>0$.  (a) Our results for the non-SUSY interface, dual to non-SUSY Janus.  (b) Our results for a SUSY interface, computed using a string with $y=0$ in SUSY Janus. (c) Our results for a SUSY interface, computed using a string with $y=\pi/2$ in SUSY Janus.  (d) Close-up of (c).}\label{fig:straightaxion}
\end{figure}

In fig.~\ref{fig:straightaxion}  we present our numerical results for $\N=4$ SYM with a jumping $\theta$-angle.  Specifically, we plot $VL_3/\sqrt{\lambda}$ as a function of $\frac{g^4 \left(\Delta \theta\right)^2}{64\pi^4+g^4 \left(\Delta \theta\right)^2}$, which is the natural quantity to consider from the point of view of electromagnetism with a jumping $\theta$-angle: the image charge in eq.~\eqref{tildeqdef} depends linearly on this quantity.
In the 't Hooft limit $g^2=\mathcal{O}\left(1/N_c\right)$ and we implicitly take $\theta_+=\mathcal{O}\left(N_c\right)$, so that $g^4 \left(\Delta \theta\right)^2=\mathcal{O}\left(1\right)$, as shown in fig.~\ref{fig:straightaxion}.
\begin{itemize}[leftmargin=0.2in]
\item \textbf{Non-SUSY, jumping $\theta$-angle (fig.~\ref{fig:straightaxion} (a)):} For any $\Delta \theta$ we find a non-zero $V$, indicating that a non-zero image charge is present. The qualitative behavior is similar to electromagnetism: the test charge is always attracted towards the interface, however, in contrast with electromagnetism, $V L_3$ is not linear in $\frac{g^4 \left(\Delta \theta\right)^2}{64\pi^4+g^4 \left(\Delta \theta\right)^2}$.
\item \textbf{SUSY, jumping $\theta$-angle (fig.~\ref{fig:straightaxion} (b)):} As in the jumping coupling case, the result for $V$ depends sensitively on $y$. For $y=0$, we find that $V$ is non-zero for any finite $\Delta \theta$.  Two curious features also appear, compared to the non-SUSY interface. First, in contrast to electromagnetism, a test charge is always attracted to the side with \textit{larger} $\theta$: the image charge has the ``wrong'' sign. Second, for a test charge on either side of the interface, we find that as $\Delta \theta$ increases, $VL_3$ increases, reaches a maximum, and then eventually goes to zero as $\Delta \theta \to \infty$, suggesting that the image charge vanishes in this limit. Apparently, in our interacting theory, the dyonic potential generated by a jumping $\theta$-angle can be screened by additional forces, in this case from exchange of adjoint scalars. Such behavior is dramatically different from that proposed for conventional TIs~\cite{Qi27022009}.

For the $y=\pi/2$ case, shown in figs.~\ref{fig:straightaxion} (c) and (d), we find that $V$ is non-zero for any finite $\Delta \theta$, but now a test charge is always attracted to the side with \textit{smaller} $\theta$.  When the charge is on the side with smaller $\theta$-angle (the blue dots in fig.~\ref{fig:straightaxion} (c) and (d)), $V$ decreases, reaches a minimum, and then increases to zero as $\Delta \theta \to \infty$. When the charge is on the side with larger $\theta$-angle, $V$ decreases monotonically.
\end{itemize}

As shown in ref.~\cite{Drukker:2010jp}, for Euclidean $\N=4$ SYM on $S^4$ with a SUSY interface where the coupling jumps, the matrix model for the circular Wilson loop on the interface is of exactly the same form as the pure (no interface) $\N=4$ SYM theory. The authors of ref.~\cite{Drukker:2010jp} also claimed that the same is not true for an interface where the $\theta$-angle jumps. The fact that we find vanishing self-energy in the case with a jumping coupling and $y=0$, while we find a non-vanishing self-energy in the case with a jumping $\theta$-angle and $y=0$, is nicely consistent with the claim made in ref.~\cite{Drukker:2010jp}.

\hiddensubsubsection{Exact Results for Straight Strings in SUSY Janus}

In fig.~\ref{fig:straight} (b) we presented numerical evidence that a straight string with $y=0$ in jumping-dilaton SUSY Janus gives rise to a trivial expectation value for a straight, timelike Wilson line, indicating that the associated test charge has no image charge.  In this subsection we will consider a small jump in the coupling or $\theta$-angle, meaning $\delta \phi \ll 1$, and work analytically to leading order in perturbation theory in that small parameter.  Within this approximation, we will confirm the numerical results.

To proceed for the jumping-dilaton case, we use the equation of motion for the perpendicular case, eq.~\eqref{eq:eomxperp}, with the gauge $\sigma = u$, so that we need to solve for $x(u)$. When the jump in the dilaton vanishes, $\delta\phi=0$, the background solution is $AdS_5 \times S^5$, in which case a straight string sits at constant $(x_1,x_2,x_3)$, and is spatially extended in the radial direction. Placing the string at $(x_1,x_2,x_3)=(0,0,L_3)$, and using eq.~\eqref{uxcoords}, the solution for the straight string is
\beq
x(u) = \tanh^{-1}(L_3u).
\eeq
Now introducing a small jump in the dilaton, $\delta \phi \ll1$, we expand the equation of motion in $\delta \phi$. The solution for the straight string is then straightforward to obtain to order $\delta \phi$,
\begin{align}
x(u) &= \tanh^{-1}(L_3u) - \delta \phi \left(\frac{c_1u}{1-L_3^2 u^2} + \frac{c_2\sqrt{1-L_3^2 u^2}}{3u^2} + \frac{1+L_3^2 u^2}{2(1-L_3^2 u^2)} \right) + \mathcal{O}\left(\delta \phi^2\right),&  &y = 0,& \cr
x(u) &= \tanh^{-1}(L_3u) - \delta \phi \left(\frac{d_1u}{1-L_3^2 u^2} + \frac{d_2\sqrt{1-L_3^2 u^2}}{3u^2} - \frac{1-3L_3^2 u^2}{6 L_3^2 u^2(1-L_3^2 u^2)} \right) + \mathcal{O}\left(\delta \phi^2\right),& &y = \frac{\pi}{2},&
\end{align}
where $c_1$, $c_2$, $d_1$, and $d_2$ are integration constants, which we fix as follows. At the Poincar\'e horizon, $u=0$, we find a quadratic pole in $u$. If the coefficient of that pole is non-zero, then the solution will describe a string that bends and returns to the boundary, rather than a straight string. We thus set the coefficient of that pole to zero by choosing $c_2 = 0$ in the case $y=0$ and $d_2 = 1/(2L_3^2)$ in the case $y = \pi/2$. To fix $c_1$ and $d_1$, we require the string to intersect the boundary at $x_3 = L_3$, which means $c_1 = -|L_3|$ and $d_1 = - L_3/3$. We next insert these solutions into eq.~\eqref{area1} to obtain the renormalized on-shell action:
\begin{align}
\label{eq:straightexact}
S_{\textrm{ren}}^{\textrm{straight}} &=  \mathcal{O}\left(\delta \phi^2\right),& &y = 0,& \cr S_{\textrm{ren}}^{\textrm{straight}} &= - \frac{T}{2 \pi \alpha'} R^2 e^{\phi_{\pm}} \frac{\delta \phi}{2 L_3} +  \mathcal{O}\left(\delta \phi^2\right),& &y = \pi/2,&
\end{align}
where for $y=\pi/2$ we take $\phi_{\pm}$ when $L_3>0$ or $L_3<0$, respectively. Our numerical results in fig.~\ref{fig:straight} (b) agree with eq.~\eqref{eq:straightexact} when $\delta \phi \ll 1$. In particular, $V=0$ when $y=0$ up to $\mathcal{O}\left(\delta \phi^2\right)$.

To proceed for the jumping-axion case, we begin with a jumping-dilaton SUSY Janus solution, with some $\delta \hat{\phi} \ll 1$, and then perform an $SL(2,\mathbb{R})$ transformation with
\beq
a = \frac{1}{1+e^{2 \delta \hat{\phi}}}, \qquad b = e^{-(\hat{\phi}_+ + \hat{\phi}_-)} \frac{e^{2 \delta \hat{\phi}}}{1 + e^{2 \delta \hat{\phi}}}, \qquad c = - e^{\hat{\phi}_+ + \hat{\phi}_-}, \qquad d = 1.
\eeq
The resulting solution has zero jump in the dilaton, but a non-zero jump in the axion:
\beq
e^{2 \phi_+} = e^{2 \phi_-} = e^{2 \hat{\phi}_+} + e^{2 \hat{\phi}_-}, \qquad C_{(0)-} = 0, \qquad C_{(0)+} = e^{-(\hat{\phi}_+ + \hat{\phi}_-)} \tanh(\delta \hat{\phi}).
\eeq
We next expand the equation of motion in $\delta \hat{\phi} \ll 1$, which is equivalent to expanding in $\delta C_{(0)} = C_{(0)+} - C_{(0)-} \ll 1$. The solution for $x(u)$, to order $\delta\hat{\phi}$, is
\begin{align}
x(u) &= \tanh^{-1}(L_3u) - \delta \hat{\phi} \left(\frac{c_1 u}{1-L_3^2 u^2} + \frac{c_2\sqrt{1-L_3^2 u^2}}{3u^2} + \frac{1+3L_3^4 u^4}{12 L_3^2 u^2 (1-L_3^2 u^2)} \right) + \mathcal{O}\left(\delta\hat{\phi}^2\right),& &y = 0,& \cr
x(u) &= \tanh^{-1}(L_3u) - \delta \hat{\phi} \left(\frac{d_1 u}{1-L_3^2 u^2} + \frac{d_2\sqrt{1-L_3^2 u^2}}{3u^2} - \frac{1+3L_3^4 u^4}{12 L_3^2 u^2 (1-L_3^2 u^2)} \right) + \mathcal{O}\left(\delta\hat{\phi}^2\right),& &y = \frac{\pi}{2},&
\end{align}
where $c_1$, $c_2$, $d_1$, and $d_2$ are integration constants. To avoid a quadratic pole in the solutions at $u=0$, we choose $c_2 = - 1/(2L_3^2)$ for the $y=0$ case and $d_2 = 1/(2L_3^2)$ for the $y=\pi/2$ case. Requiring that the string intersect the boundary at $x_3=L_3$ fixes $c_1 =-|L_3|/3$ for the $y=0$ case and $d_1 =|L_3|/3$ for the $y=\pi/2$ case. Inserting these solutions into eq.~\eqref{area1}, we find
\begin{align}
\label{eq:straightaxionexact}
S_{\textrm{ren}}^{\textrm{straight}} &= \frac{T}{2 \pi \alpha'} R^2 e^{\phi} \frac{\delta \hat{\phi}}{4 L_3} + \mathcal{O}\left(\delta\hat{\phi}^2\right)= \frac{T}{2 \pi \alpha'} R^2 e^{3\phi} \frac{\delta C_{(0)}}{8 |L_3|} + \mathcal{O}\left(\delta C_{(0)}^2\right),& &y = 0,& \cr
S_{\textrm{ren}}^{\textrm{straight}} &= - \frac{T}{2 \pi \alpha'} R^2 e^{\phi} \frac{\delta \hat{\phi}}{4 L_3} + \mathcal{O}\left(\delta\hat{\phi}^2\right)= - \frac{T}{2 \pi \alpha'} R^2 e^{3\phi} \frac{\delta C_{(0)}}{8 |L_3|}+ \mathcal{O}\left(\delta C_{(0)}^2\right),& &y = \pi/2.&
\end{align}
We have confirmed that our numerical results, presented in figs.~\ref{fig:straightaxion} (b), (c), and (d), agree with eq.~\eqref{eq:straightaxionexact} in the $\delta C_{(0)} \ll 1$ regime.

In summary, we have shown that for a SUSY interface with either a small jump in the coupling or a small jump in the $\theta$-angle, in all cases the renormalized self-energy $V$ is non-zero, with one exception: a small jump in the coupling and $y=0$. Our numerical results in figs.~\ref{fig:straight} and~\ref{fig:straightaxion} suggest that the same is also true for \textit{all} finite values of the jump in the coupling or $\theta$-angle.

\subsection{Perpendicular Wilson Loops}
\label{sec:perpendicular}

In this subsection we present our numerical results for the expectation values of Wilson loops perpendicular to the interface, depicted in fig.~\ref{fig:configs} (a), representing two test charges along a line perpendicular to the interface, on opposite sides of the interface. In particular, we present our results for the interaction potential $V_{\perp}(L,L_{\textrm{av}})$ as determined from strings in Janus spacetimes via eqs.~\eqref{area1} and~\eqref{eq:vdicholo}.

In electromagnetism with an interface, the $V_{\perp}(L,L_{\textrm{av}})$ in eq.~\eqref{emperpdipolepot} is linear, \textit{i.e.}\ is a sum of the interaction energy between the two test charges and the interaction energies with the image charges. In a non-Abelian gauge theory such as $SU(N_c)$ $\N=4$ SYM, however, generically $V_{\perp}(L,L_{\textrm{av}})$ will not be linear. Indeed, strictly speaking, whether the method of image charges will be applicable in $SU(N_c)$ $\N=4$ SYM is not clear. To provide a measure of the deviation from linearity, we will compare our numerical results for $V_{\perp}(L,L_{\textrm{av}})$ to an \textit{ad hoc} linear potential of the form
\beq
\label{eq:adhoclinearpotperp}
V_{\perp}^{\textrm{lin}}(L,L_{\textrm{av}}) \equiv V^{\textrm{int}}_{\perp}(L) + V_{\textrm{self}}(L,L_{\textrm{av}}),
\eeq
where for the ``interaction potential'' $V^{\textrm{int}}_{\perp}(L)$ we use the holographic result for $V(L)$ in $\N=4$ SYM with constant coupling, eq.~\eqref{eq:maldacenapotential}, but replacing $\lambda$ with an ``effective 't Hooft coupling,''
\beq
\label{eq:adhocinteractionpot}
V^{\textrm{int}}_{\perp}(L) = - \frac{4\pi^2}{\Gamma\left(1/4\right)^4} \frac{\sqrt{2 \lambda_+}}{L} \sqrt{1+\tilde{Q}^e/Q^e},
\eeq
where $\tilde{Q}^e$ is defined in eq.~\eqref{tildeqdef}.
The self-energy $V_{\textrm{self}}(L,L_{\textrm{av}})$ is a sum of two terms, representing the interaction of each test charge with its own image charge, which we obtain from our numerical results in section~\ref{sec:straight}.\footnote{Our $V_{\perp}^{\textrm{lin}}(L,L_{\textrm{av}})$ is designed to mimic the analogous potential $V_{\perp}(L,L_{\textrm{av}})$ in electromagnetism, eq.~\eqref{emperpdipolepot}, but with modifications for the $\lambda \gg 1$ regime, in which $V(L) \propto \sqrt{\lambda}/L$ rather than $V(L)\propto \lambda/L$ as in the $\lambda \ll 1$ regime. Specifically, $V^{\textrm{int}}_{\perp}(L)$ is designed to mimic the first two terms on the right-hand-side of eq.~\eqref{emperpdipolepot}, while $V_{\textrm{self}}(L,L_{\textrm{av}})$ is designed to mimic the last two terms.}  We shall see that the holographic results are well approximated by this potential: the method of images is a good approximation even in strongly coupled $\N=4$ SYM.

\begin{figure}[ht!]
  \begin{center}
   \subfigure[Non-SUSY interface, jumping coupling.]{
        \includegraphics[width=0.45\textwidth]{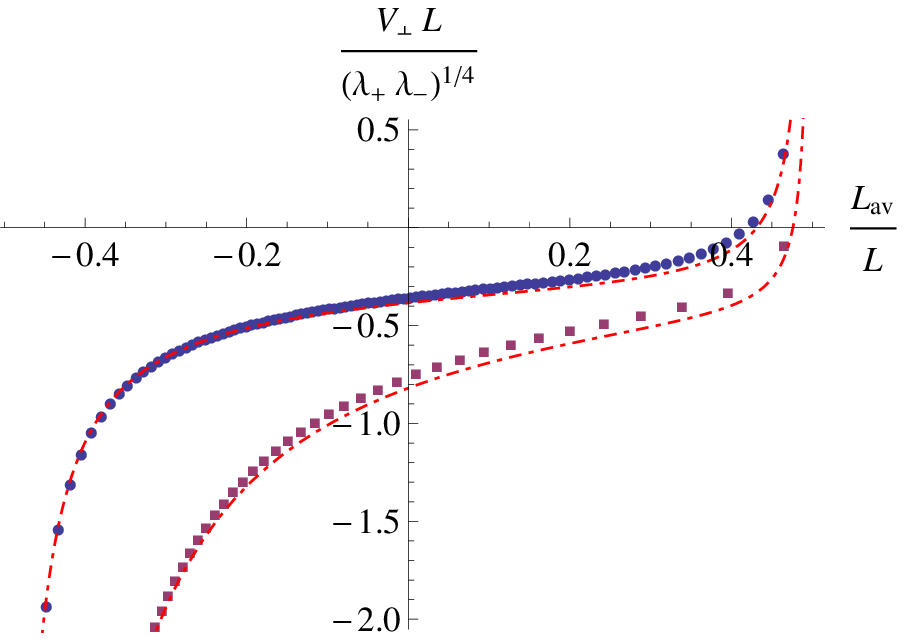}}\hspace{5mm}
     \subfigure[Non-SUSY interface, jumping $\theta$-angle.]{
        \includegraphics[width=0.45\textwidth]{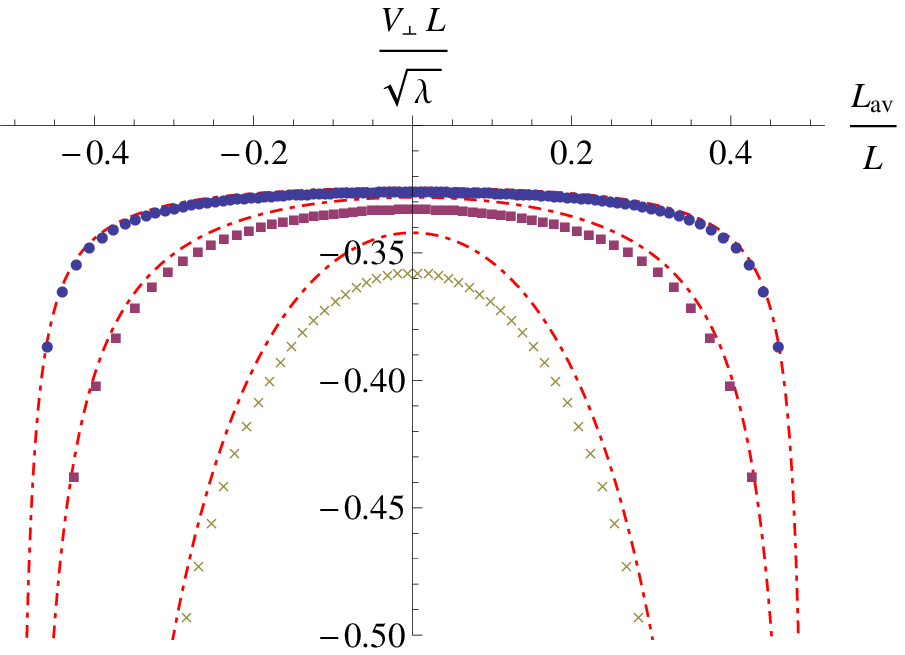}}
  \end{center}
  \caption{The potential $V_{\perp}(L,L_{\textrm{av}})$ between two test charges separated along a line perpendicular to a non-SUSY interface. (a) $V_{\perp} L / (\lambda_+\lambda_-)^{1/4}$ for a jumping coupling. The blue dots are our numerical results for a jump with $\log (\lambda_+/\lambda_-)=1$ while the purple squares are for a jump with $\log (\lambda_+/\lambda_-)=3$. (b) $V_{\perp} L / \sqrt{\lambda}$ for a jumping $\theta$-angle. The blue dots, purple squares, and golden crosses are our numerical results for $g^2 \Delta \theta=2\pi^2$, $4 \pi^2$, and $40\pi^2$, respectively. In both (a) and (b) the red dot-dashed lines are our results for the \textit{ad hoc} linear potential $V^{\textrm{lin}}_{\perp}$ defined in eq.~\eqref{eq:adhoclinearpotperp}. For both (a) and (b), our results are qualitatively similar to the analogous results in electromagnetism shown in fig.~\ref{fig:empotperpfigs}.} \label{fig:perpendicularns}
\end{figure}
\begin{figure}[ht!]
  \begin{center}
   \subfigure[SUSY interface, jumping coupling, $y=0$]{
        \includegraphics[width=0.45\textwidth]{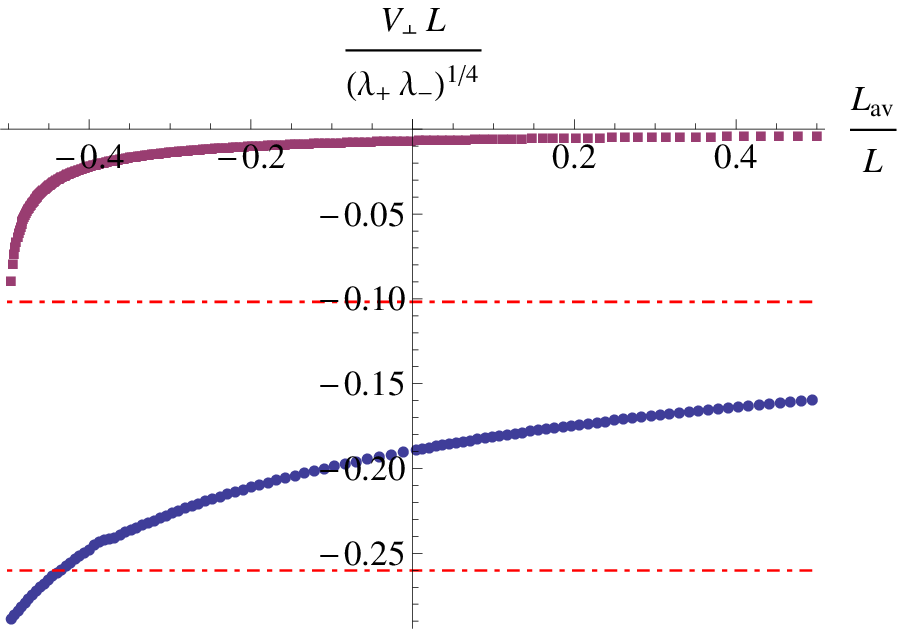}}\hspace{5mm}
   \subfigure[SUSY interface, jumping coupling, $y=\pi/2$]{
        \includegraphics[width=0.45\textwidth]{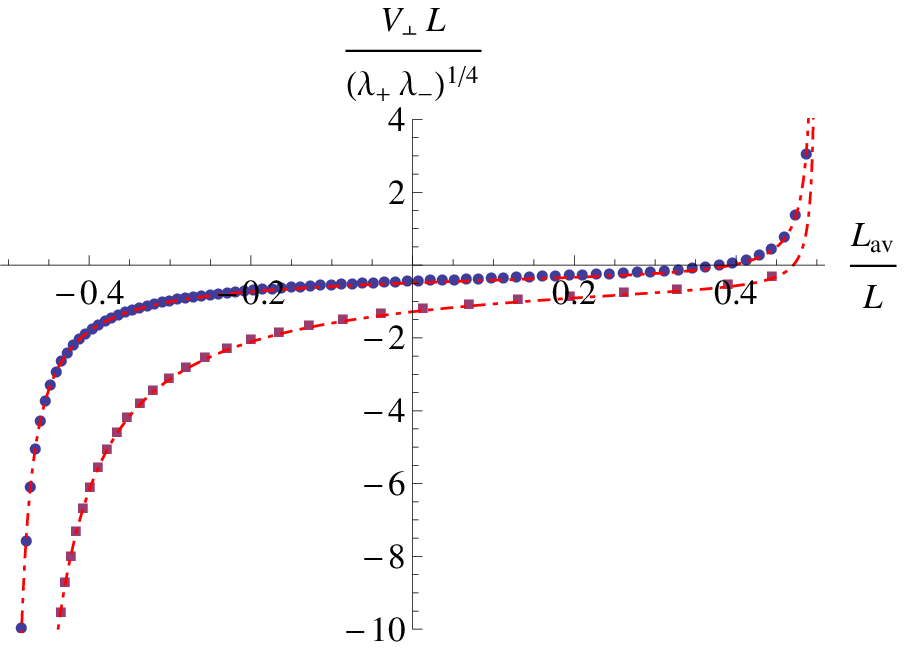}}
  \end{center}
  \caption{The potential $V_{\perp}(L,L_{\textrm{av}})$ between two test charges along a line perpendicular to a SUSY interface with jumping coupling. We plot $V_{\perp} L / (\lambda_+\lambda_-)^{1/4}$ versus $L_{\textrm{av}}/L$, as computed from a string at (a) $y=0$ or (b) $y=\pi/2$. In both (a) and (b), the blue dots and purple squares are our numerical results for a jump with $\log (\lambda_+/\lambda_-)=1$ or $3$, respectively, while the red dot-dashed lines are our numerical results for the \textit{ad hoc} linear potential $V_{\perp}^{\textrm{lin}}$ defined in eq.~\eqref{eq:adhoclinearpotperp}.} \label{fig:perpendicularsusydilaton}
\end{figure}
\begin{figure}[ht!]
  \begin{center}
   \subfigure[SUSY interface, jumping $\theta$-angle, $y=0$]{
        \includegraphics[width=0.45\textwidth]{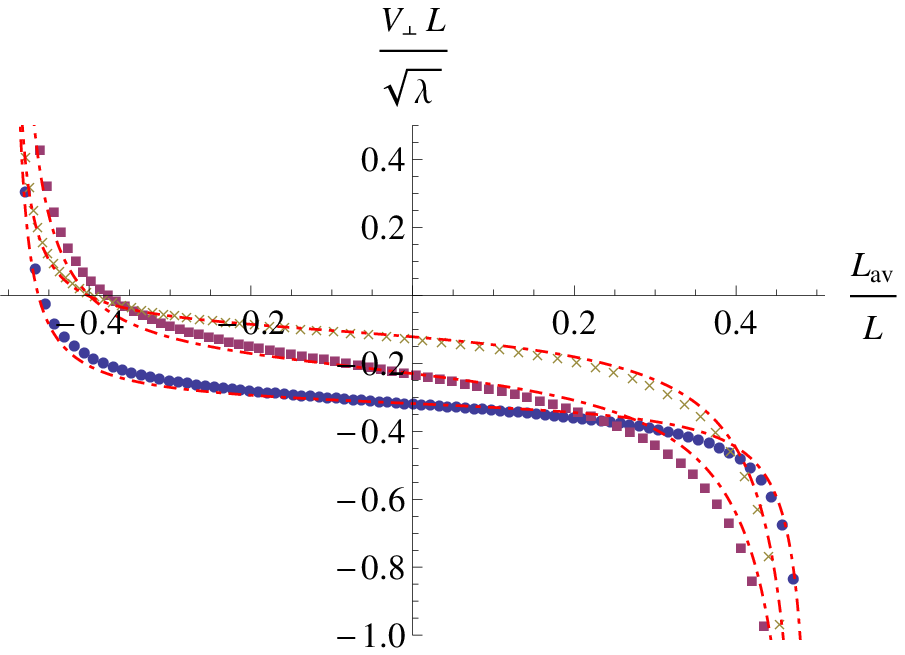}}\hspace{5mm}
   \subfigure[SUSY interface, jumping $\theta$-angle, $y=\pi/2$]{
        \includegraphics[width=0.45\textwidth]{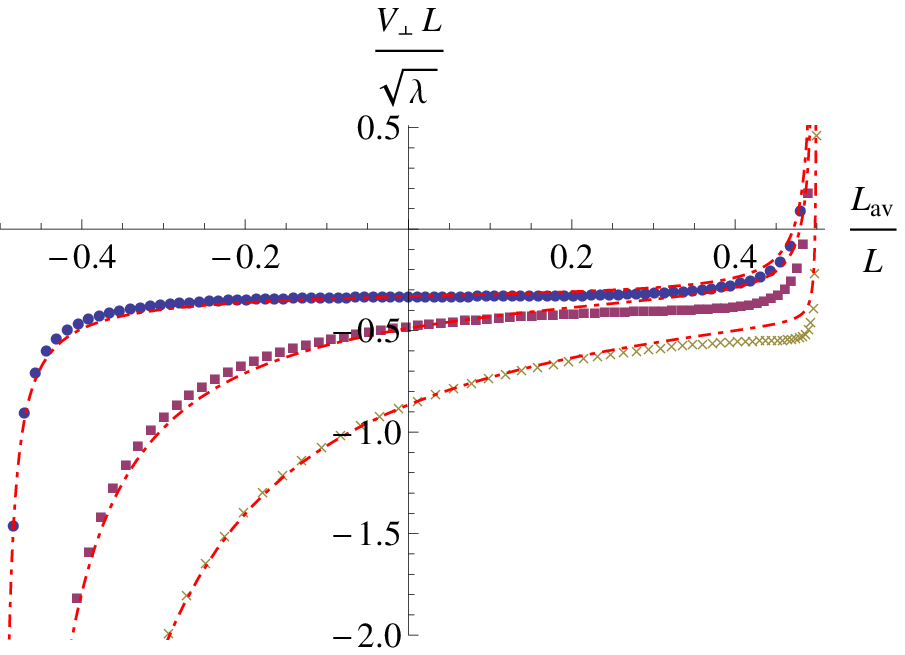}}
  \end{center}
  \caption{The potential $V_{\perp}(L,L_{\textrm{av}})$ between two test charges along a line perpendicular to a SUSY interface with jumping $\theta$-angle. More precisely we plot $V_{\perp} L / \sqrt{\lambda}$ versus $L_{\textrm{av}}/L$ as computed from a string at (a) $y=0$ or (b) $y=\pi/2$. In both (a) and (b), the blue dots, purple squares, and golden crosses are our numerical results for $g^2 \Delta \theta=2\pi^2$, $12\pi^2$, and $40\pi^2$, respectively, while the red dot-dashed lines are our results for the \textit{ad hoc} linear potential $V^{\textrm{lin}}_{\perp}$ defined in eq.~\eqref{eq:adhoclinearpotperp}.} \label{fig:perpendicularsusyaxion}
\end{figure}

\begin{itemize}[leftmargin=0.2in]
\item \textbf{Non-SUSY, jumping coupling/$\theta$-angle (fig.~\ref{fig:perpendicularns}):} In figs.~\ref{fig:perpendicularns} (a) and (b) we present our numerical results for a non-SUSY interface with jumping coupling or $\theta$-angle, respectively. In each case, notice the striking similarity to the analogous results in electromagnetism shown in fig.~\ref{fig:empotperpfigs}: for a jumping coupling, a test charge is attracted to the side with smaller coupling, while for a jumping $\theta$-angle, a test charge is attracted to the interface. We also see that in most cases our \textit{ad hoc} linear potential $V_{\perp}^{\textrm{lin}}$ provides a surprisingly good approximation to the holographic results, especially for small values in the jump.
\item \textbf{SUSY, jumping coupling (fig.~\ref{fig:perpendicularsusydilaton}):} Here we find that the results depend sensitively on $y$. For $y=0$, our holographic result for $V_{\perp} L$, depicted in fig.~\ref{fig:perpendicularsusydilaton} (a), remains finite at $L_{\textrm{av}}/L = \pm1/2$, in stark contrast to electromagnetism. This is consistent with our results in section~\ref{sec:straight} for the vanishing of the image charges. For the case with $y=\pi/2$, we see behavior qualitatively similar to that in electromagnetism: compare fig.~\ref{fig:perpendicularsusydilaton} (b) to fig.~\ref{fig:empotperpfigs} (a).
\item \textbf{SUSY, jumping $\theta$-angle (fig.~\ref{fig:perpendicularsusyaxion}):} Again the results depend on the value of $y$ and the behavior of the potential is qualitatively different than in electromagnetism: compare to fig.~\ref{fig:empotperpfigs} (b). For both $y=0$ and $y = \pi/2$, the behavior of $V_{\perp}$ is consistent with our results for a single test charge in figs.~\ref{fig:straightaxion} (b) and (c) and with the interpretation in terms of non-zero image charges.
\end{itemize}

We also see from figs.~\ref{fig:perpendicularsusydilaton} and~\ref{fig:perpendicularsusyaxion} that in all cases with a SUSY interface $V_{\perp}^{\textrm{lin}}$ (the red dot-dashed lines in the figures) again provides a surprisingly good approximation to the holographic results, with one exception: an interface with a jumping coupling and a test charge with $y=0$, fig.~\ref{fig:perpendicularsusydilaton} (a.). In that case, the fact that $V_{\textrm{self}}=0$ means $V_{\perp}^{\textrm{lin}} = V^{\textrm{int}}_{\perp}$, which is constant in $L_{\textrm{av}}/L$. The holographic results are clearly not constant in $L_{\textrm{av}}/L$.

\subsection{Parallel Wilson Loops}
\label{sec:parallel}

In this subsection we present our numerical results for the expectation values of Wilson loops parallel to the interface, depicted in fig.~\ref{fig:configs} (b), representing two test charges along a line parallel to the interface. In particular, we present our results for the interaction potential $V_{\parallel}(L,L_3)$ as determined from strings in Janus spacetimes via eqs.~\eqref{area2} and~\eqref{eq:vdicholo}.

As in the previous subsection, we will compare our holographic results to an \textit{ad hoc} linear potential, $V_{\parallel}^{\textrm{lin}}(L,L_3)$, defined similarly to $V_{\perp}^{\textrm{lin}}(L,L_{\textrm{av}})$,
\beq
\label{eq:adhoclinearpotpara}
V_{\parallel}^{\textrm{lin}}(L,L_3) \equiv V^{\textrm{int}}_{\parallel}(L,L_3) + V_{\textrm{self}}(L_3),
\eeq
where, for $L_3>0$,
\beq
V^{\textrm{int}}_{\parallel}(L,L_3) \equiv - \frac{4\pi^2}{\Gamma\left(1/4\right)^4} \frac{\sqrt{2 \lambda_+}}{L} \sqrt{1+\frac{\tilde{Q}^e/Q^e}{\sqrt{1+4L_3^2/L^2}}},
\eeq
and the same for $L_3<0$ but with $g_+ \leftrightarrow g_-$ and $\theta_+ \leftrightarrow \theta_-$. The self-energy $V_{\textrm{self}}(L_3)$ is a sum of two terms, representing the interaction of each test charge with its own image charge, which we obtain from our numerical results in section~\ref{sec:straight}. As in the previous subsection, our \textit{ad hoc} linear potential is designed to mimic the analogous potential $V_{\parallel}(L,L_3)$ in electromagnetism, eq.~\eqref{emparadipolepot}. Specifically, $V^{\textrm{int}}_{\parallel}(L,L_3)$ is designed to mimic the first and fourth terms on the right-hand-side of eq.~\eqref{emparadipolepot}, while $V_{\textrm{self}}(L_3)$ is designed to mimic the second and third terms.

\begin{figure}[ht!]
  \begin{center}
   \subfigure[Non-SUSY interface, jumping coupling]{
        \includegraphics[width=0.45\textwidth]{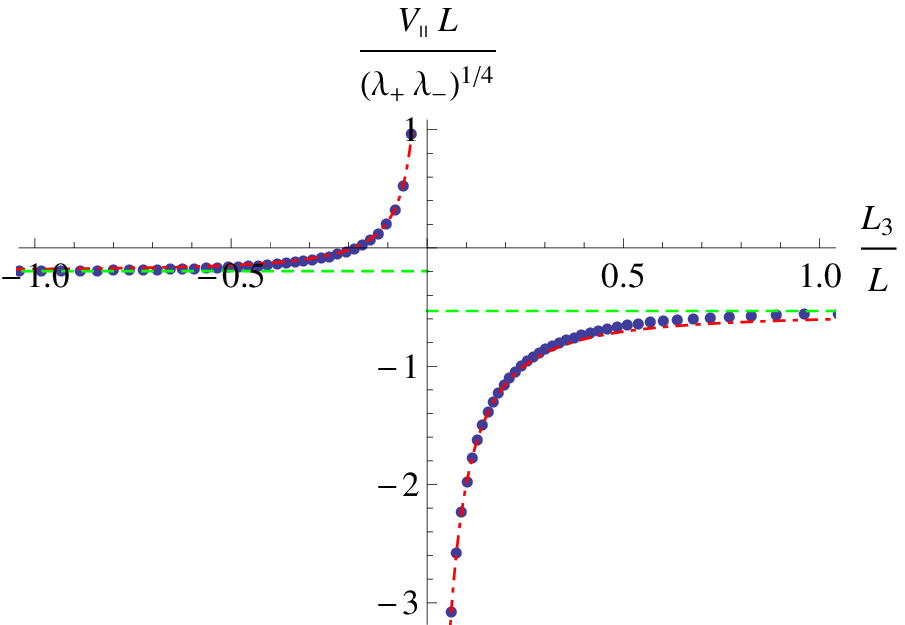}} \hspace{5mm}
          \subfigure[Non-SUSY interface, jumping coupling]{
        \includegraphics[width=0.45\textwidth]{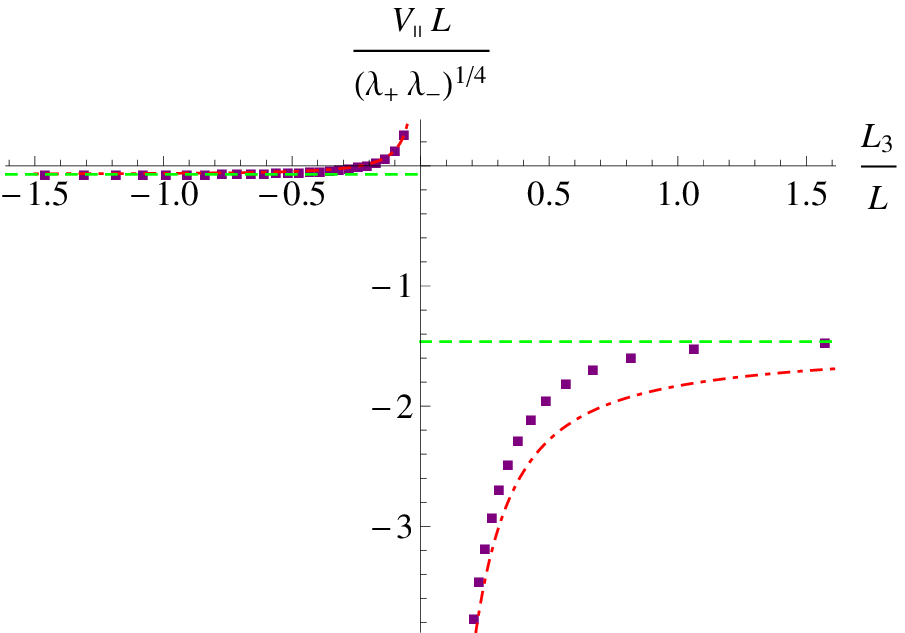}} \hspace{5mm}
  \end{center}
  \caption{The potential $V_{\parallel}(L,L_3)$ between two test charges along a line parallel to a non-SUSY interface with a jumping coupling.  (a) The blue dots are our numerical results for $\log (\lambda_+/\lambda_-)=1$. (b) The purple squares are our numerical results for $\log (\lambda_+/\lambda_-)=3$. In both (a) and (b), the horizontal green dashed lines are the result for $\N=4$ SYM without an interface, eq.~\eqref{eq:maldacenapotential}. In both (a) and (b) the red dot-dashed lines are the \textit{ad hoc} linear potential $V_{\parallel}^{\textrm{lin}}$ defined in eq.~\eqref{eq:adhoclinearpotpara}.} \label{fig:paralleldilaton}
\end{figure}
\begin{figure}[ht!]
  \begin{center}
           \includegraphics[width=0.5\textwidth]{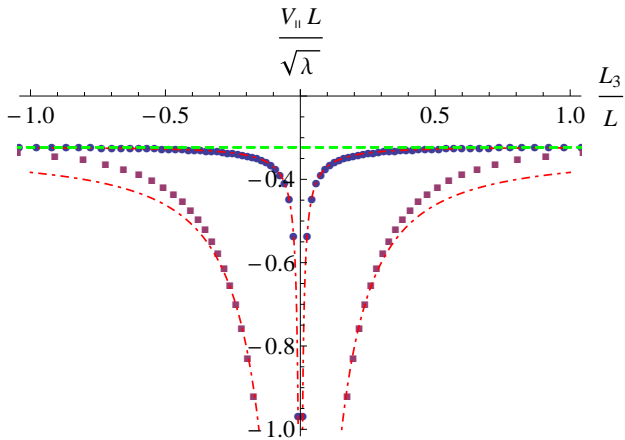}
  \end{center}
  \caption{The potential $V_{\parallel}(L,L_3)$ between two test charges along a line parallel to a non-SUSY interface with jumping $\theta$-angle. The blue dots and purple squares are our numerical results for $g^2 \Delta \theta=2\pi^2$ and $40\pi^2$, respectively. The horizontal green dashed line is the result for $\N=4$ SYM without an interface, eq.~\eqref{eq:maldacenapotential}. The red dot-dashed lines are our results for the \textit{ad hoc} linear potential $V_{\parallel}^{\textrm{lin}}$ defined in eq.~\eqref{eq:adhoclinearpotpara}.} \label{fig:parallelaxion}
\end{figure}

\begin{itemize}[leftmargin=0.2in]
\item \textbf{Non-SUSY, jumping coupling (fig.~\ref{fig:paralleldilaton}):}  As in the previous cases with non-SUSY interfaces, we find a striking similarity to the analogous results in electromagnetism, fig.~\ref{fig:empotparafigs} (a). Moreover, the fact that the results in fig.~\ref{fig:paralleldilaton} are qualitatively different from the potential obtained from the sum of ladder diagrams when $\kappa=0$, shown in fig.~\ref{fig:ladder}, provides additional  evidence that the field theory dual to non-SUSY, jumping-dilaton Janus has non-zero $\kappa$.\footnote{Of course, we must bear in mind that even in $\N=4$ SYM without an interface, other diagrams besides the ladders contribute to $V_{\parallel}(L,L_3)$~\cite{Erickson:1999qv,Erickson:2000af}, so the comparison between figs.~\ref{fig:ladder} and~\ref{fig:paralleldilaton}, though suggestive, is not conclusive.} If we make the distance between the charges much less than the distance to the interface, $L_3/L \to \pm \infty$, then we should recover the result for $\N=4$ SYM without an interface, eq.~\eqref{eq:maldacenapotential}, with the appropriate value of $\lambda$ for that side of the interface, $\lambda_{\pm}$. In fig.~\ref{fig:paralleldilaton} we depict the result of eq.~\eqref{eq:maldacenapotential} for the two sides of the interface as horizontal green dashed lines. Our holographic results for $V_{\parallel}(L,L_3)$ indeed approach these limiting values as $L_3/L \to \pm\infty$.
\item \textbf{Non-SUSY, jumping $\theta$-angle (fig.~\ref{fig:parallelaxion}):}  Once again we find a striking similarity to the analogous results in electromagnetism, fig.~\ref{fig:empotparafigs} (b). In particular, we see that the dipole is attracted to the interface. Again we see that our holographic results approach the result in eq.~\eqref{eq:maldacenapotential}, depicted as the horizontal green dashed line in fig.~\ref{fig:parallelaxion}, in the limits $L_3/L \to \pm \infty$.
\end{itemize}

From figs.~\ref{fig:paralleldilaton} and~\ref{fig:parallelaxion} we see that $V_{\parallel}^{\textrm{lin}}(L,L_3)$ again provides a surprisingly good approximation to the holographic results, and that, as also occurred for $V_{\perp}(L,L_{\textrm{av}})$ for non-SUSY interfaces, that approximation grows worse as the jump in the coupling or $\theta$-angle increases.

We will not present results for $V_{\parallel}(L,L_3)$ for SUSY interfaces. In contrast to all previous cases, here we found the matching between our numerical string solutions in SUSY Janus and the near-boundary asymptotics in eq.~\eqref{asymsol} prohibitively difficult.

\subsection{Wilson Loops on the Interface}
\label{intloops}

In this subsection, we consider rectangular Wilson loops located precisely on the interface, at $x_3=0$. We can obtain these by taking the $L_3\to0$ limit of the parallel Wilson loops we considered in the previous subsection. In electromagnetism, in such a limit the potential between two test charges diverges because the test charges approach the interface and eventually coincide with their image charges: recall fig.~\ref{fig:empotparafigs}. In other words, in that limit the self-energy of each test charge diverges. Our results of the previous subsections for the self-energy of a single test charge, figs.~\ref{fig:straight} and~\ref{fig:straightaxion}, show that the same divergence occurs for either a non-SUSY or SUSY interface in $\N=4$ SYM at large $N_c$ and large coupling, with one exception, a test charge with $y=0$ in the presence of a SUSY interface where the coupling jumps.

The divergence of the self-energy at the interface has a simple, intuitive explanation from the gravity point of view. As mentioned in section~\ref{topo}, the boundary of the Janus solutions is not globally flat, but rather has an angular excess at the point where the axio-dilaton jumps. As a result, the Janus metrics do not admit a globally well-defined Fefferman-Graham expansion, so holographic renormalization must be performed separately in each of the two asymptotically $AdS_5 \times S^5$ regions. In each region, the Fefferman-Graham expansion is an expansion in $z/x_3$, as explained in appendix~\ref{holorg} (specifically eqs.~\eqref{eq:nonsusyjanusg11g33expansion} and~\eqref{eq:susyjanusg11g33expansions}), which clearly must break down in the limit $x_3 \to 0$.\footnote{For a detailed analysis of the breakdown of the Fefferman-Graham expansion in non-SUSY Janus in the $x_3\to0$ limit, see appendix B of ref.~\cite{Papadimitriou:2004rz}.} Consequently, the entire holographic renormalization procedure breaks down, and hence we can obtain a divergent result for the self-energy.

To obtain a finite result for the expectation values of rectangular Wilson loops precisely on the interface, we need a prescription to subtract the divergent self-energies of the test charges. The dual gravity description suggests a natural prescription. The strings describing the rectangular Wilson loops will sit at fixed $x$, and so will effectively behave as strings in $AdS_4$. We can thus perform holographic renormalization directly in that $AdS_4$, that is, using the counterterm of eq.~\eqref{eq:ctaction}, but for $AdS_4$ rather than $AdS_5$.

We are considering conformal interfaces in $\N=4$ SYM, so after renormalization the expectation value of a rectangular Wilson loop sitting on the interface will depend on only one scale, the separation $L$ between the test charges. As a result, the potential $V(L) \propto 1/L$, with a proportionality constant that will depend on the values of the coupling or $\theta$-angle on the two sides of the interface, $\lambda_{\pm}$ and $\theta_{\pm}$. Indeed, the holographic result will provide us with a value of the ``effective coupling'' precisely on the interface, as defined within our renormalization prescription.

To compute $V(L)$ holographically, we need solutions of the string equations of motion, eqs.~\eqref{eq:eomu}, \eqref{eq:eomx}, and~\eqref{eq:eomy}, describing strings with both endpoints on the boundary at $x_3=0$. From eq.~\eqref{uxcoords}, we see that the simplest such string will sit at fixed $x$ and $y$ and hence will move only in $x_1$ and $u$, or in other words will move only along an $AdS_4$ slice. The endpoints of such a string will thus end at the $AdS_4$ boundary, which is $x_3=0$. The equations of motion for such a string require
\beq
\label{eq:defectconst}
\p_x (e^\phi f_4^2) = 0, \qquad \p_y (e^\phi f_4^2) = 0.
\eeq
If we can solve eq.~\eqref{eq:defectconst}, then we obtain a string that moves only in $AdS_4$, hence we can compute $S^{\parallel}_{\textrm{ren}}$ following the procedure of refs.~\cite{Rey:1998ik,Maldacena:1998im}. The resulting renormalized on-shell string action is
\beq
\label{eq:renactiononinterface}
S^{\textrm{int}}_{\textrm{ren}} = - \frac{T}{2 \pi \alpha^\prime} \frac{4 \ {\cal I}^2 e^\phi f_4^2}{L}.
\eeq
Using eq.~\eqref{eq:vdicholo}, we then find the potential $V(L)$, which is almost identical in form to that of $\N=4$ SYM without an interface, eq.~\eqref{eq:maldacenapotential}, except for the value of the ``effective 't Hooft coupling'' on the interface, $\lambda_\textrm{eff}$,
\beq
\label{eq:lambdaeffdef}
V(L) = - \frac{4\pi^2}{\Gamma\left(1/4\right)^4} \frac{\sqrt{2 \lambda_{\textrm{eff}}}}{L}, \qquad \lambda_{\textrm{eff}} \equiv 2 \pi \, e^{2 \phi} N_c \,\frac{f_4^4}{R^4}.
\eeq

\begin{figure}[ht!]
  \begin{center}
   \subfigure[Non-SUSY interface, coupling]{
        \includegraphics[width=0.45\textwidth]{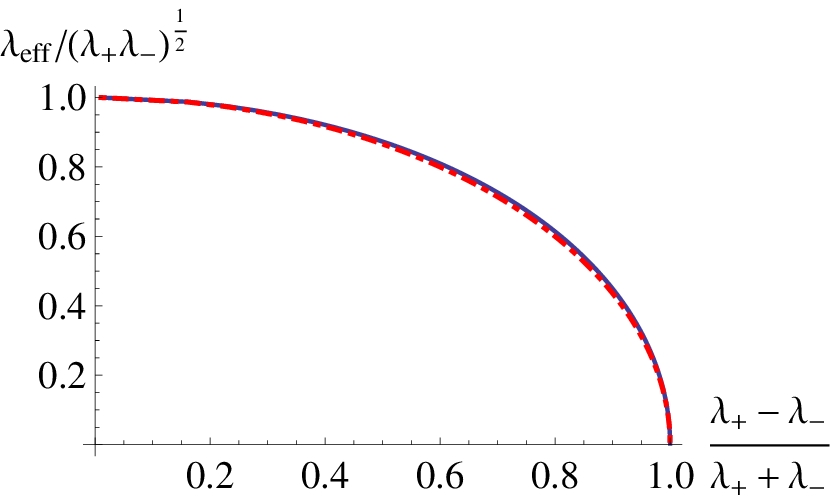}}\hspace{5mm}
   \subfigure[SUSY interface, jumping coupling]{
        \includegraphics[width=0.45\textwidth]{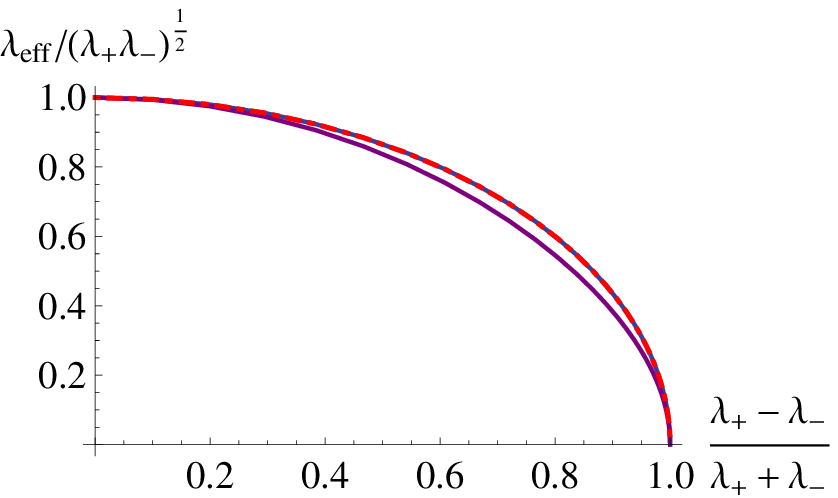}} \\
   \subfigure[Non-SUSY interface, jumping $\theta$-angle]{
        \includegraphics[width=0.45\textwidth]{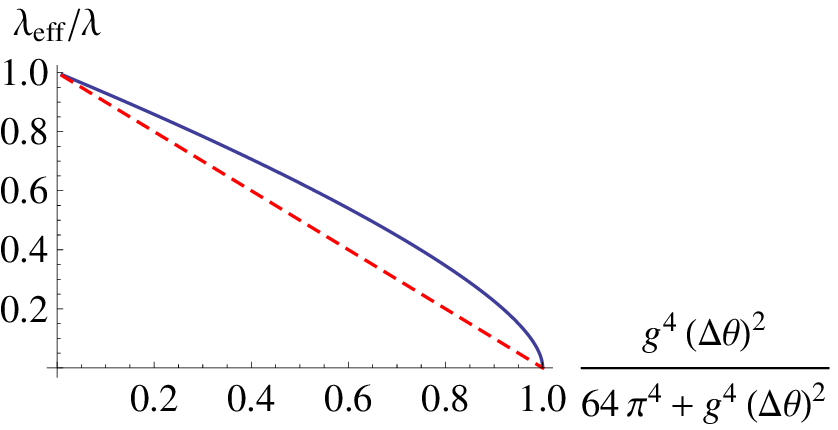}} \hspace{5mm}
   \subfigure[SUSY interface, jumping $\theta$-angle]
        {\includegraphics[width=0.45\textwidth]{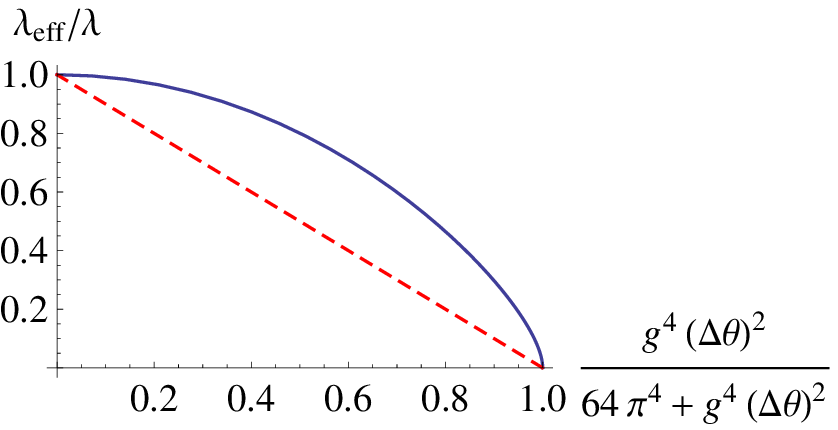}}
  \end{center}
\caption{The effective 't Hooft coupling $\lambda_{\textrm{eff}}$, defined in eq.~\eqref{eq:lambdaeffdef}. (a) $\lambda_{\textrm{eff}}/\sqrt{\lambda_+ \lambda_-}$ as a function of $\frac{\lambda_+ - \lambda_-}{\lambda_+ + \lambda_-}$ for a non-SUSY interface where the coupling jumps.  The purple solid line is our numerical result, while the red dashed line is  $\lambda_+( 1 + \tilde{Q}^e/Q^e)/\sqrt{\lambda_+ \lambda_-} $ with $\tilde{Q}^e$ defined in eq.~\eqref{tildeqdef}. The two lines are nearly coincident. (b) The same as (a), but for a SUSY interface where the coupling jumps. Here the upper blue solid line is our numerical result for $y=0$ while the lower purple solid line is our numerical result for $y=\pi/2$. The upper blue line is coincident with the red dashed line. (c) $\lambda_{\textrm{eff}}/\lambda$ as a function of $\frac{g^4\left(\Delta \theta\right)^2}{64\pi^4 + g^4\left(\Delta \theta\right)^2}$ for a non-SUSY interface where the $\theta$-angle jumps. The blue solid and red dashed lines have the same meaning as in (a). (d) The same as (c), but for a SUSY interface where the $\theta$-angle jumps. Our numerical results for $y=0$ and $y=\pi/2$, the blue solid lines, are identical.}
\label{fig:defcoups}
\end{figure}

To determine $\lambda_{\textrm{eff}}$ requires solving eq.~\eqref{eq:defectconst}, which we do numerically. Our results appear in fig.~\ref{fig:defcoups}. We compare our numerical results against the effective 't Hooft coupling $\lambda_+ ( 1 + \tilde{Q}^e/Q^e)$ that we used in eq.~\eqref{eq:adhocinteractionpot}, and whose form is motivated by the result for the potential in electromagnetism, eq.~\eqref{emparadipolepot}, ignoring the self-energy contributions and setting $L_3=0$.

For either a non-SUSY or SUSY interface where the coupling jumps, shown in figs.~\ref{fig:defcoups} (a) and (b), we find that our numerical results for $\lambda_{\textrm{eff}}$ agree remarkably well with $\lambda_+ ( 1 + \tilde{Q}^e/Q^e)$. Indeed, for the SUSY interface and a test charge with $y=0$, the agreement appears to be exact. For either non-SUSY or SUSY interfaces where the $\theta$-angle jumps, however, the holographic results only approach $\lambda_+ ( 1 + \tilde{Q}^e/Q^e)$ in the limits $\Delta \theta \to 0$ or $\to \infty$. For any finite $\Delta \theta$, the holographic result is larger than $\lambda_+( 1 + \tilde{Q}^e/Q^e)$.

\section*{Acknowledgements}

We thank C.~Hoyos, E.~Kiritsis, M.~Lippert, H.~Osborn, I.~Papadimitriou, V.~Pestun, A.~Petkou, G.~Semenoff, K. Sfetsos, K.~Skenderis, D.~Tong, E.~Witten and J.~Zaanen for useful conversations and correspondence. We especially thank A. Karch for reading and commenting on a preliminary draft of the manuscript, and A. Rebhan and M. Jech for catching typographical errors in the first version of the paper. J.~E. was supported by the FWO - Vlaanderen, Project No. G.0651.11, the ``Federal Office for Scientific, Technical and Cultural Affairs through the Inter-University Attraction Poles Programme,''  Belgian Science Policy P6/11-P, as well as the European Science Foundation Holograv Network, and is currently supported in part by STFC grant ST/J0003533/1. The work of A.O'B. was supported in part by the European Union grant FP7-REGPOT-2008-1-CreteHEPCosmo-228644. The research leading to these results has received funding from the European Research Council under the European Community's Seventh Framework Programme (FP7/2007-2013) / ERC grant agreement no. 247252. The work of T.W. is supported by a Research Fellowship (Grant number WR 166/1-1) of the German Research Foundation (DFG).

\appendix

\section{String Equations of Motion and Solutions}
\label{stringeoms}

To accommodate both non-SUSY and SUSY Janus, we write both metrics in the same general form as that of SUSY Janus, eq.~\eqref{eq:susyjanusmetric},
\beq
\label{eq:generalmetric}
ds^2 =  G_{MN} dx^M dx^N = f_4^2 \left(\frac{du^2}{u^2} +  u^2(dt^2 + dx_1^2 + dx_2^2) \right)+ \rho_1^2 dx^2 + \rho_2^2 dy^2  + f_1^2 ds^2_{S^2} + f_2^2 ds^2_{S^2},
\eeq
with $M,N = 0, \ldots, 9$, and where for SUSY Janus the warp factors $f_4^2$, $f_1^2$, $f_2^2$, $\rho_1^2$, and $\rho_2^2$ are functions of $x$ and $y$, while for non-SUSY Janus $f_4^2$ and $\rho_1^2$ depend only on $x$ while $f_1^2$ and $f_2^2$ depend only on $y$, and $\rho_2^2 = R^2$, as we can see from eqs.~\eqref{eq:nonsusyjanusmetric} and \eqref{s5metric}.

At leading order in $\alpha'$, the string action includes two terms. The first is the Nambu-Goto term, representing the area of the string worldsheet times the string tension. The second is a term involving the pull-back of the NS two-form. The non-SUSY Janus solution has a vanishing NS two-form, but the SUSY Janus solution includes a non-trivial NS two-form. We will argue below that for our string solutions in SUSY Janus, the pull-back of the NS two-form vanishes. We thus only present the Nambu-Goto term explicitly, which in Euclidean signature and Einstein frame is
\beq
\label{eq:NGaction}
S=\frac{1}{2\pi\alpha'} \int d\tau \, d\sigma \, \sqrt{\det{e^{\phi}  \, G_{MN} \, \partial_m X^M \partial_n X^N}},
\eeq
where $X^M(\tau,\sigma)$ are the worldsheet scalars, and the determinant is over the worldsheet coordinates $m,n=\tau,\sigma$. We employ static gauge, $\tau = t$ in all that follows. In that gauge, the integration over $\tau$ in eq.~\eqref{eq:NGaction} always trivially produces a factor of $T$.

As explained in section~\ref{topo}, we will use $SL(2,\mathbb{R})$ transformations to convert jumping-dilaton Janus into jumping-axion Janus. The only way that our strings will ``know'' about these $SL(2,\mathbb{R})$ transformations is via their coupling to $\phi$ in eq.~\eqref{eq:NGaction}, since they do not couple directly to $C_{(0)}$ and the Einstein-frame metric $G_{MN}$ is $SL(2,\mathbb{R})$-invariant.

We can use the isometries of the bulk metric to simplify the string action. Using the $SO(3) \times SO(3)$ symmetry of the internal space, we may place the string at a fixed point on each $S^2$. For a string in SUSY Janus, which has NS two-form flux on one $S^2$, the pull-back of the NS two-form to the worldsheet then vanishes, as advertised. Using the symmetries of the $(x_1,x_2)$ plane, we may place the string at $x_2=0$. After these simplifications, the most general Ansatz we can write is $u = u(\s)$, $x = x(\s)$, $y=y(\s)$, $x_1 = x_1(\s)$, in which case the Nambu-Goto action becomes
\beq
\label{eq:action}
S = \frac{T}{2\pi\alpha'} \int d\s \, e^{\phi}  \sqrt{f_4^4 u^4 (\partial_{\sigma}x_1)^2 + f_4^4 (\partial_{\sigma} u)^2 +f_4^2 \, u^2 \left[ \rho_1^2 (\partial_{\sigma} x)^2 + \rho_2^2 (\partial_{\sigma} y)^2 \right]}.
\eeq
This action depends only on the derivative of $x_1(\s)$, hence the system has a conserved charge, $C$, obeying $\partial_{\s} C = 0$,
\beq
 \label{eq:C}
C \equiv -\frac{e^{\phi} f_4^4 \, u^4 \partial_{\sigma} x_1}{\sqrt{f_4^4 u^4 (\partial_{\sigma}x_1)^2 + f_4^4 (\partial_\sigma u)^2  + f_4^2 \,  u^2 \left[ \rho_1^2 (\partial_\sigma x)^2 + \rho_2^2 (\partial_\sigma y)^2 \right]}}.
\eeq
The following rescaling of the coordinates is an isometry of the bulk metric:
\beq
\label{eq:rescaling}
u \to \mu^{-1} \, u, \qquad (t,x_1,x_2) \to \mu \, (t, x_1, x_2), \qquad (x,y) \to (x,y),
\eeq
with $\mu$ a real, positive constant. This rescaling isometry is dual to the field theory dilatation symmetry. Letting $\sigma$ transform with arbitrary weight $\alpha$ under this rescaling, $\sigma \to \mu^{\alpha} \sigma$, the action transforms only by an overall constant, $S \to \mu^{-1} S$, hence this rescaling will be a symmetry of the string equations of motion. Notice that $C \to \mu^{-2} C$ under this rescaling.

The straight timelike Wilson line and the rectangular Wilson loop perpendicular to the interface are both localized in the $(x_1,x_2)$ plane, and the corresponding strings both preserve the same subgroup of the isometries, while the rectangular Wilson loop parallel to the interface is extended in one direction in the $(x_1,x_2)$ plane, and so the corresponding string preserves a different subgroup of the isometries. To simplify the string action and equations of motion any further we must treat these two cases separately.

For the timelike line and the perpendicular loop, depicted in fig.~\ref{fig:configs} (a), we can choose $x_1$ to be constant, $\partial_{\sigma} x_1=0$. In that case the conserved charge $C=0$ trivially. To write the equations of motion in a simple form, let us define
\beq
\label{eq:Fdef}
F \equiv \sqrt{(\p_\sigma u)^2 + u^2 \left[ \rho_1^2 (\p_\sigma x)^2 + \rho_2^2 (\p_\sigma y)^2 \right] / f_4^2}.
\eeq
The equations of motion arising from the action in eq.~\eqref{eq:action} then take the form
\begin{subequations}
\beq
\label{eq:eomuperp}
\p_\sigma \left( \frac{e^{\phi} f_4^2}{F} \p_\sigma u \right) - \frac{e^\phi \,u}{F} \left[ \rho_1^2 (\partial_\sigma x)^2 + \rho_2^2 (\partial_\sigma y)^2 \right] =0,
\eeq
\beq
\label{eq:eomxperp}
\p_\sigma \left[ \frac{u^2 \rho_1^2 e^\phi}{F} \p_\sigma x \right] - \p_x (e^\phi f_4^2 F)=0,
\eeq
\beq
\label{eq:eomyperp}
\p_\sigma \left[ \frac{u^2 \rho_2^2 e^\phi}{F} \p_\sigma y  \right] - \p_y (e^\phi f_4^2 F)=0.
\eeq
\end{subequations}

We will first solve $y$'s equation of motion, eq.~\eqref{eq:eomyperp}, although not in complete generality. We will restrict to a special subset of solutions that are particularly simple: we consider constant $y$, meaning $\partial_{\sigma} y=0$. Eq.~\eqref{eq:eomyperp} then reduces to $\p_y (e^\phi f_4^2 F)=0$. For non-SUSY Janus, this condition is satisfied for any constant $y$. For SUSY Janus this condition is only satisfied for $y=0$ or $y=\pi/2$, meaning the string sits at a point where one or the other $S^2$ collapses to zero size.

Strings sitting at different values of $y$ translate into heavy test charges that couple to different subsets of the adjoint scalars, \textit{i.e.}\ different values of the $\theta^I$ in eq.~\eqref{defwilson}. In non-SUSY Janus, a choice of $y$ breaks the $SO(6)$ isometry but does not change the solutions for the remaining fields, $u(\sigma)$ and $x(\sigma)$, or the on-shell action. The corresponding field theory statements are that a choice of the $\theta^I$ breaks $SO(6)$ but does not otherwise affect the Wilson loop expectation value. In SUSY Janus, where the $SO(6)$ is broken to $SO(3) \times SO(3)$, each of our $y$ solutions, $y=0$ or $y=\pi/2$, breaks one $SO(3)$ but preserves the other. Here the choice of $y$ will affect the solutions for $u(\sigma)$ and $x(\sigma)$, since the metric and dilaton factors appearing in their equations of motion depend on the value of $y$. The on-shell action will also depend on the value of $y$. In the field theory, we divide the six adjoint scalars into two subsets of three, those that appear in the interface-localized term in eq.~\eqref{susyint} and those that do not. The choice of $y$ corresponds to a choice of the $\theta^I$ such that the test charge couples to one subset but not the other, and hence breaks one $SO(3)$ but preserves the other. Clearly the associated Wilson loop expectation value will depend on that choice.

What remains is to solve the equations of motion for $u(\sigma)$ and $x(\sigma)$. As a gauge choice we take $x(\sigma)=\sigma$, in which case a straightforward exercise shows that eqs.~\eqref{eq:eomuperp} and \eqref{eq:eomxperp} are equivalent. We thus obtain a single second-order, non-linear, ordinary differential equation, eq.~\eqref{eq:eomuperp}, for a single function, $u(\sigma) = u(x)$.

We solve eq.~\eqref{eq:eomuperp} numerically as follows. Strings describing a perpendicular loop will have a ``turn-around point'' at some value of $x$, which we call $\overline{x}$, where $\partial_x u(\overline{x})=0$. The existence of a turn-around point is intuitively obvious from fig.~\ref{fig:configs} (a): the string will emerge from the boundary at the point $x_3^{\textrm{R}}$ and extend in the $x_3$ direction as it dips into the bulk, eventually turning around and rising back up to intersect the boundary at the point $x_3^{\textrm{L}}$. A solution of the second-order eq.~\eqref{eq:eomuperp} will be completely specified by two boundary conditions, which we take to be the values of $\overline{x}$ and $u(\overline{x})$. Thanks to the rescaling symmetry of eq.~\eqref{eq:rescaling}, under which $u$ rescales but $x$ does not, we can obtain all physically inequivalent solutions by fixing $u(\overline{x})$ and scaning through values of $\overline{x}$. The solutions are thus distinguished only by a single number, $\overline{x}$, which is consistent with the parameter counting in the field theory: as explained in the introduction, for the perpendicular loop the potential $V$ depends only on the single dimensionless ratio $L_{\textrm{av}}/L$. For a given $\overline{x}$, we numerically integrate eq.~\eqref{eq:eomuperp} up to some values of $u$ and $x$ near the boundary, and extract the $x_3$ values of the endpoints (in units of the asymptotic $AdS_5 \times S^5$ radius) using either eq.~\eqref{uxcoords} or eqs.~\eqref{eq:nonsusyjanuszandxconversion} and \eqref{eq:susyjanuszandxconversion}. Those $x_3$ values then give us $L_{\textrm{av}}/L$.

For the straight Wilson line, we look for solutions with the turn-around point at the Poincar\'e horizon, \textit{i.e.}\ $\partial_x u(\overline{x})=0$ occurs at $u(\overline{x})=0$. These describe ``straight strings'' that extend from a point at the boundary to the Poincar\'e horizon. Intuitively, such strings represent perpendicular Wilson loops with one of the test charges sent to the ``point at infinity,'' leaving behind a single test charge. In practice, we ``shoot from the boundary": we pick a value of $u(x)$ near the boundary, which fixes the $x_3$ value of the endpoint at the boundary, pick a value of $\partial_x u(x)$ at that point, integrate eq.~\eqref{eq:eomuperp} up to $x=\overline{x}$ such that $\partial_x u(\overline{x})=0$, and then check whether $u(\overline{x})=0$.

Now let us consider the parallel loop, depicted in fig.~\ref{fig:configs} (b). Using $x_1(\sigma)$'s equation of motion, $\partial_{\s} C = 0$, and making the gauge choice $x_1(\s)= \s$, in which case $C$ is nonzero, we can write the remaining equations of motion as
\begin{subequations}
\beq
\label{eq:eomu}
\partial^2_\sigma u -3\frac{(\partial_\sigma u)^2}{u} - \frac{e^{2 \phi} f_4^4 u^7}{C^2} - u^3  = 0,
\eeq
\beq
\label{eq:eomx}
\partial_\sigma \left(\frac{\rho_1^2}{f_4^2} \frac{\partial_\sigma x}{u^2} \right) - \frac{u^4}{2 C^2} \partial_x \left(e^{2\phi} f_4^4 \right) -\frac{1}{2u^2} \bigg[ \left( \p_x \frac{\rho_1^2}{f_4^2} \right) (\p_\sigma x)^2 + \left( \p_x \frac{\rho_2^2}{f_4^2} \right) (\p_\sigma y)^2 \bigg] = 0,
\eeq
\beq
\label{eq:eomy}
\partial_\sigma \left(\frac{\rho_2^2}{f_4^2} \frac{\partial_\sigma y}{u^2} \right) - \frac{u^4}{2 C^2} \partial_y \left(e^{2\phi} f_4^4 \right) -\frac{1}{2u^2} \bigg[ \left( \p_y \frac{\rho_1^2}{f_4^2} \right) (\p_\sigma x)^2 + \left( \p_y \frac{\rho_2^2}{f_4^2} \right) (\p_\sigma y)^2 \bigg] = 0.
\eeq
\end{subequations}

Again we will only consider $y$ solutions with $\partial_{\sigma} y=0$, in which case in eq.~\eqref{eq:eomy} two terms remain. We will only consider solutions for which each of these terms vanishes independently,
\beq
\label{eq:yconds}
\partial_y \left(e^{2\phi} f_4^4 \right)=0, \qquad \partial_y \left( \frac{\rho_1^2}{f_4^2} \right) = 0.
\eeq
For non-SUSY Janus, the conditions in eq.~\eqref{eq:yconds} are satisfied for any constant value of $y$. For SUSY Janus, the conditions in eq.~\eqref{eq:yconds} are only satisfied for $y=0$ or $y=\pi/2$. All of our comments above about constant $y$ solutions apply to these solutions as well.

We solve eqs.~\eqref{eq:eomu} and \eqref{eq:eomx} for $u(\sigma)=u(x_1)$ and $x(\sigma)=x(x_1)$ numerically as follows. Strings describing a parallel loop will have a turn-around point at some $x_1$, which we can take to be $x_1=0$. To guarantee that both ends of the string have the same position in $x_3$, we impose $\partial_{x_1} u(0)=0$ and $\partial_{x_1} x(0)=0$. The existence of a turn-around point is obvious from fig.~\ref{fig:configs} (b.): the string will emerge from the boundary and extend in the $x_1$ direction as it dips into the bulk before turning around and rising up to intersect the boundary again. Notice that with our gauge choice $x_1 \in [-L/2,+L/2]$. Solutions of the second-order eqs.~\eqref{eq:eomu} and \eqref{eq:eomx} are completely specified by four boundary conditions: $\partial_{x_1} u(0)=0$ and $\partial_{x_1} x(0)=0$ and the values of $u(0)$ and $x(0)$. Thanks to the rescaling symmetry of eq.~\eqref{eq:rescaling} we can obtain all physically inequivalent solutions by fixing $u(0)$ and scanning through values of $x(0)$. The solutions are thus distinguished only by a single number, $x(0)$, which is consistent with the parameter counting in the field theory: as explained in the introduction, for the parallel loop the potential $V$ depends only on the single dimensionless ratio $L_3/L$. Notice that eqs.~\eqref{eq:eomu} and \eqref{eq:eomx} depend on the value of $C$. Once we pick a gauge, solve for $y$, and pick our values of $u(0)$ and $x(0)$, the value of $C$ is fixed. For a given $x(0)$, we numerically integrate eqs.~\eqref{eq:eomu} and \eqref{eq:eomx} up to values of $u$ and $x$ near the boundary, and then extract the $x_1$ and $x_3$ values of the endpoints (in units of the asymptotic $AdS_5 \times S^5$ radius) using either eq.~\eqref{uxcoords} or eqs.~\eqref{eq:nonsusyjanuszandxconversion} and \eqref{eq:susyjanuszandxconversion}. These $x_1$ and $x_3$ values then give us $L_3/L$.

\section{Holographic Renormalization}
\label{holorg}

In this appendix, we show how to remove divergences of the on-shell string action. Three methods exist to accomplish that. The first method is holographic renormalization, in which we add counterterms to the string action to cancel divergences~\cite{Henningson:1998gx,Balasubramanian:1999re,deHaro:2000xn,Skenderis:2002wp}. The second method, proposed in ref.~\cite{Drukker:1999zq}, is to perform a Legendre transform of the string action. The third method is to subtract from the on-shell action the (divergent) action of a straight string extending from the boundary to the Poincar\'e horizon~\cite{Rey:1998ik,Maldacena:1998im}. In field theory terms, that means computing, not the renormalized expectation value of a Wilson loop, but the difference between the expectation values of a Wilson loop and straight timelike Wilson line(s). Although these three methods are distinct in principle, for strings in $AdS_5 \times S^5$ they all produce the same result: they all cancel the divergence of the on-shell action, and nothing more. In particular they make no finite contribution to the on-shell action. We will consider only holographic renormalization, although at the end of this appendix we compare to the other two methods.

In principle we could work with the coordinates $u$ and $x$, however, holographic renormalization is dramatically simpler in Fefferman-Graham coordinates. Any asymptotically $AdS_5$ metric can, near the boundary, be written in Fefferman-Graham form,
\beq
ds^2 = \frac{R^2}{z^2} \left( dz^2 + g_{\mu\nu}(z,x^{\rho}) \, dx^{\mu} dx^{\nu} \right),
\eeq
where $z$ is the radial coordinate and the boundary is at $z=0$. The coordinate $z$ is related to the radial coordinate $r$ of the Poincar\'e slicing in eq.~\eqref{poincareslicing} by $z = R^2/r$. Expanding the metric near the boundary, \textit{i.e.}\ in powers of $z$, the leading term, $g_{\mu\nu}(z=0,x^{\rho})$, corresponds to the metric of the spacetime in which the field theory ``lives''. We use Euclidean-signature Poincar\'e slicing, so $g_{\mu\nu}(z=0,x^{\rho})=\delta_{\mu\nu}$. The metric is the source for the the stress-energy tensor. In most cases, holographic renormalization reveals that the coefficient of the order $z^4$ term is proportional to the expectation value of the stress-energy tensor~\cite{Henningson:1998gx,Balasubramanian:1999re,deHaro:2000xn,Skenderis:2002wp}. Similarly, the leading term in the near-boundary expansion of the dilaton corresponds to the source for the Lagrangian, $1/g^2$, while the coefficient of the order $z^4$ term corresponds to the expectation value of the Lagrangian~\cite{Skenderis:2002wp}.

Crucially, the above statements assume that the $AdS_5$ solution arises in a consistent Kaluza-Klein truncation of a ten-dimensional solution. As emphasized in ref.~\cite{Skenderis:2006uy}, extracting field theory expectation values directly from a ten-dimensional solution is often subtle, and may require more than just the lowest Kaluza-Klein modes retained in a consistent truncation. In particular, given a ten-dimensional solution for the metric and dilaton, the field theory expectation values may not be given simply by the coefficients of their order $z^4$ terms in a near-boundary expansion~\cite{Skenderis:2006uy}.

The holographic renormalization of non-SUSY Janus was performed in ref.~\cite{Papadimitriou:2004rz}. We will not attempt the holographic renormalization of SUSY Janus here. Instead, in section~\ref{holorgjanus} we will just present the change of coordinates from those of the Janus metrics in eq.~\eqref{eq:generalmetric} to Fefferman-Graham coordinates. Along the way, we will make some comments about holographic renormalization of the expectation values of the stress-energy tensor and Lagrangian for these cases. We then perform the holographic renormalization of the string action in section~\ref{holorgstring}.

\subsection{On Holographic Renormalization in Janus Spacetimes}
\label{holorgjanus}
In going from $AdS_5$ to the Janus spacetimes, translations in $x_3$ are broken, hence unlike $AdS_5$ the near-boundary Fefferman-Graham forms of the Janus metrics can depend on $x_3$. The rescaling isometry of eq.~\eqref{eq:rescaling} remains unbroken in Janus space times, however, so the Fefferman-Graham form of the Janus metric must depend on $x_3$ only in the combination $z/x_3$. We can thus write the Fefferman-Graham form of the Janus metrics as
\beq
\label{FGmetricAppendix}
ds^2 = \frac{R^2}{z^2} \left( d z^2 + g_{11}(z/x_3) (- dt^2 + dx_1^2 + dx_2^2) + g_{33}(z/x_3) dx_3^2 \right).
\eeq
In the coordinates of the metric in eq.~\eqref{eq:generalmetric}, all the string solutions we study in section \ref{wilsonholo} have constant $y = 0,\pi/2$. In our conversion to Fefferman-Graham coordinates, we will thus restrict to the cases $y=0,\pi/2$ in which case we can write the metric of eq.~\eqref{eq:generalmetric} in the form
\beq
\label{f4rhometric}
ds^2 = f_4(x)^2 \left( \frac{d u^2}{u^2} + u^2(- dt^2 + dx_1^2 + dx_2^2) \right) + \rho^2(x) dx^2.
\eeq
The change of coordinates
\begin{subequations}
\beq
\label{FGcoords}
z \equiv \frac{k_1(x)}{u}, \qquad \qquad x_3 \equiv \frac{k_2(x)}{u},
\eeq
\beq
\label{FGtrans}
k_1(x) \equiv \exp \left[ \mp \int dx \frac{\rho(x)}{f_4(x)} \sqrt{\frac{f_4(x)^2}{R^2} - 1}  \right], \quad k_2(x) \equiv \exp \left[ \pm \int dx \frac{\rho(x)}{f_4(x)}\frac{1}{ \sqrt{\frac{f_4(x)^2}{R^2} - 1}} \right],
\eeq
\end{subequations}
puts the metric in the Fefferman-Graham form of eq.~\eqref{FGmetricAppendix}, with
\beq
\label{eq:g11g33def}
g_{11}(z/x_3) = \frac{f_4(x)^2}{R^2} k_1(x)^2, \qquad \qquad g_{33}(z/x_3) = \left(\frac{f_4(x)^2}{R^2} - 1\right) \frac{k_1(x)^2}{k_2(x)^2}.
\eeq
To recover a flat boundary metric, we impose $g_{11}(0)=g_{33}(0) = 1$. What remains is to express $x$ in terms of $z/x_3$.

For the non-SUSY Janus metric in eq.~\eqref{eq:nonsusyjanusmetric}, near the boundaries $x \rightarrow \pm x_0$ the expansions of the $z$ and $x_3$ in eq.~\eqref{FGcoords} are
\begin{subequations}
\label{eq:nonsusyjanuszandxconversion}
\bea
\label{eq:nonsusyjanuszexpansion}
z &=& \frac{1}{u} \left(\frac{1}{\sqrt{h(x)}} + \sqrt{2}(\gamma - 1) \gamma^{\frac{3}{4}} (x_0 \mp x)^{\frac{9}{2}} + \mathcal{O}\left(\left(x_0\mp x\right)^{\frac{11}{2}}\right) \right), \\
\label{eq:nonsusyjanusxexpansion}
x_3 &=& \pm \frac{1}{u} \left(1 - \frac{1}{\sqrt{\gamma}} (x_0 \mp x) + \frac{4}{15} (\gamma-1) (x_0 \mp x)^6 + \mathcal{O}\left(\left(x_0\mp x\right)^7\right) \right).
\eea
\end{subequations}
Inverting these, we find $u$ and $x$ in terms of $z/x_3$,
\begin{subequations}
\label{eq:nonsusyjanusuxexp}
\bea
u &=& \pm \frac{1}{x_3} \left[1 - \frac{1}{2} \left( \frac{z}{x_3} \right)^2 + \frac{3}{8} \left( \frac{z}{x_3} \right)^4 - \frac{5}{16} \left( \frac{z}{x_3} \right)^6 + \frac{35}{128} \left( \frac{z}{x_3} \right)^8 \right. \nonumber\\
& & \left. - \frac{315 + 16 \gamma^3 - 16 \gamma^4}{1280} \left( \frac{z}{x_3} \right)^{10} + \mathcal{O}\left((z/x_3)^{12}\right)
\right], \\
x &=& \pm \left[ x_0 - \frac{\sqrt{\gamma}}{2} \left( \frac{z}{x_3} \right)^2 + \frac{3\sqrt{\gamma}}{8} \left( \frac{z}{x_3} \right)^4 - \frac{5\sqrt{\gamma}}{16} \left( \frac{z}{x_3} \right)^6 + \frac{35\sqrt{\gamma}}{128} \left( \frac{z}{x_3} \right)^8 \right.
\nonumber\\ & & \left.
+ \frac{\sqrt{\gamma}(16 \gamma^4 -16 \gamma^3 - 315)}{1280} \left( \frac{z}{x_3} \right)^{10}  + \mathcal{O}\left((z/x_3)^{12}\right) \right].
\eea
\end{subequations}
and hence the expansions of the metric factors in eq.~\eqref{eq:g11g33def} are
\begin{subequations}
\label{eq:nonsusyjanusg11g33expansion}
\bea
g_{11}(z/x_3) &=& 1 + \frac{(\gamma - 1) \gamma^3}{8} \left(\frac{z}{x_3}\right)^8 + \mathcal{O}\left((z/x_3)^{10}\right), \\ g_{33}(z/x_3) &=& 1 + \frac{(\gamma - 1) \gamma^3}{8} \left(\frac{z}{x_3}\right)^8 + \mathcal{O}\left((z/x_3)^{10}\right).
\eea
\end{subequations}
The Fefferman-Graham expansion of the non-SUSY Janus dilaton in eq.~\eqref{eq:nonsusyjanusdilaton} is
\beq
\label{eq:nonsusyjanusdilatonexpansion}
\phi(z/x_3) = \phi_{\pm} \mp \sqrt{\frac{3}{2}} \frac{\gamma^{\frac{3}{2}}\sqrt{1 - \gamma}}{2} \left(\frac{z}{x_3}\right)^4 + \mathcal{O}\left((z/x_3)^6\right).
\eeq

In the field theory, $SO(3,2)$ conformal symmetry forbids the field theory stress-energy tensor from acquiring an expectation value, but allows a scalar field of conformal dimension $\Delta$ to acquire an expectation value proportional to $1/|x_3|^{\Delta}$~\cite{1995NuPhB.455..522M}. Notice that the expansions of $g_{11}(z/x_3)$ and $g_{33}(z/x_3)$ in eq.~\eqref{eq:nonsusyjanusg11g33expansion} have no $z^4$ term, indicating that the one-point function of the stress-energy tensor is indeed zero~\cite{Papadimitriou:2004rz,Clark:2004sb}. In the expansion of the dilaton in eq.~\eqref{eq:nonsusyjanusdilatonexpansion}, the $z^4$ term has a nonzero coefficient, indicating that the Lagrangian acquires a nonzero expectation value that goes as $1/|x_3|^4$~\cite{Papadimitriou:2004rz,Clark:2004sb}. Such a straightforward identification of the coefficients of the order $z^4$ terms with field theory one-point functions is possible because the non-SUSY Janus metric and dilaton are solutions of a consistent trunctation of type IIB supergravity on $S^5$~\cite{Bak:2003jk,Freedman:2003ax}.

For the SUSY Janus metric in eq.~\eqref{eq:susyjanusmetric}, near the boundaries $x \rightarrow \pm \infty$ the expansions of the $z$ and $x_3$ in eq.~\eqref{FGcoords} are
\begin{subequations}
\label{eq:susyjanuszandxconversion}
\bea
z &=& \frac{1}{u} \left[ \frac{1}{\cosh(x \mp \frac{1}{2} \ln \cosh \dph)} \pm \sqrt{\cosh \dph} \left( \frac{\cos 2y \sinh \dph}{2} e^{\mp 3x} \right. \right. \nonumber \\& & \left . \left .
- \frac{8 \cos 2y \sinh 2 \dph \pm (\cos 4y -3) \sin^2 \dph}{16} e^{\mp 5x} + \mathcal{O}\left(e^{\mp 7x}\right) \right) \right], \\
x_3 &= &\frac{1}{u} \left[ \tanh(x \mp \frac{1}{2} \ln \cosh \dph) + \left( \frac{\cos 2y \sinh 2 \dph}{4} e^{\mp 4 x} \right. \right . \\ &&
\left. \left . - \frac{\cosh \dph \sinh \dph [24 \cos 2y \cosh \dph \pm (13 + 5 \cos 4y) \sinh \dph]}{24} e^{\mp 6 x} + \mathcal{O}\left(e^{\mp 8x}\right) \right) \right].  \nonumber
\eea
\end{subequations}
Inverting these, we find $u$ and $x$ in terms of $z/x_3$,
\begin{subequations}
\label{eq:susyjanusuxexp}
\bea
u &=& \pm \frac{1}{x_3} \left[1 - \frac{1}{2} \left( \frac{z}{x_3} \right)^2 + \frac{3(4 \pm \cos 2y \tanh \dph)}{32} \left( \frac{z}{x_3} \right)^4 \right. \\
& & -\left. \frac{919+1001 \cosh 2 \dph + 98 \cos 4y \sinh^2 \dph \pm 672 \cos 2y \sinh 2 \dph}{6144 \cosh^2 \dph} \left( \frac{z}{x_3} \right)^6
+ \mathcal{O}\left((z/x_3)^8\right)\right], \nonumber  \\
x &=& \pm \frac{1}{2} \ln \left(4 \cosh \dph \left(\frac{x_3}{z}\right)^2 \right) \pm \frac{4\pm\cos 2y \tanh \dph}{16} \left( \frac{z}{x_3} \right)^2
\\& &
 \mp \frac{96 + \tanh \dph(7 \cos 4y \tanh \dph \pm 112 \cos 2y - \tanh \dph)}{1024} \left( \frac{z}{x_3} \right)^4 + \mathcal{O}\left((z/x_3)^6\right)
,  \nonumber
\eea
\end{subequations}
and hence the expansions of the metric factors in eq.~\eqref{eq:g11g33def} are
\begin{subequations}
\label{eq:susyjanusg11g33expansions}
\bea
g_{11}(z/x_3) & = &1 \pm \frac{3 \cos 2y \tanh \dph}{8} \left( \frac{z}{x_3} \right)^2 \\ & & \mp \frac{\tanh \dph [240 \cos 2y \pm (5+13 \cos 4y)\tanh \dph]}{512} \left( \frac{z}{x_3} \right)^4 + \mathcal{O}\left((z/x_3)^6\right), \nonumber  \\
g_{33}(z/x_3) & = &1 \pm \frac{3 \cos 2y \tanh \dph}{8} \left( \frac{z}{x_3} \right)^2  \\ & & \mp \frac{\tanh \dph [144 \cos 2y \pm (5+13 \cos 4y)\tanh \dph]}{512} \left( \frac{z}{x_3} \right)^4 + \mathcal{O}\left((z/x_3)^6\right). \nonumber
\eea
\end{subequations}
The expansion of the SUSY Janus dilaton in eq.~\eqref{eq:susyjanusdilaton} is
\beq
\label{eq:susyjanusdilatonexpansion}
\phi = \phi_{\pm} \mp \frac{3 \tanh \dph}{16} \left( \frac{z}{x_3} \right)^4 \pm \frac{\tanh \dph (20 \pm 11 \cos 2y  \tanh \dph)}{64} \left( \frac{z}{x_3} \right)^6 + \mathcal{O}\left((z/x_3)^8\right).
\eeq
In eqs.~\eqref{eq:susyjanuszandxconversion}, \eqref{eq:nonsusyjanusuxexp}, \eqref{eq:susyjanusg11g33expansions}, and \eqref{eq:susyjanusdilatonexpansion}, $y$ only takes the values $y=0,\pi/2$ to which we restricted.

The expansions of $g_{11}(z/x_3)$ and $g_{33}(z/x_3)$ in eq.~\eqref{eq:susyjanusg11g33expansions} include nonzero coefficients for the $z^4$ terms. Na\"ively, that suggests the field theory stress-energy tensor has a nonzero expectation value. Notice, however, that the coefficients of the order $z^4$ terms depend explicitly on $y$: they are different when $y=0$ or $y=\pi/2$. In the field theory that suggests the expectation value of the stress-energy tensor has some R-charge, which is clearly unphysical: the stress-energy tensor is invariant under all global symmetries. Clearly, we cannot simply identify the coefficient of the order $z^4$ term in the metric with the stress-energy tensor one-point function. Indeed, the SUSY Janus metric is genuinely ten-dimensional, \textit{i.e.}\ is not obtained as the consistent truncation of a ten-dimensional solution, so extracting the one-point function of the stress-energy tensor may take more work, as emphasized in ref.~\cite{Skenderis:2006uy}. Similar statements apply for the expansion of the dilaton in eq.~\eqref{eq:susyjanusdilatonexpansion}.

\subsection{Holographic Renormalization of the String Action}
\label{holorgstring}

The near-boundary, small $z/x_3$ expansions of the functions $g_{11}(z/x_3)$ and $g_{33}(z/x_3)$ in eqs.~\eqref{eq:nonsusyjanusg11g33expansion} and~\eqref{eq:susyjanusg11g33expansions} are of the form
\beq
\label{eq:g11g33exp}
g_{11}(z/x_3) = 1 + \left( \frac{z}{x_3} \right)^n \sum_{i=0}^\infty \mathcal{A}^{(\pm)}_i \left( \frac{z}{x_3} \right)^i, \quad g_{33}(z/x_3) = 1 + \left( \frac{z}{x_3} \right)^n \sum_{i=0}^\infty \mathcal{B}^{(\pm)}_i \left( \frac{z}{x_3} \right)^i,
\eeq
with coefficients $\mathcal{A}^{(\pm)}_i $ and $\mathcal{B}^{(\pm)}_i$ that depend on the size of the jump in the axio-dilaton, and that go to zero if the jump in the axio-dilaton goes to zero. The superscripts $\pm$ on $\mathcal{A}^{(\pm)}_i $ and $\mathcal{B}^{(\pm)}_i$ indicate that in general these coefficients take different values in the two distinct asymptotic $AdS_5 \times S^5$ regions of a Janus spacetime, in which the dilaton approaches the values $\phi_+$ and $\phi_-$. The first non-trivial power $n$ of $z/x_3$ in the expansions is $n=8$ for non-SUSY Janus and $n=2$ for SUSY Janus. The near-boundary expansion of the dilaton takes the form (see eqs.~\eqref{eq:nonsusyjanusdilatonexpansion} and~\eqref{eq:susyjanusdilatonexpansion})
\beq
\label{eq:dilexp}
\phi\left(z/x_3\right) = \phi_{\pm} + \left( \frac{z}{x_3} \right)^4 \sum_{i=0}^\infty \mathcal{C}^{(\pm)}_i \left( \frac{z}{x_3} \right)^i,
\eeq
where again the coefficients $\mathcal{C}^{(\pm)}_i$ depend on the size of the jump in the axio-dilaton, and go to zero if the jump in the axio-dilaton goes to zero. Notice that the first non-trivial power of $z/x_3$ in eq.~\eqref{eq:dilexp} is four for both non-SUSY and SUSY Janus.

Let us again simplify the string action of eq.~\eqref{eq:NGaction} using isometries, but now using Fefferman-Graham coordinates. We use the symmetries of the $(x_1,x_2)$ plane to set $x_2=0$. The most general Ansatz we can then write is $z(\sigma)$, $x_1(\sigma)$, $x_3(\sigma)$. In static gauge, $\tau = t$, the string action then reduces to
\beq
\label{eq:actionFG}
S=\frac{T}{2\pi\alpha'} \int d\sigma \, e^\phi \, \frac{R^2}{z^2} \sqrt{g_{11}} \sqrt{(\partial_{\sigma} z)^2 + g_{11} (\partial_{\sigma} x_1)^2 + g_{33} (\partial_{\sigma} x_3)^2}.
\eeq
Notice that if the jump in the axio-dilaton is zero, in which case $g_{11}(z/x_3)=1$ and $g_{33}(z/x_3)=1$, then our string action reduces to that of refs.~\cite{Rey:1998ik,Maldacena:1998im}, for a string in $AdS_5 \times S^5$. As a gauge choice we now take $z(\sigma)=\sigma$. The equations of motion for $x_1(\sigma)=x_1(z)$ and $x_3(\sigma)=x_3(z)$ are straightforward to obtain but unilluminating, so we will omit them. What we need for holographic renormalization are the small-$z$ asymptotic expansions of the on-shell $x_1(z)$ and $x_3(z)$,
\beq
\label{asymsol}
x_1(z) = x_1^{(0)} + x_1^{(1)} z^3 + ... , \qquad x_3(z) = x_3^{(0)} + x_3^{(1)} z^3 + ...,
\eeq
where $x_1^{(0)}$, $x_1^{(1)}$, $x_3^{(0)}$, and $x_3^{(1)}$ are independent of $z$. In each case, the $\ldots$ represent terms of higher order in $z$. In $AdS_5$, for both $x_1(z)$ and $x_3(z)$ the first sub-leading term is order $z^7$. In non-SUSY Janus, for $x_1(z)$ the first sub-leading term is again order $z^7$, but for $x_3(z)$ the first sub-leading term is order $z^6$, with a coefficient proportional to the coefficient $\mathcal{C}_0^{\pm}$ from eq.~\eqref{eq:dilexp}. In SUSY Janus, for $x_1(z)$ the first sub-leading term is order $z^5$ and for $x_3(z)$ the first sub-leading term is order $z^4$, both with coefficients proportional to the coefficient $\mathcal{A}_0^{\pm}$ from eq.~\eqref{eq:g11g33exp}.

In holographic renormalization, we proceed as follows~\cite{Henningson:1998gx,Balasubramanian:1999re,deHaro:2000xn,Skenderis:2002wp}. First, in the action we introduce a cutoff on the $z$ integration to regulate any divergences: we integrate not to $z=0$ but to some small but finite $z=\varepsilon$. The result is a regulated on-shell action, $S_{\textrm{reg}}$. Next, we insert the solutions for $x_1(z)$ and $x_3(z)$ into the action, expand the integrand in powers of $z$, perform the integration in $z$, and then isolate all terms that diverge in the $\varepsilon \to 0$ limit. Finally, we add a counterterm action, $S_{CT}$, consisting of terms localized at $z=\varepsilon$, built from the induced metric on the $z=\varepsilon$ surface and designed to cancel all divergences in the $\varepsilon \to 0$ limit. The renormalized action is then
\beq
\label{eq:sren}
S_{\textrm{ren}} = \lim_{\varepsilon \to 0} \left ( S_{\textrm{reg}} + S_{CT} \right).
\eeq

Plugging eq.~\eqref{asymsol} into eq.~\eqref{eq:actionFG} and expanding in $z$, we find
\beq
\label{eq:regaction}
S_{\textrm{reg}} = \frac{T}{2\pi\alpha'} \int_{\varepsilon} dz \, \left [ e^{\phi_{\pm}} \frac{R^2}{z^2} + \ldots \right] = + \frac{T}{2\pi\alpha'} \, e^{\phi_{\pm}} \frac{R^2}{\varepsilon} + \ldots,
\eeq
where the $\ldots$ represents all non-divergent terms. After the $z$ integration (\textit{i.e.}\ in the second equality), that includes terms that remain finite or that vanish as $\varepsilon \to 0$. For both $AdS_5 \times S^5$ and non-SUSY Janus, among the terms that vanish as $\varepsilon \to 0$, the leading term is $\mathcal{O}\left(\varepsilon^3\right)$, while for SUSY Janus, the leading term is $\mathcal{O}\left(\varepsilon\right)$, with a coefficient proportional to the coefficient $\mathcal{A}_0^{\pm}$ from eq.~\eqref{eq:g11g33exp}.

Now we need counterterms. In our case, these are built from the induced metric on the $z = \varepsilon$ surface, which we denote $\gamma_{tt}$,
\beq
\gamma_{tt} \, dt^2 = + \frac{R^2}{\varepsilon^2} \, g_{11}(\varepsilon/x_3(\varepsilon)) \, dt^2 = +\frac{R^2}{\varepsilon^2} dt^2 + \mathcal{O}(\varepsilon^{n-2}),
\eeq
where $n$ was defined in eq.~\eqref{eq:g11g33exp}. The counterterm we need is then
\beq
\label{eq:ctaction}
S_{CT} = - \frac{R}{2\pi\alpha'} \, e^{\phi(\varepsilon/x_3)} \int dt \, \sqrt{\gamma_{tt}} = - \frac{T}{2\pi\alpha'} e^{\phi_{\pm}}\frac{R^2}{\varepsilon} + \ldots.
\eeq
We introduced an overall factor of $R$ to make $S_{CT}$ dimensionless, like $S_{\textrm{reg}}$. In our case $S_{CT}$ involves an integral over $dt$, but more generally the integral will be over the worldline of the endpoint of the string. In $AdS_5 \times S^5$, the $1/\varepsilon$ divergent term is the only contribution to $S_{CT}$. In the Janus spacetimes, additional terms appear, but these vanish as $\varepsilon \to 0$. These terms are represented by the $\ldots$ in eq.~\eqref{eq:ctaction}. For non-SUSY Janus, the first sub-leading term is $\mathcal{O}(\varepsilon^3)$, with coefficient proportional to the coefficient $\mathcal{C}_0^{\pm}$ from eq.~\eqref{eq:dilexp}. For SUSY Janus, the first sub-leading term is $\mathcal{O}(\varepsilon)$, with coefficient proportional to the coefficient $\mathcal{A}_0^{\pm}$ from eq.~\eqref{eq:g11g33exp}. With our counterterm, the $S_{\textrm{ren}}$ defined in eq.~\eqref{eq:sren} will be finite.

Crucially, notice that the divergence in $S_{\textrm{reg}}$ and the counterterm are not only independent of the solutions $x_1(z)$ and $x_3(z)$, but are also independent of the size of the jump in the axio-dilaton: the divergence or counterterm in each asymptotically $AdS_5 \times S^5$ region depends on the value of the dilaton $\phi_{\pm}$ in that region, but not on the difference $\phi_+ - \phi_-$. The holographic renormalization of the string action in Janus spacetimes is identical to that in $AdS_5 \times S^5$, up to terms that vanish as $\varepsilon \to 0$. \textit{A priori} that is intuitively obvious, if we recall that the $AdS_5$ radial direction corresponds to the field theory energy scale, with the boundary corresponding to the UV: deforming $\N=4$ SYM by a conformal interface should not affect any field theory physics in the extreme UV, including in particular UV divergences of Wilson loop expectation values. Translating that statement into the bulk, we expect the divergences of the on-shell string action, and hence its holographic renormalization, to be identical in the Janus and $AdS_5 \times S^5$ spacetimes.

Additionally, no finite counterterms are possible in our cases. Finite counterterms can produce ambiguities in the value of the on-shell action, since we cannot fix their coefficients by demanding cancellation of divergences. If we think of $x_1(z)$ and $x_3(z)$ as scalar fields in AdS, then in our cases the only candidates for finite counterterms that are allowed by covariance are of the form
\beq
e^{\phi(\varepsilon/x_3)} \int dt \, \sqrt{\gamma_{tt}} \, x_1(\varepsilon)^p \, x_3(\varepsilon)^q = T\, e^{\phi_{\pm}} \frac{R}{\varepsilon} (x_1^{(0)})^p (x_3^{(0)})^q + \mathcal{O}(\varepsilon^2),
\eeq
where $p$ and $q$ are non-negative integers, and never both zero. Clearly, in cases where $x_1^{(0)}$ and/or $x_3^{(0)}$ are nonzero, which includes all the cases we consider in subsections \ref{sec:straight} to \ref{sec:parallel}, such counterterms will introduce divergences beyond those in $S_{\textrm{reg}}$, and hence the coefficients of these counterterms must be set to zero. The upshot is that in our cases covariance forbids any finite counterterms.

Using the counterterm in eq.~\eqref{eq:ctaction} we obtain the renormalized string actions,
\begin{subequations}
\begin{align}
\label{area1}
S^{\perp}_{\textrm{ren}} &= \frac{T}{2\pi\alpha'} \, 3 R^2 \left( e^{\phi_-} x_3^{(0)} x_3^{(1)} \big |_{x_{(1)}} + e^{\phi_+}x_3^{(0)} x_3^{(1)} \big |_{x_{(2)}} \right),
\end{align}
\begin{align}
\label{area2}
S^{\parallel}_{\textrm{ren}} &= 2 \, \frac{T}{2\pi\alpha'} \, e^{\phi_{\pm}} \, 3 R^2 \, (x_1^{(0)} x_1^{(1)} + x_3^{(0)} x_3^{(1)}) \big |_{+L/2}.
\end{align}
\end{subequations}

Our procedure to calculate $S^{\perp}_{\textrm{ren}}$ and $S^{\parallel}_{\textrm{ren}}$ numerically is the following. We find string solutions as explained in appendix~\ref{stringeoms}, using $u(\sigma)$ and $x(\sigma)$, up to some finite cutoff near the asymptotic $AdS_5$ boundary. We then convert those solutions for $u(\sigma)$ and $x(\sigma)$ into solutions for $x_1(z)$ and $x_3(z)$ using eqs.~\eqref{eq:nonsusyjanuszandxconversion} and \eqref{eq:susyjanuszandxconversion}. In Fefferman-Graham coordinates, the cutoff is simply $z=\varepsilon$. We then fit the numerical solutions for $x_1(z)$ and $x_3(z)$ to the asymptotic solution given in eq.~\eqref{asymsol}. (In fact, in our numerics we work to higher order in $z$ than in eq.~\eqref{asymsol}, to improve the the quality of the fits.) From these fits we obtain the values of $x_1^{(0)}$, $x_1^{(1)}$, $x_3^{(0)}$ and $x_3^{(1)}$ for each endpoint of the string. We then extract from these constants the values of $L$, $L_{\textrm{av}}$, and $S^{\perp}_{\textrm{ren}}$ for the perpendicular string, and $L$, $L_3$, and $S^{\parallel}_{\textrm{ren}}$ for the parallel string.

For the straight string dual to the straight Wilson line, one subtlety arises: in this case, only one endpoint of the string reaches the boundary, while the other endpoint lies on the Poincar\'e horizon, $u=0$. We have numerically confirmed that the contribution from the Poincar\'e horizon vanishes. For these strings the renormalized action is thus the $S^{\perp}_{\textrm{ren}}$ in eq.~\eqref{area1}, with $x_{(1)}$ corresponding to the boundary endpoint and the contribution from $x_{(2)}$ set to zero.

Finally, let us compare holographic renormalization to the Legendre transform and to subtracting a straight string. A straightforward exercise shows that for both non-SUSY and SUSY Janus spacetimes, as in $AdS_5 \times S^5$, the Legendre transform has the same effect as holographic renormalization: each subtracts the divergence in eq.~\eqref{eq:regaction}, and nothing more. Subtracting a straight string has dramatically different consequences in $AdS_5 \times S^5$ and Janus spacetimes, however. In $AdS_5 \times S^5$, subtracting a straight string again simply subtracts the divergence in eq.~\eqref{eq:regaction}, and nothing more. In Janus spacetimes, however, subtracting the straight string additionally subtracts a finite term. In field theory language, holographic renormalization or the Legendre transform each correspond to subtracting the infinite self-energy of a test charge, while subtracting a straight string corresponds to subtracting not only the infinite self-energy of a test charge but also the finite interaction energy of the test charge with its image. In all of our holographic calculations of Wilson loop expectation values from strings in Janus spacetimes, we used holographic renormalization or equivalently the Legendre transform to obtain a finite on-shell string action.

\bibliography{wilsoninterfacev2}

\providecommand{\href}[2]{#2}\begingroup\raggedright\begin{thebibliography}{10}

\bibitem{Bak:2003jk}
D.~Bak, M.~Gutperle, and S.~Hirano, {\it {A Dilatonic Deformation of AdS(5) and
  its Field Theory Dual}},  {\em JHEP} {\bf 0305} (2003) 072,
  [\href{http://xxx.lanl.gov/abs/hep-th/0304129}{{\tt hep-th/0304129}}].

\bibitem{Clark:2004sb}
A.~Clark, D.~Freedman, A.~Karch, and M.~Schnabl, {\it {The Dual of Janus: an
  Interface CFT}},  {\em Phys.Rev.} {\bf D71} (2005) 066003,
  [\href{http://xxx.lanl.gov/abs/hep-th/0407073}{{\tt hep-th/0407073}}].

\bibitem{D'Hoker:2006uv}
E.~D'Hoker, J.~Estes, and M.~Gutperle, {\it {Interface Yang-Mills,
  Supersymmetry, and Janus}},  {\em Nucl.Phys.} {\bf B753} (2006) 16--41,
  [\href{http://xxx.lanl.gov/abs/hep-th/0603013}{{\tt hep-th/0603013}}].

\bibitem{Azeyanagi:2007qj}
T.~Azeyanagi, A.~Karch, T.~Takayanagi, and E.~Thompson, {\it {Holographic
  Calculation of Boundary Entropy}},  {\em JHEP} {\bf 0803} (2008) 054--054,
  [\href{http://xxx.lanl.gov/abs/0712.1850}{{\tt arXiv:0712.1850}}].

\bibitem{Gaiotto:2008sd}
D.~Gaiotto and E.~Witten, {\it {Janus Configurations, Chern-Simons Couplings,
  and the $\theta$-Angle in N=4 Super Yang-Mills Theory}},  {\em JHEP} {\bf
  1006} (2010) 097, [\href{http://xxx.lanl.gov/abs/0804.2907}{{\tt
  arXiv:0804.2907}}].

\bibitem{Karch:2000gx}
A.~Karch and L.~Randall, {\it {Open and Closed String Interpretation of SUSY
  CFT's on Branes with Boundaries}},  {\em JHEP} {\bf 0106} (2001) 063,
  [\href{http://xxx.lanl.gov/abs/hep-th/0105132}{{\tt hep-th/0105132}}].

\bibitem{DeWolfe:2001pq}
O.~DeWolfe, D.~Freedman, and H.~Ooguri, {\it {Holography and Defect Conformal
  Field Theories}},  {\em Phys.Rev.} {\bf D66} (2002) 025009,
  [\href{http://xxx.lanl.gov/abs/hep-th/0111135}{{\tt hep-th/0111135}}].

\bibitem{Erdmenger:2002ex}
J.~Erdmenger, Z.~Guralnik, and I.~Kirsch, {\it {Four-dimensional Superconformal
  Theories with Interacting Boundaries or Defects}},  {\em Phys.Rev.} {\bf D66}
  (2002) 025020, [\href{http://xxx.lanl.gov/abs/hep-th/0203020}{{\tt
  hep-th/0203020}}].

\bibitem{D'Hoker:2007xz}
E.~D'Hoker, J.~Estes, and M.~Gutperle, {\it {Exact Half-BPS Type IIB Interface
  Solutions. II. Flux Solutions and Multi-Janus}},  {\em JHEP} {\bf 0706}
  (2007) 022, [\href{http://xxx.lanl.gov/abs/0705.0024}{{\tt
  arXiv:0705.0024}}].

\bibitem{Kim:2008dj}
C.~Kim, E.~Koh, and K.-M. Lee, {\it {Janus and Multifaced Supersymmetric
  Theories}},  {\em JHEP} {\bf 0806} (2008) 040,
  [\href{http://xxx.lanl.gov/abs/0802.2143}{{\tt arXiv:0802.2143}}].

\bibitem{Kim:2009wv}
C.~Kim, E.~Koh, and K.-M. Lee, {\it {Janus and Multifaced Supersymmetric
  Theories II}},  {\em Phys.Rev.} {\bf D79} (2009) 126013,
  [\href{http://xxx.lanl.gov/abs/0901.0506}{{\tt arXiv:0901.0506}}].

\bibitem{Clark:2005te}
A.~Clark and A.~Karch, {\it {Super Janus}},  {\em JHEP} {\bf 0510} (2005) 094,
  [\href{http://xxx.lanl.gov/abs/hep-th/0506265}{{\tt hep-th/0506265}}].

\bibitem{D'Hoker:2006uu}
E.~D'Hoker, J.~Estes, and M.~Gutperle, {\it {Ten-dimensional Supersymmetric
  Janus Solutions}},  {\em Nucl.Phys.} {\bf B757} (2006) 79--116,
  [\href{http://xxx.lanl.gov/abs/hep-th/0603012}{{\tt hep-th/0603012}}].

\bibitem{Gomis:2006cu}
J.~Gomis and C.~Romelsberger, {\it {Bubbling Defect CFT's}},  {\em JHEP} {\bf
  0608} (2006) 050, [\href{http://xxx.lanl.gov/abs/hep-th/0604155}{{\tt
  hep-th/0604155}}].

\bibitem{D'Hoker:2007xy}
E.~D'Hoker, J.~Estes, and M.~Gutperle, {\it {Exact Half-BPS Type IIB Interface
  Solutions. I. Local Solution and Supersymmetric Janus}},  {\em JHEP} {\bf
  0706} (2007) 021, [\href{http://xxx.lanl.gov/abs/0705.0022}{{\tt
  arXiv:0705.0022}}].

\bibitem{Aharony:2011yc}
O.~Aharony, L.~Berdichevsky, M.~Berkooz, and I.~Shamir, {\it {Near-horizon
  Solutions for D3-branes Ending on 5-branes}},  {\em Phys.Rev.} {\bf D84}
  (2011) 126003, [\href{http://xxx.lanl.gov/abs/1106.1870}{{\tt
  arXiv:1106.1870}}].

\bibitem{Assel:2011xz}
B.~Assel, C.~Bachas, J.~Estes, and J.~Gomis, {\it {Holographic Duals of D=3 N=4
  Superconformal Field Theories}},  {\em JHEP} {\bf 1108} (2011) 087,
  [\href{http://xxx.lanl.gov/abs/1106.4253}{{\tt arXiv:1106.4253}}].

\bibitem{Suh:2011xc}
M.-W. Suh, {\it {Supersymmetric Janus Solutions in Five and Ten Dimensions}},
  {\em JHEP} {\bf 1109} (2011) 064,
  [\href{http://xxx.lanl.gov/abs/1107.2796}{{\tt arXiv:1107.2796}}].

\bibitem{Maldacena:1997re}
J.~M. Maldacena, {\it {The Large N limit of Superconformal Field Theories and
  Supergravity}},  {\em Adv.Theor.Math.Phys.} {\bf 2} (1998) 231--252,
  [\href{http://xxx.lanl.gov/abs/hep-th/9711200}{{\tt hep-th/9711200}}].

\bibitem{Witten:1998qj}
E.~Witten, {\it {Anti-de Sitter Space and Holography}},  {\em
  Adv.Theor.Math.Phys.} {\bf 2} (1998) 253--291,
  [\href{http://xxx.lanl.gov/abs/hep-th/9802150}{{\tt hep-th/9802150}}].

\bibitem{Gubser:1998bc}
S.~Gubser, I.~R. Klebanov, and A.~M. Polyakov, {\it {Gauge Theory Correlators
  from Noncritical String Theory}},  {\em Phys.Lett.} {\bf B428} (1998)
  105--114, [\href{http://xxx.lanl.gov/abs/hep-th/9802109}{{\tt
  hep-th/9802109}}].

\bibitem{Rey:1998ik}
S.-J. Rey and J.-T. Yee, {\it {Macroscopic Strings as Heavy Quarks in Large N
  Gauge Theory and Anti-de Sitter Supergravity}},  {\em Eur.Phys.J.} {\bf C22}
  (2001) 379--394, [\href{http://xxx.lanl.gov/abs/hep-th/9803001}{{\tt
  hep-th/9803001}}].

\bibitem{Maldacena:1998im}
J.~Maldacena, {\it {Wilson loops in Large N Field Theories}},  {\em
  Phys.Rev.Lett.} {\bf 80} (1998) 4859--4862,
  [\href{http://xxx.lanl.gov/abs/hep-th/9803002}{{\tt hep-th/9803002}}].

\bibitem{Drukker:1999zq}
N.~Drukker, D.~Gross, and H.~Ooguri, {\it {Wilson Loops and Minimal Surfaces}},
   {\em Phys.Rev.} {\bf D60} (1999) 125006,
  [\href{http://xxx.lanl.gov/abs/hep-th/9904191}{{\tt hep-th/9904191}}].

\bibitem{Erickson:2000af}
J.~Erickson, G.~Semenoff, and K.~Zarembo, {\it {Wilson Loops in N=4
  Supersymmetric Yang-Mills Theory}},  {\em Nucl.Phys.} {\bf B582} (2000)
  155--175, [\href{http://xxx.lanl.gov/abs/hep-th/0003055}{{\tt
  hep-th/0003055}}].

\bibitem{Erickson:1999qv}
J.~Erickson, G.~Semenoff, R.~Szabo, and K.~Zarembo, {\it {Static Potential in
  N=4 Supersymmetric Yang-Mills Theory}},  {\em Phys.Rev.} {\bf D61} (2000)
  105006, [\href{http://xxx.lanl.gov/abs/hep-th/9911088}{{\tt
  hep-th/9911088}}].

\bibitem{Drukker:2010jp}
N.~Drukker, D.~Gaiotto, and J.~Gomis, {\it {The Virtue of Defects in 4D Gauge
  Theories and 2D CFTs}},  {\em JHEP} {\bf 1106} (2011) 025,
  [\href{http://xxx.lanl.gov/abs/1003.1112}{{\tt arXiv:1003.1112}}].

\bibitem{Nagasaki:2011ue}
K.~Nagasaki, H.~Tanida, and S.~Yamaguchi, {\it {Holographic Interface-Particle
  Potential}},  {\em JHEP} {\bf 1201} (2012) 139,
  [\href{http://xxx.lanl.gov/abs/1109.1927}{{\tt arXiv:1109.1927}}].

\bibitem{Drukker:2000rr}
N.~Drukker and D.~Gross, {\it {An Exact Prediction of N=4 SUSYM Theory for
  String Theory}},  {\em J.Math.Phys.} {\bf 42} (2001) 2896--2914,
  [\href{http://xxx.lanl.gov/abs/hep-th/0010274}{{\tt hep-th/0010274}}].

\bibitem{Pestun:2007rz}
V.~Pestun, {\it {Localization of Gauge Theory on a Four-sphere and
  Supersymmetric Wilson Loops}},  {\em Commun.Math.Phys.} {\bf 313} (2012)
  71--129, [\href{http://xxx.lanl.gov/abs/0712.2824}{{\tt arXiv:0712.2824}}].

\bibitem{Berenstein:1998ij}
D.~Berenstein, R.~Corrado, W.~Fischler, and J.~Maldacena, {\it {The Operator
  Product Expansion for Wilson Loops and Surfaces in the Large N limit}},  {\em
  Phys.Rev.} {\bf D59} (1999) 105023,
  [\href{http://xxx.lanl.gov/abs/hep-th/9809188}{{\tt hep-th/9809188}}].

\bibitem{Hasan:2010xy}
M.~Hasan and C.~Kane, {\it {Topological Insulators}},  {\em Rev.Mod.Phys.} {\bf
  82} (2010) 3045, [\href{http://xxx.lanl.gov/abs/1002.3895}{{\tt
  arXiv:1002.3895}}].

\bibitem{Moore:2010rev}
J.~E. Moore, {\it {The Birth of Topological Insulators}},  {\em Nature} {\bf
  464} (Mar, 2010) 194.

\bibitem{2011RvMP...83.1057Q}
X.-L. {Qi} and S.-C. {Zhang}, {\it {Topological Insulators and
  Superconductors}},  {\em Rev.Mod.Phys.} {\bf 83} (Oct., 2011) 1057--1110,
  [\href{http://xxx.lanl.gov/abs/1008.2026}{{\tt arXiv:1008.2026}}].

\bibitem{2011ARCMP...2...55H}
M.~Z. {Hasan} and J.~E. {Moore}, {\it {Three-Dimensional Topological
  Insulators}},  {\em Ann.Rev.Cond.Matt.Phys.} {\bf 2} (Mar., 2011) 55--78,
  [\href{http://xxx.lanl.gov/abs/1011.5462}{{\tt arXiv:1011.5462}}].

\bibitem{doi:10.1142/S0217979292000037}
S.~C. Zhang, {\it {The Chern-Simons Landau-Ginzburg Theory of the Fractional
  Quantum Hall Effect}},  {\em Int.Jour.Mod.Phys.} {\bf {B06}} (1992), no.~01
  25--58.

\bibitem{2008PhRvB..78s5424Q}
X.-L. {Qi}, T.~L. {Hughes}, and S.-C. {Zhang}, {\it {Topological Field Theory
  of Time-reversal Invariant Insulators}},  {\em Phys.Rev.} {\bf B78} (Nov.,
  2008) 195424, [\href{http://xxx.lanl.gov/abs/0802.3537}{{\tt
  arXiv:0802.3537}}].

\bibitem{2008arXiv0810.2998E}
A.~M. {Essin}, J.~E. {Moore}, and D.~{Vanderbilt}, {\it {Magnetoelectric
  Polarizability and Axion Electrodynamics in Crystalline Insulators}},  {\em
  Phys.Rev.Lett.} {\bf 102} (Oct., 2008) 146805,
  [\href{http://xxx.lanl.gov/abs/0810.2998}{{\tt arXiv:0810.2998}}].

\bibitem{Qi27022009}
X.-L. Qi, R.~Li, J.~Zang, and S.-C. Zhang, {\it {Inducing a Magnetic Monopole
  with Topological Surface States}},  {\em Science} {\bf 323} (2009), no.~5918
  1184--1187.

\bibitem{Maciejko:2010tx}
J.~Maciejko, X.-L. Qi, A.~Karch, and S.-C. Zhang, {\it {Fractional Topological
  Insulators in Three Dimensions}},  {\em Phys.Rev.Lett.} {\bf 105} (2010)
  246809, [\href{http://xxx.lanl.gov/abs/1004.3628}{{\tt arXiv:1004.3628}}].

\bibitem{Swingle:2010rf}
B.~Swingle, M.~Barkeshli, J.~McGreevy, and T.~Senthil, {\it {Correlated
  Topological Insulators and the Fractional Magnetoelectric Effect}},  {\em
  Phys.Rev.} {\bf B83} (2011) 195139,
  [\href{http://xxx.lanl.gov/abs/1005.1076}{{\tt arXiv:1005.1076}}].

\bibitem{Maciejko:2011ed}
J.~Maciejko, X.-L. Qi, A.~Karch, and S.-C. Zhang, {\it {Models of
  Three-dimensional Fractional Topological Insulators}},  {\em Phys.Rev.} {\bf
  B86} (2012) 235128, [\href{http://xxx.lanl.gov/abs/1111.6816}{{\tt
  arXiv:1111.6816}}].

\bibitem{Karch:2009sy}
A.~Karch, {\it {Electric-Magnetic Duality and Topological Insulators}},  {\em
  Phys.Rev.Lett.} {\bf 103} (2009) 171601,
  [\href{http://xxx.lanl.gov/abs/0907.1528}{{\tt arXiv:0907.1528}}].

\bibitem{Thouless:1982zz}
D.~Thouless, M.~Kohmoto, M.~Nightingale, and M.~den Nijs, {\it {Quantized Hall
  Conductance in a Two-Dimensional Periodic Potential}},  {\em Phys.Rev.Lett.}
  {\bf 49} (1982) 405--408.

\bibitem{2007PhRvB..75l1306M}
J.~E. {Moore} and L.~{Balents}, {\it {Topological Invariants of
  Time-reversal-invariant Band Structures}},  {\em Phys.Rev.} {\bf B75} (Mar.,
  2007) 121306, [\href{http://xxx.lanl.gov/abs/{cond-mat/0607314}}{{\tt
  {cond-mat/0607314}}}].

\bibitem{2009PhRvB..79s5322R}
R.~{Roy}, {\it {Topological Phases and the Quantum Spin Hall Effect in Three
  Dimensions}},  {\em Phys.Rev.} {\bf {B79}} (May, 2009) 195322,
  [\href{http://xxx.lanl.gov/abs/{cond-mat/0607531}}{{\tt
  {cond-mat/0607531}}}].

\bibitem{2007PhRvL..98j6803F}
L.~{Fu}, C.~L. {Kane}, and E.~J. {Mele}, {\it {Topological Insulators in Three
  Dimensions}},  {\em Phys.Rev.Lett.} {\bf 98} (Mar., 2007) 106803,
  [\href{http://xxx.lanl.gov/abs/{cond-mat/0607699}}{{\tt
  {cond-mat/0607699}}}].

\bibitem{PhysRevLett.51.2077}
A.~J. Niemi and G.~W. Semenoff, {\it {Axial-Anomaly-Induced Fermion
  Fractionization and Effective Gauge-Theory Actions in Odd-Dimensional
  Space-Times}},  {\em Phys.Rev.Lett.} {\bf 51} (Dec, 1983) 2077--2080.

\bibitem{PhysRevLett.52.18}
A.~N. Redlich, {\it {Gauge Noninvariance and Parity Nonconservation of
  Three-Dimensional Fermions}},  {\em Phys.Rev.Lett.} {\bf 52} (Jan, 1984)
  18--21.

\bibitem{Ryu:2010ah}
S.~Ryu, J.~E. Moore, and A.~W. Ludwig, {\it {Electromagnetic and Gravitational
  Responses and Anomalies in Topological Insulators and Superconductors}},
  {\em Phys.Rev.} {\bf B85} (2012) 045104,
  [\href{http://xxx.lanl.gov/abs/1010.0936}{{\tt arXiv:1010.0936}}].

\bibitem{PhysRevLett.99.146806}
K.~Nomura, M.~Koshino, and S.~Ryu, {\it {Topological Delocalization of
  Two-Dimensional Massless Dirac Fermions}},  {\em Phys. Rev. Lett.} {\bf 99}
  (Oct, 2007) 146806.

\bibitem{PhysRevB.23.5632}
R.~B. Laughlin, {\it {Quantized Hall Conductivity in Two Dimensions}},  {\em
  Phys. Rev.} {\bf {B23}} (May, 1981) 5632--5633.

\bibitem{PhysRevB.25.2185}
B.~I. Halperin, {\it {Quantized Hall Conductance, Current-carrying Edge States,
  and the Existence of Extended States in a Two-dimensional Disordered
  Potential}},  {\em Phys. Rev.} {\bf {B25}} (Feb, 1982) 2185--2190.

\bibitem{2007PhRvB..76d5302F}
L.~{Fu} and C.~L. {Kane}, {\it {Topological Insulators with Inversion
  Symmetry}},  {\em Phys.Rev.} {\bf {B76}} (July, 2007) 045302,
  [\href{http://xxx.lanl.gov/abs/{cond-mat/0611341}}{{\tt
  {cond-mat/0611341}}}].

\bibitem{PhysRevB.78.195125}
A.~P. Schnyder, S.~Ryu, A.~Furusaki, and A.~W.~W. Ludwig, {\it {Classification
  of Topological Insulators and Superconductors in Three Spatial Dimensions}},
  {\em Phys. Rev.} {\bf {B78}} (Nov, 2008) 195125.

\bibitem{Ryuetalrev}
A.~P. Schnyder, S.~Ryu, A.~Furusaki, and A.~W.~W. Ludwig, {\it {Classification
  of Topological Insulators and Superconductors}},  {\em AIP Conf.Proc.} {\bf
  1134} (2009) 10--22.

\bibitem{Kitaev:2009mg}
A.~Kitaev, {\it {Periodic Table for Topological Insulators and
  Superconductors}},  {\em AIP Conf.Proc.} {\bf 1134} (2009) 22--30,
  [\href{http://xxx.lanl.gov/abs/0901.2686}{{\tt arXiv:0901.2686}}].

\bibitem{Ryu:2010zza}
S.~Ryu, A.~P. Schnyder, A.~Furusaki, and A.~Ludwig, {\it {Topological
  Insulators and Superconductors: Ten-fold Way and Dimensional Hierarchy}},
  {\em New J.Phys.} {\bf 12} (2010) 065010.

\bibitem{LeClair:2012qc}
A.~LeClair and D.~Bernard, {\it {Holographic classification of Topological
  Insulators and its 8-fold periodicity}},  {\em J.Phys.} {\bf A45} (2012)
  435203, [\href{http://xxx.lanl.gov/abs/1205.3810}{{\tt arXiv:1205.3810}}].

\bibitem{Hartnoll:2012ux}
S.~Hartnoll and D.~Radicevic, {\it {Holographic Order Parameter for Charge
  Fractionalization}},  {\em Phys.Rev.} {\bf D86} (2012) 066001,
  [\href{http://xxx.lanl.gov/abs/1205.5291}{{\tt arXiv:1205.5291}}].

\bibitem{1995RPPh...58..977V}
J.~{Voit}, {\it {One-dimensional Fermi Liquids}},  {\em Rep.Prog.Phys.} {\bf
  58} (Sept., 1995) 977--1116,
  [\href{http://xxx.lanl.gov/abs/{cond-mat/9510014}}{{\tt
  {cond-mat/9510014}}}].

\bibitem{1983PhRvL..50.1395L}
R.~B. {Laughlin}, {\it {Anomalous Quantum Hall Effect - An Incompressible
  Quantum Fluid with Fractionally Charged Excitations}},  {\em Phys.Rev.Lett.}
  {\bf 50} (May, 1983) 1395--1398.

\bibitem{PhysRevB.40.8079}
J.~K. Jain, {\it {Incompressible Quantum Hall States}},  {\em Phys. Rev.} {\bf
  {B40}} (Oct, 1989) 8079--8082.

\bibitem{doi:10.1142/S0217984991000058}
X.-G. Wen, {\it {Edge Excitations in the Fractional Quantum Hall States at
  General Filling Fractions}},  {\em Mod.Phys.Lett.} {\bf {B05}} (1991), no.~01
  39--46.

\bibitem{PhysRevLett.66.802}
X.-G. Wen, {\it {Non-Abelian Statistics in the Fractional Quantum Hall
  States}},  {\em Phys.Rev.Lett.} {\bf 66} (Feb, 1991) 802--805.

\bibitem{Blok1992615}
B.~Blok and X.-G. Wen, {\it {Many-body Systems with Non-Abelian Statistics}},
  {\em Nucl.Phys.} {\bf B374} (1992), no.~3 615 -- 646.

\bibitem{doi:10.1142/S0217979292000840}
X.-G. Wen, {\it {Theory of the Edge States in Fractional Quantum Hall
  Effects}},  {\em Int.J.Mod.Phys.} {\bf {B06}} (1992), no.~10 1711--1762.

\bibitem{PhysRevB.60.8827}
X.-G. Wen, {\it {Projective Construction of Non-Abelian Quantum Hall Liquids}},
   {\em Phys. Rev.} {\bf {B60}} (Sep, 1999) 8827--8838.

\bibitem{2010PhRvB..81o5302B}
M.~{Barkeshli} and X.-G. {Wen}, {\it {Effective Field Theory and Projective
  Construction for Z$_{k}$ Parafermion Fractional Quantum Hall States}},  {\em
  Phys.Rev.} {\bf B81} (Apr., 2010) 155302,
  [\href{http://xxx.lanl.gov/abs/0910.2483}{{\tt arXiv:0910.2483}}].

\bibitem{HoyosBadajoz:2010ac}
C.~Hoyos-Badajoz, K.~Jensen, and A.~Karch, {\it {A Holographic Fractional
  Topological Insulator}},  {\em Phys.Rev.} {\bf D82} (2010) 086001,
  [\href{http://xxx.lanl.gov/abs/1007.3253}{{\tt arXiv:1007.3253}}].

\bibitem{Ammon:2012dd}
M.~Ammon and M.~Gutperle, {\it {A Supersymmetric Holographic Dual of a
  Fractional topological insulator}},  {\em Phys.Rev.} {\bf D86} (2012) 025018,
  [\href{http://xxx.lanl.gov/abs/1204.2217}{{\tt arXiv:1204.2217}}].

\bibitem{Karch:2002sh}
A.~Karch and E.~Katz, {\it {Adding Flavor to AdS / CFT}},  {\em JHEP} {\bf
  0206} (2002) 043, [\href{http://xxx.lanl.gov/abs/hep-th/0205236}{{\tt
  hep-th/0205236}}].

\bibitem{Nakamura:2006xk}
S.~Nakamura, Y.~Seo, S.-J. Sin, and K.~Yogendran, {\it {A New Phase at Finite
  Quark Density from AdS/CFT}},  {\em J.Korean Phys.Soc.} {\bf 52} (2008)
  1734--1739, [\href{http://xxx.lanl.gov/abs/hep-th/0611021}{{\tt
  hep-th/0611021}}].

\bibitem{Kobayashi:2006sb}
S.~Kobayashi, D.~Mateos, S.~Matsuura, R.~Myers, and R.~Thomson, {\it
  {Holographic Phase Transitions at Finite Baryon Density}},  {\em JHEP} {\bf
  0702} (2007) 016, [\href{http://xxx.lanl.gov/abs/hep-th/0611099}{{\tt
  hep-th/0611099}}].

\bibitem{Kruczenski:2003be}
M.~Kruczenski, D.~Mateos, R.~Myers, and D.~Winters, {\it {Meson Spectroscopy in
  AdS / CFT with Flavor}},  {\em JHEP} {\bf 0307} (2003) 049,
  [\href{http://xxx.lanl.gov/abs/hep-th/0304032}{{\tt hep-th/0304032}}].

\bibitem{Davis:2008nv}
J.~Davis, P.~Kraus, and A.~Shah, {\it {Gravity Dual of a Quantum Hall Plateau
  Transition}},  {\em JHEP} {\bf 0811} (2008) 020,
  [\href{http://xxx.lanl.gov/abs/0809.1876}{{\tt arXiv:0809.1876}}].

\bibitem{Myers:2008me}
R.~Myers and M.~Wapler, {\it {Transport Properties of Holographic Defects}},
  {\em JHEP} {\bf 0812} (2008) 115,
  [\href{http://xxx.lanl.gov/abs/0811.0480}{{\tt arXiv:0811.0480}}].

\bibitem{PTPS.177.128}
S.-J. Rey, {\it {String Theory on Thin Semiconductors}},  {\em
  Prog.Theor.Phys.Supp.} {\bf 177} (2009) 128--142.

\bibitem{Wapler:2009tr}
M.~Wapler, {\it {Holographic Experiments on Defects}},  {\em Int.J.Mod.Phys.}
  {\bf A25} (2010) 4397--4473, [\href{http://xxx.lanl.gov/abs/0909.1698}{{\tt
  arXiv:0909.1698}}].

\bibitem{Wapler:2009rf}
M.~Wapler, {\it {Thermodynamics of Holographic Defects}},  {\em JHEP} {\bf
  1001} (2010) 056, [\href{http://xxx.lanl.gov/abs/0911.2943}{{\tt
  arXiv:0911.2943}}].

\bibitem{Kutasov:2011fr}
D.~Kutasov, J.~Lin, and A.~Parnachev, {\it {Conformal Phase Transitions at Weak
  and Strong Coupling}},  {\em Nucl.Phys.} {\bf B858} (2012) 155--195,
  [\href{http://xxx.lanl.gov/abs/1107.2324}{{\tt arXiv:1107.2324}}].

\bibitem{Aharony:1999ti}
O.~Aharony, S.~Gubser, J.~Maldacena, H.~Ooguri, and Y.~Oz, {\it {Large N Field
  Theories, String Theory and Gravity}},  {\em Phys.Rept.} {\bf 323} (2000)
  183--386, [\href{http://xxx.lanl.gov/abs/hep-th/9905111}{{\tt
  hep-th/9905111}}].

\bibitem{Henningson:1998gx}
M.~Henningson and K.~Skenderis, {\it {The Holographic Weyl Anomaly}},  {\em
  JHEP} {\bf 9807} (1998) 023,
  [\href{http://xxx.lanl.gov/abs/hep-th/9806087}{{\tt hep-th/9806087}}].

\bibitem{Balasubramanian:1999re}
V.~Balasubramanian and P.~Kraus, {\it {A Stress Tensor for Anti-de Sitter
  Gravity}},  {\em Commun.Math.Phys.} {\bf 208} (1999) 413--428,
  [\href{http://xxx.lanl.gov/abs/hep-th/9902121}{{\tt hep-th/9902121}}].

\bibitem{deHaro:2000xn}
S.~de~Haro, S.~Solodukhin, and K.~Skenderis, {\it {Holographic Reconstruction
  of Space-time and Renormalization in the AdS / CFT Correspondence}},  {\em
  Commun.Math.Phys.} {\bf 217} (2001) 595--622,
  [\href{http://xxx.lanl.gov/abs/hep-th/0002230}{{\tt hep-th/0002230}}].

\bibitem{Skenderis:2002wp}
K.~Skenderis, {\it {Lecture Notes on Holographic Renormalization}},  {\em
  Class.Quant.Grav.} {\bf 19} (2002) 5849--5876,
  [\href{http://xxx.lanl.gov/abs/hep-th/0209067}{{\tt hep-th/0209067}}].

\bibitem{Papadimitriou:2004rz}
I.~Papadimitriou and K.~Skenderis, {\it {Correlation Functions in Holographic
  RG Flows}},  {\em JHEP} {\bf 0410} (2004) 075,
  [\href{http://xxx.lanl.gov/abs/hep-th/0407071}{{\tt hep-th/0407071}}].

\bibitem{Skenderis:2006uy}
K.~Skenderis and M.~Taylor, {\it {Kaluza-Klein Holography}},  {\em JHEP} {\bf
  0605} (2006) 057, [\href{http://xxx.lanl.gov/abs/hep-th/0603016}{{\tt
  hep-th/0603016}}].

\bibitem{1995NuPhB.455..522M}
D.~M. {McAvity} and H.~{Osborn}, {\it {Conformal Field Theories Near a Boundary
  in General Dimensions}},  {\em Nucl.Phys.} {\bf B455} (Feb., 1995) 522--576,
  [\href{http://xxx.lanl.gov/abs/{cond-mat/9505127}}{{\tt
  {cond-mat/9505127}}}].

\bibitem{Freedman:2003ax}
D.~Freedman, C.~Nunez, M.~Schnabl, and K.~Skenderis, {\it {Fake Supergravity
  and Domain Wall Stability}},  {\em Phys.Rev.} {\bf D69} (2004) 104027,
  [\href{http://xxx.lanl.gov/abs/hep-th/0312055}{{\tt hep-th/0312055}}].

\end{thebibliography}\endgroup
\bibliographystyle{JHEP}

\end{document}